\documentclass{aa}  

\usepackage{graphicx}
\usepackage[varg]{txfonts}
\usepackage{natbib}
\bibpunct{(}{)}{;}{a}{}{,}

\usepackage{lscape}

\begin{document} 

   \title{Properties of galaxies at the faint end \\ of the H$\alpha$ luminosity function at $z\sim0.62$
   \thanks{Based on observations collected at the European Organisation for Astronomical Research in the Southern Hemisphere (ESO), Chile, Prog-Id 181.A-0485(A).}}
  
   \author{Carlos G{\'o}mez-Guijarro\inst{1,2}, Jes{\'u}s Gallego\inst{1}, V{\'i}ctor Villar\inst{1,3}, Luc{\'i}a Rodr{\'i}guez-Mu{\~n}oz\inst{1,4}, \\Benjamin Cl{\'e}ment\inst{5} \and Jean-Gabriel Cuby\inst{6}
   }

   \institute{Departamento de Astrof{\'i}sica y CC. de la Atm{\'o}sfera, Facultad de CC. F{\'i}sicas, Universidad Complutense de Madrid, Av. Complutense s/n, E-28040 Madrid, Spain\\ \email{cgguijarro@ucm.es}
   \and
   Dark Cosmology Centre, Niels Bohr Institute, University of Copenhagen, Juliane Maries Vej 30, DK-2100 Copenhagen, Denmark
   \and
   European Space Astronomy Centre, PO Box 78, E-28691 Villanueva de la Ca{\~n}ada, Madrid, Spain
   \and
   Dipartimento di Fisica e Astronomia "G. Galilei", Universit{\'a} di Padova, Vicolo dell'Osservatorio 3, I-35122, Italy
   \and
   CRAL, Observatoire de Lyon, Universit{\'e} Lyon 1, 9 Avenue Ch. Andr{\'e}, F-69561 Saint Genis Laval Cedex, France
   \and
   Aix Marseille Universit{\'e}, CNRS, LAM (Laboratoire d'Astrophysique de Marseille) UMR 7326, 13388, Marseille, France
   }

   \date{Received 15 June 2015; accepted 14 April 2016}

 
  \abstract
   {Studies measuring the star formation rate density, luminosity function, and properties of star-forming galaxies are numerous. However, it exists a gap at $0.5<z<0.8$ in H$\alpha$-based studies.}
   {Our main goal is to study the properties of a sample of faint H$\alpha$ emitters at $z\sim0.62$. We focus on their contribution to the faint end of the luminosity function and derived star formation rate density, characterising their morphologies and basic photometric and spectroscopic properties.}
   {We use a narrow-band technique in the near-infrared, with a filter centred at 1.06\,$\mu$m. The data come from ultra-deep VLT/HAWK-I observations in the GOODS-S field with a total of 31.9\,h in the narrow-band filter. In addition to our survey, we mainly make use of ancillary data coming from the CANDELS and Rainbow Cosmological Surveys Database, from the 3D-HST for comparison, and also spectra from the literature. We perform a visual classification of the sample and study their morphologies from structural parameters available in CANDELS. In order to obtain the luminosity function, we apply a traditional $V/V_{\mathrm{max}}$ method and perform individual extinction corrections for each object to accurately trace the shape of the function.}
   {Our 28 H$\alpha$-selected sample of faint star-forming galaxies reveals a robust faint-end slope of the luminosity function $\alpha=-1.46_{-0.08}^{+0.16}$. The derived star formation rate density at $z\sim0.62$ is $\rho_\mathrm{SFR} = 0.036_{-0.008}^{+0.012}\,M_{\odot}~\mathrm{yr^{-1}~Mpc^{-3}}$. The sample is mainly composed of disks, but an important contribution of compact galaxies with S{\'e}rsic indexes $n\sim2$ display the highest specific star formation rates.}
    {The luminosity function at $z\sim0.62$ from our ultra-deep data points towards a steeper $\alpha$ when an individual extinction correction for each object is applied. Compact galaxies are low-mass, low-luminosity, and starburst-dominated objects with a light profile in an intermediate stage from early to late types.}

   \keywords{galaxies: fundamental parameters -- galaxies: evolution -- galaxies: star formation -- galaxies: luminosity function, mass function}
   
   \titlerunning{Properties of galaxies at the faint end of the H$\alpha$ luminosity function at $z\sim0.62$}
   \authorrunning{G{\'o}mez-Guijarro et al.}

   \maketitle
%

\section{Introduction}
\label{sec:intro}
The evolution of the luminosity function (LF) and the star formation rate (SFR) density across cosmic time has been an object of study over the last two decades. Since it was first addressed by the pioneering studies of \citet{1995ApJ...455L...1G}; \citet{1996ApJ...460L...1L}; and \citet{1996MNRAS.283.1388M}, a wide view of the cosmic star formation history (SFH) has been revealed. The SFR density (SFRD) is now traced out to $z\sim10$. The star formation activity grew from the formation of the first galaxies and reached a peak at $z\sim2$. Then, the activity declined gradually until $z\sim0$, when it reached values a factor of 10 smaller. Extensive compilations and reviews on this topic have been published over the years \citep{2004ApJ...615..209H,2006ApJ...651..142H,2013ApJ...770...57B,2014ARA&A..52..415M}.

Diverse tracers have been used to infer SFRs \citep{1998ARA&A..36..189K,2012ARA&A..50..531K}. These indicators span a very broad wavelength range from gamma-ray/X-ray to radio. Among them, nebular emission lines such as [\ion{O}{ii}]$\lambda$3727, [\ion{O}{iii}]$\lambda$5007, H$\alpha$, and H$\beta$ have been frequently employed. The SFH based on all these tracers contribute to a comprehensive and consistent view on the evolution of star formation. However, there are discrepancies between the results obtained with different indicators at the same redshift, which are due to the intrinsic differences between galaxies selected with distinct criteria and between calibrations, the impact of dust extinction and metallicities, or observational biases. Particularly, the H$\alpha$ line has been widely used as an excellent method for measuring star formation \citep{1995ApJ...455L...1G,1998ARA&A..36..189K,2004MNRAS.351.1151B,2012ARA&A..50..531K}. Because it is sensitive to stellar masses $\geq15\,M_{\odot}$ with ages $\sim$ 3--10\,Myr, H$\alpha$ is considered a direct tracer of the instantaneous SFR in a galaxy \citep{2012ARA&A..50..531K}. The H$\alpha$ line is strong in the optical/near-infrared (NIR) range, it is less affected by dust extinction than other indicators like the ultraviolet (UV) continuum, and it is not very sensitive to metallicity. Nevertheless, this tracer presents some drawbacks. It is currently restricted to redshifts $z\leq2.23$. Over this value the line is redshifted beyond the wavelength limit of NIR spectrographs. In addition, the low efficiency in the transition from optical to NIR detectors makes the emission line virtually unreachable in the redshift range $0.5<z<0.8$. On the other hand, extinction-corrected H$\alpha$ SFRs are consistent with those derived from the total infrared (TIR) to far-ultraviolet (FUV) luminosity ratio (IRX) \citep{2011ApJ...741..124H}. The H$\alpha$ emission is tested as an exceptional way to measure star formation in the limit $z\sim2$ \citep{2015MNRAS.452.2018O}.

The existing literature in SFRD and LF measurements from H$\alpha$ is extensive. \citet{2013MNRAS.433.2764G} shows an exhaustive compilation of the SFRD results that use this tracer at different redshifts. Since then, new values of H$\alpha$-based SFRDs have been published \citep{2013MNRAS.428.1128S,2013MNRAS.433..796D,2015MNRAS.447..875G,2015MNRAS.451.2303S,2015MNRAS.453..242S}.

Among these studies, two different techniques are employed: spectroscopic and narrow-band surveys. These techniques have advantages and drawbacks. Narrow-band surveys provide deep imaging in a narrow region in terms of redshift, selecting the candidates directly from the emission line under study. In addition, these filters are more effective in the detection of weak sources. On the other hand, there are disadvantages, such as dust obscuration, active galactic nuclei (AGN) contamination, insensibility to low equivalent widths, and narrow redshift ranges that suffer from density variations due to different effects in the line of sight. Multislit spectroscopy can provide wide sky coverage and so it does not suffer from effects of cosmic variance or small statistics. Nevertheless, it is subject to slit signal losses and atmospheric effects that can limit the redshift access and can introduce undesirable selection effects. Slitless spectroscopy can avoid these concerns, but proper observations are time expensive.

In the optical regime, the work by \citet[see also \citealt{2003ApJ...591..827P}]{1995ApJ...455L...1G} first measured the local SFRD and LF using the Universidad Complutense de Madrid (UCM) Survey \citep{1994ApJS...95..387Z,1996ApJS..105..343Z}, an optical low-dispersion objective-prism survey for low-redshift emission-line galaxies (ELGs). Similar values were calculated at $z\sim0$ from the SDSS \citep{2004MNRAS.351.1151B,2004AJ....127.2511N} and SINGG \citep{2006ApJ...649..150H} projects, and from local volume studies \citep{2008A&A...482..507J,2010AJ....140.1241K}. Several authors extended local studies to redshifts $z<0.3$ using optical spectroscopy \citep{1998ApJ...495..691T,2000MNRAS.312..442S}. The SHELS project analysed several slices in a redshift range up to $z<0.4$ with a great number of spectra \citep{2010ApJ...708..534W}. \citet{2013MNRAS.433.2764G,2015MNRAS.447..875G} have recently combined SDSS data with deeper optical spectroscopy from the GAMA survey to $z<0.35$. Analogously, other authors obtained SFRDs with a narrow-band technique at $z\sim0.24$ \citep{2001A&A...379..798P,2003ApJ...586L.115F,2003A&A...402...65H,2005PASP..117..120P,2008MNRAS.383..339W,2008ApJS..175..128S,2008PASJ...60.1219M,2010ApJ...708..534W} and $z\sim0.4$ \citep{2004AJ....128.2652G,2005PASP..117..120P}. Other surveys collected large samples of objects with this technique within redshift slices up to $z<0.4$ \citep{2007ApJ...657..738L,2010ApJ...712L.189D}. \citet{2013MNRAS.433..796D} has recently studied the cosmic SFH from H$\alpha$ in the redshift bins $z\sim0.25,0.4,0.5$ with a total of 564 objects in the SXDS-UDS field.

In the NIR regime, object-by-object spectroscopic observations have provided important results in the ranges $0.79<z<1.1$ with 13 galaxies \citep{1999MNRAS.306..843G}, $0.5<z<1.1$ VLT/ISAAC observations of 30 galaxies \citep{2002MNRAS.337..369T}, or $0.77<z<1$ spectra of 38 objects \citep{2006MNRAS.370..331D}. Several studies employed \emph{HST}/NICMOS grism spectroscopy at $0.7<z<1.9$ \citep{1999ApJ...519L..47Y,2000AJ....120.2843H,2009ApJ...696..785S}. Narrow-band NIR works include redshift windows that are less affected by atmospheric features, such as $z\sim0.84$ \citep{2008ApJ...677..169V,2009MNRAS.398...75S,2011ApJ...726..109L}, $z\sim1.47$ \citep{2012MNRAS.420.1926S}, and $z\sim2.23$ \citep{2000A&A...362....9M,2008MNRAS.388.1473G,2010A&A...509L...5H,2011PASJ...63S.437T}. The HiZELS survey has recently combined narrow-band data at common redshift windows $z\sim0.4,0.84,1.47,2.23$ and has obtained a view of the last 11\,Gyr of evolution of star-forming galaxies (SFGs) by employing a homogeneously selected sample of H$\alpha$ emitters \citep{2013MNRAS.428.1128S}. They have measured a constant value of the faint-end slope of the LF with redshift $\alpha\sim-1.6$ and a steady increase of the characteristic luminosity of the LF with redshift. The CF-HiZELS survey has surveyed $\sim10$\,deg$^2$ deriving a sample of 3471 H$\alpha$ emitters at $z\sim0.8$ \citep{2015MNRAS.451.2303S}.

Our group measured the SFRD and LF locally \citep{1995ApJ...455L...1G,2003ApJ...591..827P} from a sample of H$\alpha$ emitters selected with an objective-prism survey, the UCM Survey \citep{1994ApJS...95..387Z,1996ApJS..105..343Z}. We then extended the study to $z\sim0.24$ and $z\sim0.4$ with an optical narrow-band technique \citep{2001A&A...379..798P,2005PASP..117..120P}. More recently, we applied this technique in the NIR at $z\sim0.84$ \citep{2008ApJ...677..169V}. In the mentioned redshift gap $0.5<z<0.8$ in H$\alpha$-based studies, we only find a few objects in the lower limits of the NIR spectroscopic surveys indicated above. These limits are $z>0.5$ \citep{2002MNRAS.337..369T}, $z>0.77$ \citep{2006MNRAS.370..331D}, and $z>0.79$ \citep{1999MNRAS.306..843G}. No narrow-band surveys cover this gap. In this work we use a narrow-band filter centred at 1.061\,$\mu$m with the main goal of filling the $z\sim0.6$ gap.

The layout of the paper is as follows. We describe the data and introduce the sample in Sect.~\ref{sec:data}. In Sect.~\ref{sec:prop} we characterise the sample properties: morphologies and stellar masses, followed by the calculation of line fluxes, luminosities, dust extinction, and a spectrocopic study. We obtain the H$\alpha$-based LF for our sample at $z\sim0.62$ in Sect.~\ref{sec:half}. We include a discussion in Sect.~\ref{sec:disc} and we summarise in Sect.~\ref{sec:sum}.

Throughout this work we adopt a concordance cosmology $[\Omega_\Lambda,\Omega_M,h]=[0.7,0.3,0.7]$. Considering these parameters, the age of the universe at $z=0.62$ is 7.63\,Gyr, the luminosity distance is 3673.5\,Mpc, and the scale 6.79\,kpc/\arcsec. An AB magnitude system is employed over the whole study \citep{1974ApJS...27...21O}.

\section{Data and sample selection}
\label{sec:data}

\subsection{Narrow-band data: the VLT/HAWK-I survey}
\label{sec:nbdata}
The main data used in this work correspond primarily to very deep NIR imaging observations with the HAWK-I instrument at the VLT. HAWK-I \citep{2004SPIE.5492.1763P,2006SPIE.6269E..0WC,2008A&A...491..941K,2011Msngr.144....9S} is a NIR (0.85--2.5\,$\mu$m) wide-field imager installed at the Nasmyth A focus of ESO VLT UT4. The field of view is 7.5\arcmin$\times$7.5\arcmin. It is composed of four HAWAII 2RG detectors of 2048$\times$2048 pixels. The pixel scale is 0.106\arcsec/pix.

A narrow-band (NB) filter centred at 1.061\,$\mu$m (NB1060) was employed. The data was obtained as part of an ESO large programme between September 2008 and April 2010 (project 181.A-0485(A), \citet{2012A&A...538A..66C}), that was devised to detect Ly$\alpha$ emitters (LAEs) at $z=7.7$ in order to constrain the epoch of reionization. In this programme, four deep fields corresponding to blank and cluster regions were selected. Among them, the GOODS-S blank field was chosen because of the great amount of multiwavelength data publicly available. For GOODS-S, the survey took a total integration time in the NB of 31.9\,h and 3.3\,h in the $J$ band. \citet{2015A&A...580A..42K} has recently published a study focused on the main sequence of star-formation using ELGs at $z=0.6-2$ within this large programme data. In our work, we make use of these deep GOODS-S images to extract a sample of H$\alpha$ emitters at $z\sim0.62$. We then combine our NB photometry with ancillary data to study the properties of these systems.

\subsection{Sample selection}
\label{sec:selection}
We selected emission-line objects by comparing apparent fluxes in narrow-band and broad-band (BB) images. When an emission line falls within the NB filter, it produces an excess in this filter respect to the BB flux.

The same approach that was used to detect LAEs at $z=7.7$ employing the NB1060 filter was found to be useful when selecting H$\alpha$ emission-line objects at $z\sim0.62$. In particular, given the width of the NB filter, we cover a redshift range $0.6098<z<0.6263$. Because the NB1060 central wavelength 1.061\,$\mu$m is placed in the $Y$ band, we employed $Y$ band as our BB filter. However, the transmittance of the $Y$-band filter starts to decrease at 1.06\,$\mu$m. In order to provide a proper account of the continuum flux, we interpolated its value with the contiguous $J$-band filter that was also available.

In detail, the selection process of the candidates was performed through a colour-magnitude diagram using the technique described in \citet{2007PASP..119...30P} and applied by our group in \citet{2008ApJ...677..169V}. The criterion used was
\begin{equation}
 \left(m_{\mathrm{BB}} - m_{\mathrm{NB}}\right) > \mu\left(m_{\mathrm{BB}} - m_{\mathrm{NB}}\right) + n_{\sigma}\sigma\left(m_{\mathrm{BB}} - m_{\mathrm{NB}}\right) , \label{eq:color_sel}
\end{equation}
\noindent where $m_{\mathrm{NB}}$ and $m_{\mathrm{BB}}$ are the apparent magnitudes of the NB and BB filters, respectively, $\mu$ is an offset parameter that indicates the average deviation from the zero colour, $\sigma$ is the standard deviation of the colour distribution, and $n_{\sigma}$ is the level of significance. The values of $\mu$ and $\sigma$ were calculated from the distribution of objects. These parameters are functions of the NB magnitude. Therefore, we obtained a curve that depends on the NB magnitude and the objects above it were selected as emission-line candidates. The fluxes in each band were measured using circular apertures of different sizes. The sample selection employed 9-10 apertures ranging from the PSF FWHM to 5 times this value.

We wanted to select emission-line objects at $z\sim0.62$ avoiding non-emission line and redshift interlopers as much as possible. Thus, we needed to set the level of significance and the range of apertures that returned the best results.

We created several selection curves with $n_{\sigma}$ from 1.5 to 3.0 in steps of 0.25. In each case, the defined sample was different. If the lowest $n_{\sigma}$ was employed, the largest number of candidates was recovered, but many of them could be non-emission interlopers. On the other hand, if the level of significance grew, the returned sample was smaller, because the selection criterion became more restricted. We wanted to obtain the maximum number of candidates while maintaining the accuracy. Finally, we chose $n_{\sigma}=2.5$. \citet{2008ApJ...677..169V} demostrated that a good compromise between the number of objects and accuracy is obtained with this level of significance, a value also applied in other studies \citep{2009MNRAS.398...75S}.

In order to include as many emission-line emitters as possible, we needed to use several apertures. On the one hand, the smallest apertures detected small, low-luminosity emission-line objects and bright candidates with high nuclear star formation. The fluxes for these candidates were less affected by sky emission. On the other hand, the largest apertures were better at selecting large, low surface-brightness objects with extended star formation. Large apertures for small emitters were not considered since the sky noise was high in these cases.

Once we selected the candidate emission-line emitters, we determined whether the object is a galaxy or a star. This segregation was carried out with the stellarity parameter from SExtractor. Objects with a stellarity parameter greater than 0.95 were considered stars.

We recovered a total raw sample of 46 candidate emission-line emitters. The lowest emission line equivalent width (EW) that we selected was $\sim6$\,\AA. This selection result is in agreement with the sample of 58 candidates in \citet{2015A&A...580A..42K}.

\subsection{Incompleteness}
\label{sec:incompleteness}
The selection process described above suffers from selection effects, and recovers a different number of objects depending on the emission-line flux. In this kind of study this effect is commonly known as the incompleteness factor. The real number of emission-line emitters is higher than the number detected, and even higher for the faintest objects. In other words, the number of emission-line emitters is underestimated, and this effect is particularly important for the faintest sources. To take this effect into account, we used the results that our group obtained in \citet{2008ApJ...677..169V}. In this work, a sample of simulated objects was introduced in the data, applying the same selection technique and measuring the fraction of objects recovered over the total as a function of the emission-line flux \citep[see][for details]{2008ApJ...677..169V}.

\subsection{Photometric ancillary data}
\label{sec:photdata}
The CANDELS survey \citep{2011ApJS..197...35G,2011ApJS..197...36K} is a powerful imaging survey of the distant universe carried out with the \emph{HST}. It is designed to study galactic evolution from $z=8$ to 1.5. To accomplish this objective it employs deep imaging of more than 250\,000 galaxies with the WFC3/IR and ACS instruments.

For the present work we were interested in the advanced data products that the CANDELS project had released in the GOODS-S field. The CANDELS GOODS-S Multiwavelength catalogue \citep{2013ApJS..207...24G} presents the multiwavelength photometry available in this field. It combines the newly obtained data from \emph{HST}/WFC3 F105W, F125W, and F160W bands with previous ground-based public data. This catalogue is based on source detection in the WFC3 F160W mosaic, that includes the data from the CANDELS deep and wide programmes. The result is a catalogue of 34\,930 sources in an area of 173\,arcmin$^2$. In addition to the WFC3 bands, this catalogue includes UV data ($U$ band from CTIO/MOSAIC and VLT/VIMOS), optical (\emph{HST}/ACS F435W, F606W, F775W, F814W and F850LP), and infrared (IR) (\emph{HST}/WFC3 F098M, VLT/ISAAC $Ks$, VLT/HAWK-I $Ks$, and \emph{Spitzer}/IRAC 3.6, 4.5, 5.8, 8.0\,$\mu$m). In addition to the photometric catalogue, the stellar masses have been recently published \citep{2015ApJ...801...97S} and the structural parameters are available \citep{2012ApJS..203...24V}.

Furthermore, we used complementary mid-IR photometry in \emph{Spitzer}/MIPS 24 and 70\,$\mu$m (30\,$\mu$Jy and 1\,mJy, $5\sigma$) from \citet{2005ApJ...630...82P,2008ApJ...687...50P}, and far-IR photometry from \citet{2010A&A...518L..15P}, the GOODS-\emph{Herschel} survey \citep{2011A&A...533A.119E}, and the PEP survey \citep{2013A&A...553A.132M} to characterise the TIR luminosity of the galaxies. A description of the method used to derive consistent mid-to-far IR spectral energy distributions (SEDs) and rest-frame luminosities was presented in \citet{2008ApJ...675..234P} and \citet{2011ApJS..193...30B}. The mid- and far-IR fluxes and derived TIR luminosities for the sources in the CANDELS catalogue presented in \citet{2013ApJS..207...24G} are publicly available through the Rainbow Cosmological Surveys Database \citep{2008ApJ...675..234P,2011ApJS..193...13B,2011ApJS..193...30B}\footnote{https://rainbowx.fis.ucm.es/Rainbow\_navigator\_public/}. The Rainbow Database also includes rest-frame near-ultraviolet (NUV) and FUV luminosities based on synthetic photometry \citep[see][for details]{2008ApJ...675..234P,2011ApJS..193...13B,2011ApJS..193...30B}.

We used these sources to collect several measurements required for our study: astrometry, photometric redshifts (photo-$z$'s), stellar masses, structural parameters, photometry, rest-frame absolute magnitudes, synthetic TIR luminosity, and synthetic FUV and NUV magnitudes. We refer to TIR synthetic luminosity as that calculated via integration of SED from 8.0\,$\mu$m to 1000\,$\mu$m. The FUV and NUV synthetic magnitudes are those obtained integrating the SED over the filter profiles.

We found counterparts in the CANDELS database for 42 objects out of the 46 candidate emission-line emitters in the original raw sample. They were confirmed as ELGs by the stellarity parameter, photo-$z$'s, and visual inspection of the postage stamps. The four objects with no CANDELS identification are those with reference numbers HAWKI0003148, HAWKI0003151, HAWKI0003361, and HAWKI0003606 in our original catalogue. If we try to visualise them in the CANDELS multicolour Interactive Display\footnote{http://archive.stsci.edu/prepds/candels/display\_gs-tot\_v1.0.html}, we find that HAWKI0003148 and HAWKI0003151 are located beyond the images edges. HAWKI0003361 is either a star or it is contaminated by a neighbouring object. HAWKI0003606 does not appear on the CANDELS images, thus, it is an artefact or a very low-luminosity line emitter with no obvious continuum (apparent line flux from the NB image was $4.5 \times 10^{-18}$\,erg s$^{-1}$ cm$^{-2}$). This last object is an extremely interesting candidate low-luminosity SFG with a very high EW. A more detailed analysis in the future will be needed to reveal its nature.

\subsection{Spectroscopic ancillary data}
\label{sec:specdata}
In order to complete the information of the sample, we carried out a search of the spectroscopic data available in existing published studies. First, we performed a complete search in the ESO\footnote{http://archive.eso.org/cms.html}, Centre de Donn{\'e}es astronomiques de Strasbourg (CDS)\footnote{http://vizier.u-strasbg.fr/viz-bin/VizieR}, and The VIMOS VLT Deep Survey (VVDS)\footnote{http://cesam.lam.fr/vvdsproject/} archives, and also throughout the literature in the GOODS-S field.

Among the 46 objects selected in the raw sample, 13(28\%) have a spectroscopic redshift (spec-$z$) determined. Seven of these targets have a FITS optical spectrum available in the archives mentioned above \citep{2004A&A...428.1043L,2008A&A...478...83V,2010A&A...512A..12B}. Five more galaxies have spectra measured from the ACES survey \citep{2012MNRAS.425.2116C}. We analysed the FITS spectra and we remeasured the emission lines that we can identify. In Fig.~\ref{fig:spectra} we show the collected spectra along with the identified emission lines and some basic data about the galaxies, including the spec-$z$'s.

In addition, we looked for additional spectroscopic data in the 3D-HST survey \citep{2014ApJS..214...24S}, a NIR spectroscopic survey designed to study the physical processes that shape galaxies in the universe at $z>1$. Although the instrumental set-up of this programme is very badly suited to observe emission-line emitters at 1.06\,$\mu$m, its depth was able to provide spec-$z$'s for some objects in our sample. We found six counterparts for which we had already found a spec-$z$. Four of them show exactly the same value and the other two correspond to ancillary spectra with high quality flag, so we decided to maintain the values of the ground-based spectra and to discard the 3D-HST data. Moreover, 3D-HST provides a determination of the photo-$z$'s and stellar masses. Comparing them with the CANDELS data we did not find deviations out of the uncertainties except for two objects: CANDELS 12039 = 3D-HST 23397 and CANDELS 18444 = 3D-HST 3408. We decided to keep the CANDELS values except in the two discrepant cases for which the 3D-HST values appeared more suitable to the colours and morphologies these galaxies show in their postage stamps. Actually, these two systems are characterised by blue colours and low masses, properties that make the photo-$z$ estimation more difficult. In particular, they show none or a small 4000\,\AA break feature and the photometry beyond the F160W band is close to detection limits. It is worth looking at grism data for this kind of objects with signs of emission lines with high EW.

\setcounter{figure}{0}
\begin{figure*}
\centering
\includegraphics[width=0.95\textwidth]{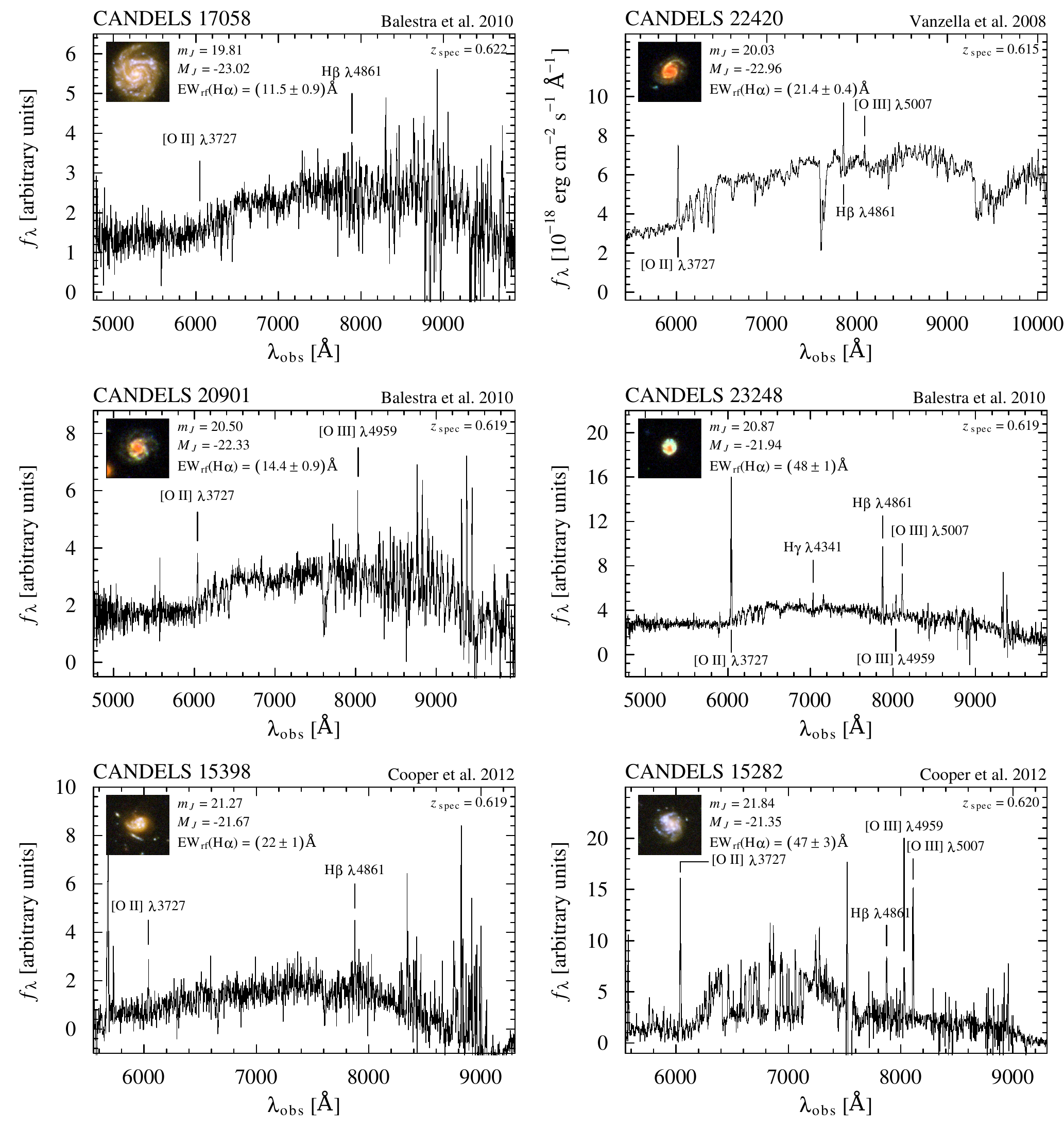}
\caption{Spectra collected from the literature. Among the 46 objects selected in the raw sample, 12(26\%) have a FITS optical spectrum available. Each spectrum shows the CANDELS ID and the reference on top; an RGB image built using ACS $b$, $v$, and $z$ bands with north up, east left, and a size of 5\arcsec$\times$5\arcsec; the spec-$z$ taken from the reference; observed $J$-band apparent magnitude; rest-frame $J$-band absolute magnitude; H$\alpha$ rest-frame EW (observed EW is 1+$z$ times the rest-frame value); and the identification of the emission lines that we can measure.}
\label{fig:spectra}
\end{figure*}

\setcounter{figure}{0}
\begin{figure*}
\centering
\includegraphics[width=0.95\textwidth]{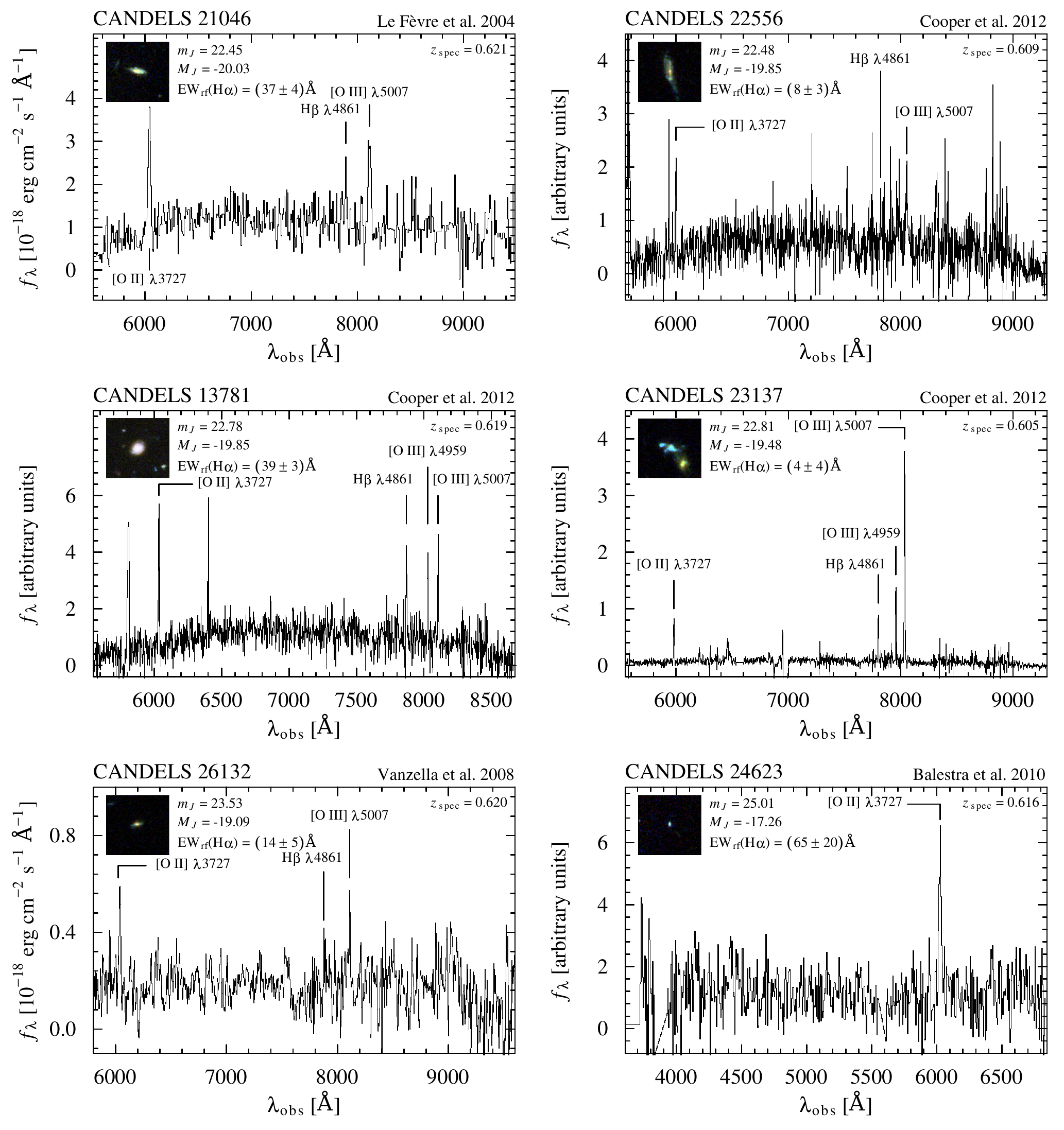}
\caption{continued.}
\label{fig:spectra}
\end{figure*}

\subsection{Photometric redshifts}
\label{sec:photoz}
One of the products we took from CANDELS were the photo-$z$'s. We accessed them through the stellar masses catalogue \citep{2015ApJ...801...97S}. The final CANDELS photo-$z$ catalogue in the GOODS-S field is not available yet, although a preliminary study on photo-$z$ estimation methods has been published \citep{2013ApJ...775...93D}. This work consists of a comparison between several photo-$z$ codes using a control sample of accurate spec-$z$'s. In practice, the authors test the dependence of photo-$z$'s on code and SED templates. They conclude that there is no particular code or SED template library that provides a better determination of the photo-$z$'s. Finally, they combine the results and obtain lower photo-$z$ scatter. Our aim is to use the photo-$z$'s as a way to discriminate between H$\alpha$ ELGs or interlopers at other redshifts when we lack spec-$z$'s (see Sect.~\ref{sec:haemitters}).

For our sample, we compared the CANDELS photo-$z$'s with the available spec-$z$'s from the literature. Twelve out of the 13 objects with spec-$z$ are classified at $z\sim0.6$ by the photo-$z$ and the spec-$z$ confirms them as H$\alpha$ emitters at $z\sim0.62$. The object CANDELS 16483 has photo-$z=1.91$ and spec-$z=1.17$, being confirmed as an H$\beta$ emitter at $z\sim1.17$. This means that the accuracy of the photo-$z$'s is good enough for our purpose of distinguishing between H$\alpha$ emitters and interlopers at other redshifts.

\subsection{H$\alpha$ emitters and interlopers at other redshifts}
\label{sec:haemitters}
With the collected spec-$z$'s and the CANDELS photo-$z$'s, we estimated the number of different ELGs among the 42 objects with a CANDELS counterpart. Photo-$z$'s suggest that the number of H$\alpha$ emitters at $z\sim0.62$ are 28(67\%). Among them 12(29\%) are spectroscopically confirmed. In addition, one is a (2\%) [\ion{S}{iii}]$\lambda$9069 emitter at $z\sim0.2$ (CANDELS 17704), eight (19\%) H$\beta$ or [\ion{O}{iii}]$\lambda\lambda$4959,5007 at $z\sim1.1$, and three (7\%) [\ion{O}{ii}]$\lambda$3727 at $z\sim1.8$ (CANDELS 14122, 17075, and 17686). One of the H$\beta$ emitters (CANDELS 16483) is spectroscopically confirmed and it appears at the ACS images as a blue compact object (see Fig.~\ref{fig:multifp}). Finally, two objects (5\%) have photometric redshift $z>2.5$ (CANDELS 11513 and 18444). We summarise this information in Table~\ref{tab:emitters}.

\begin{table}
\caption{Summary of numbers and percentages of H$\alpha$ emitters and interlopers at other redshifts within the 42 galaxies with CANDELS counterparts.}
\label{tab:emitters}
\centering
\begin{tabular}{l c c c}
\hline\hline
Line\tablefootmark{a} & $z$ & Number & Percentage \\
\hline
	H$\alpha$ & $\sim0.62$ & 28(12) & 67\%(19\%) \\
	$[$\ion{S}{iii}$]$ & $\sim0.2$ & 1 & 2\% \\
	H$\beta$ & $\sim1.17$ & 8(1) & 19\%(2\%) \\
	$[$\ion{O}{ii}$]$ & $\sim1.8$ & 3 & 7\% \\
	(...) & $>2.5$ & 2 & 5\% \\
\hline
\end{tabular}
\tablefoot{
The numbers and percentages of the spectroscopically confirmed objects are in parentheses.\\
\tablefoottext{a}{Simplified notation: [\ion{S}{iii}]$\lambda$9069 emitters at $z\sim0.2$ are indicated as [\ion{S}{iii}]; H$\beta$ or [\ion{O}{iii}]$\lambda\lambda$4959,5007 at $z\sim1.1$ as H$\beta$; [\ion{O}{ii}]$\lambda$3727 at $z\sim1.8$ as [\ion{O}{ii}].}
}
\end{table}

\begin{figure}
\resizebox{\hsize}{!}{\includegraphics{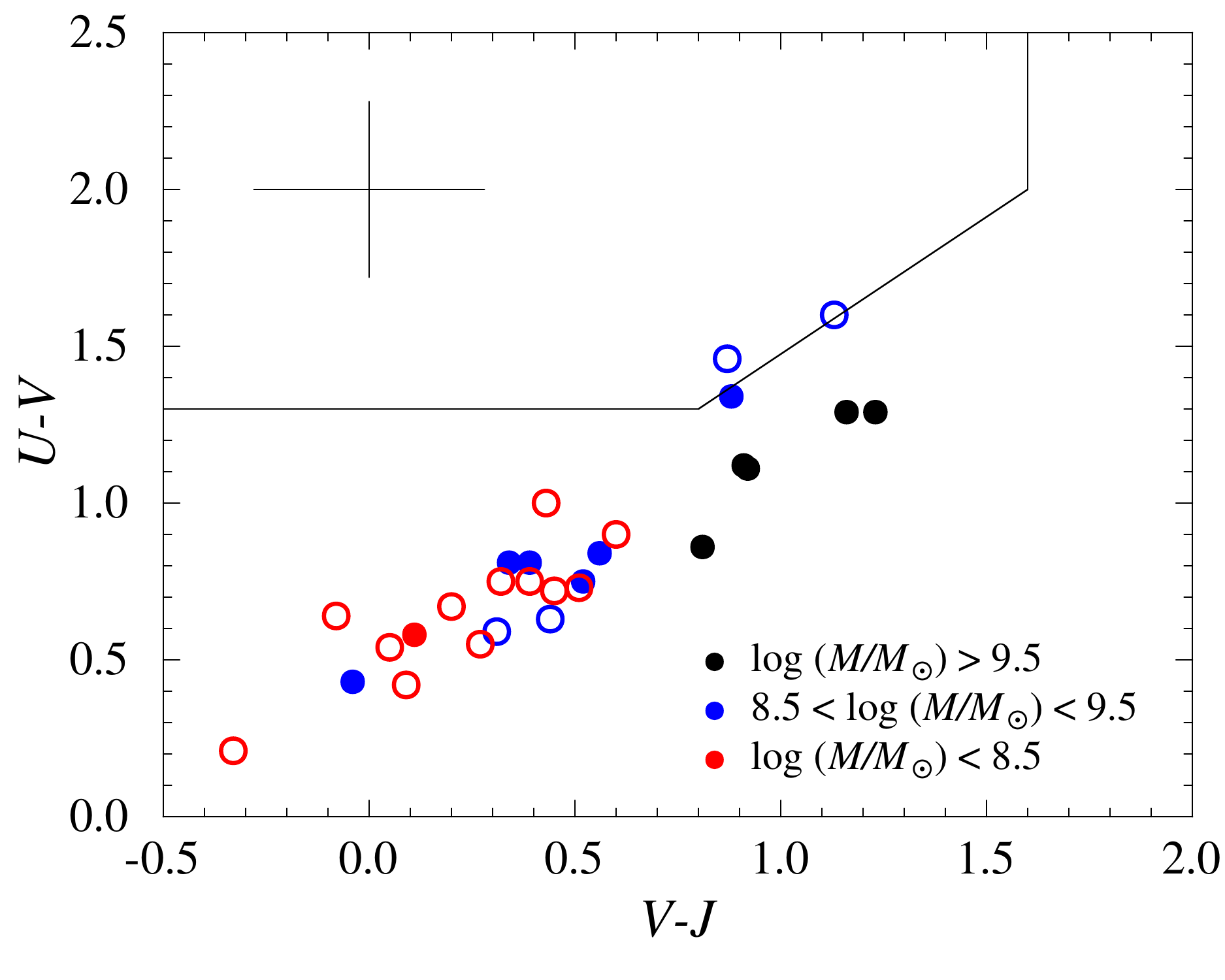}}
\caption{$UVJ$ diagram, rest-frame $U-V$ vs. $V-J$ for our sample of 28 H$\alpha$ SFGs at $z\sim0.62$. The black edge boundary from \citet{2009ApJ...691.1879W} separates quiescent galaxies (top left) from SFGs. The sample is divided into three mass categories as explained in the legend. Filled circles indicate those galaxies with available spectra. Average error bars are in the upper left corner.}
\label{fig:uvj}
\end{figure}

In Fig.~\ref{fig:uvj} we present a rest-frame $U-V$ versus $V-J$ diagram (i.e. the $UVJ$ diagram). This diagram is used in many studies to discriminate between SFGs and quiescent galaxies \citep{2009ApJ...691.1879W,2012ApJ...748L..27P}. We took the rest-frame absolute magnitudes from the CANDELS stellar masses catalogue \citep{2015ApJ...801...97S}. The catalogue does not provide uncertainties, so we assumed a 0.2\,mag photometric uncertainty in the absolute magnitudes, although in SED-fitting derived magnitudes of this kind, the main source for errors is the model assumption. With this diagram, we checked that our selection of H$\alpha$ emitters at $z\sim0.62$ lie in the SFG region within the uncertainties. We split the sample into three mass bins: $\log {(M/M_{\odot})}<8.5$, $8.5<\log {(M/M_{\odot})}<9.5$, and $\log {(M/M_{\odot})}>9.5$. Low-mass galaxies have lower $U-V$ and $V-J$ colours. In contrast, high-mass systems have higher $U-V$ and $V-J$ colours. These colours are an indication that low-mass galaxies have an important young stellar population that dominates the emission and high-mass are more dominated by an older stellar population.

\subsection{AGN contribution}
\label{sec:agn}
We checked the sample by searching for an AGN contribution. If we detected any AGN we rejected them in the final sample as we only wanted to study the contribution of SFGs. We performed a search in the CDF-S field through the 4\,Ms source catalogues \citep{2011ApJS..195...10X}, the deepest \emph{Chandra} survey to date. We only found X-ray detection for the object CANDELS 20901, with absorption-corrected rest-frame 0.5--8\,keV luminosity of $3.8 \times 10^{41}$\,erg s$^{-1}$. This object is not classified as an AGN by the 4\,Ms catalogues. Furthermore, the spectroscopic data available does not indicate an AGN contribution in this galaxy (see Fig.~\ref{fig:spectra} and Sec.~\ref{sec:spec}). Consequently, we have no apparent AGN contribution in our sample of ELGs.

\section{Properties of H$\alpha$ star-forming galaxies}
\label{sec:prop}

\subsection{Morphologies}
\label{sec:morph}
\begin{figure*}
\centering
\includegraphics[width=0.8\textwidth]{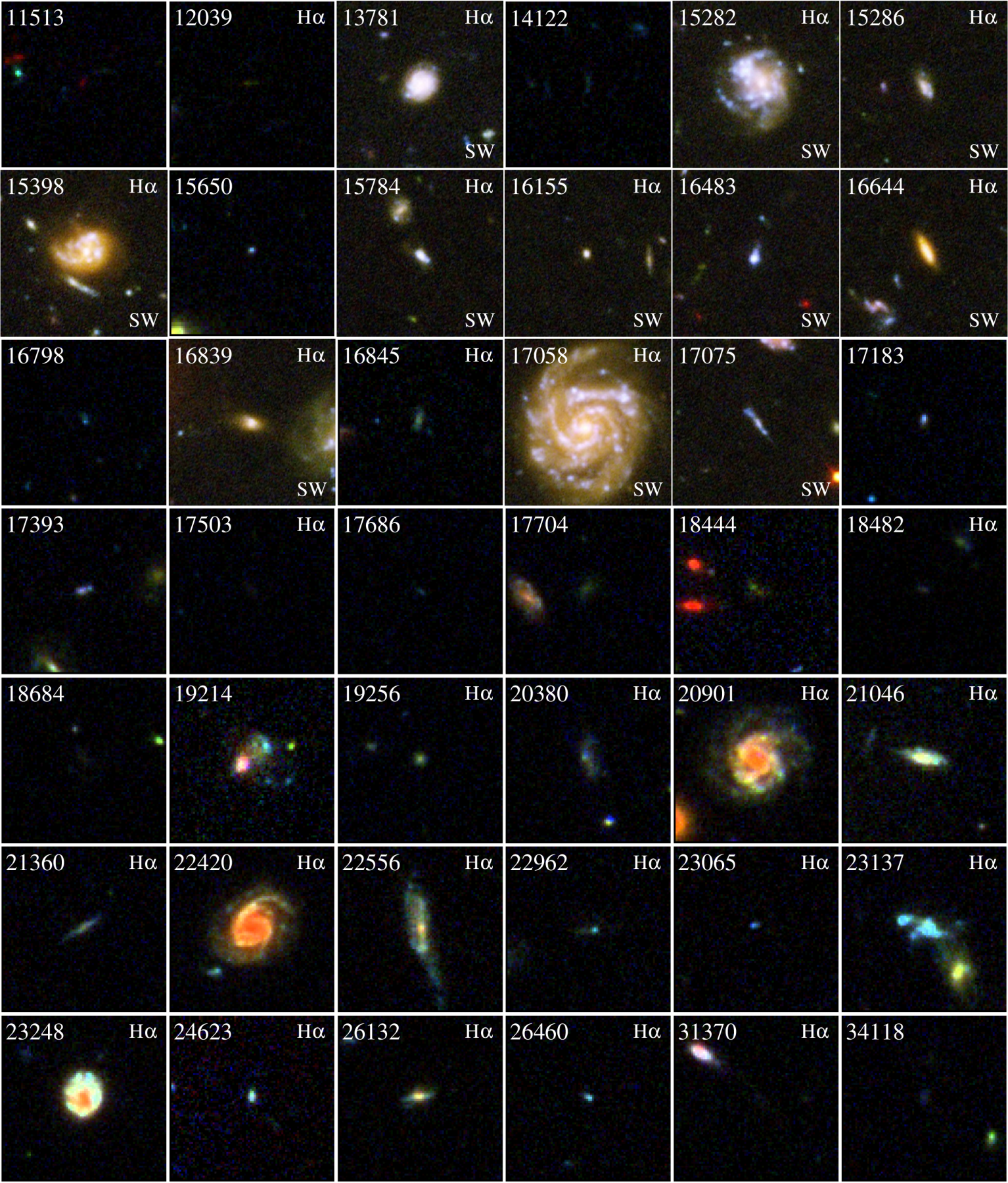}
\caption{RGB postage stamps of the 42 objects in our raw sample with CANDELS counterparts. North is up, east is to the left, and the images have a size of 5\arcsec$\times$5\arcsec. The CANDELS ID is in the upper left corner. Those classified as H$\alpha$ emitters show this label in the upper right corner. Those images built with UDF Skywalker show the label SW in the bottom right corner.}
\label{fig:multifp}
\end{figure*}

We generated RGB postage stamps of the objects in our sample from the Rainbow Database. We used three filters of \emph{HST}/ACS. The colour blue was assigned to the F435W band, green to the F606W band, and red to the F850W band. In this configuration, we observe the rest-frame spectral range of NUV, $B$, and $V$ bands for an object at $z\sim0.62$. Therefore, we show regions associated with star formation and young populations in blue and green and a more evolved stellar population in red. In addition, we searched in the II/258 Hubble Ultra Deep Field Catalog (UDF) \citep{2005yCat.2258....0B} and found 11 results. We used the UDF SkyWalker\footnote{http://www.aip.de/groups/galaxies/sw/udf/swudfV1.0.html} tool to create images with a greater exposure time and quality than the ones we obtained from the Rainbow Database for these 11 objects. The filters used in this case are similar: ACS F435W, F606W, F775W, and F850LP. We present the final RGB postage stamps of the 42 objects with CANDELS counterparts in Fig.~\ref{fig:multifp}. Thirty-one of them come from the Rainbow Database and 11 from the UDF Skywalker.

We were able to develop a visual morphological classification for our objects. We used the $z$ band ACS image in the GOODS-S field. We based it on the CANDELS morphological classification in \citet{2015ApJS..221...11K}. In this work, the authors classify the objects in five main categories: `Disk', `Spheroid', `Irregular', `Compact', and `Unclassifiable'. They add mixed classes, but we decided to maintain only the main categories.

The Disk category includes objects with a disk structure. Among the Disk objects, we classify those that show a clear spiral structure as `Spiral' and those that do not as Disk. Spheroid galaxies appear concentrated and round or ellipsoidal. Irregular objects includes galaxies that do not fall into the other categories because they show irregular/peculiar structures. Strongly disturbed objects are included here, but also those disks or spheroids with slightly disturbed morphologies. Compact objects are clear point sources, unresolved compact galaxies, or are too small to show internal features. A small but clearly resolved spheroidal galaxy is classified as a Spheroid. Unclassifiable objects cannot be placed in another main category because there is problem in the image or because they are too faint to show any structure.

We also include the interaction flags `Merger', `Interaction', and `Close Neighbour'. The Merger flag is used to signal objects that show tidal and structure features such as tails or loops. All these objects are classified as Irregular in the main category scheme. The Interaction flag refers to primary objects that appear to be interacting with a companion galaxy. Interactions show tidal features. To be assigned this flag instead of Merger, two galaxies must be visible. Close Neighbour refers to galaxies with a close visible companion in the field of view of the postage stamp. No evidence of tidal interaction or disturbed morphology is apparent.

In Table~\ref{tab:morpho} we show the results of this visual morphological classification for two groups: Total, which comprises the 42 objects with CANDELS counterparts, and H$\alpha$, which comprises the 28 H$\alpha$ emitters mentioned in Sect.~\ref{sec:haemitters}. We see that the majority of our objects are Disk or Irregular in the Total sample and Disk in the H$\alpha$ sample. In terms of interactions, 79\%(79\%) have a Close Neighbour, 10\%(11\%) show signatures of Interaction, 5\%(4\%) are Merger, and 6\%(6\%) do not show any interactions features in the Total(H$\alpha$) sample. We do not find any significant differences between the Total and H$\alpha$ samples in terms of interactions.

\begin{table*}
\caption{Visual morphological classification.}
\label{tab:morpho}
\centering
\begin{tabular}{l c c c c c c}
\hline\hline
Sample & Spiral & Disk & Spheroid & Compact & Irregular & Unclassifiable \\
\hline
	Total\tablefootmark{a} & 6(14\%) & 15(36\%) & 4(9\%) & 2(5\%) & 13(31\%) & 2(5\%) \\
	H$\alpha$\tablefootmark{b} & 6(21\%) & 13(46\%) & 3(11\%) & 1(4\%) & 3(11\%) & 2(7\%) \\
\hline
\end{tabular}
\tablefoot{
\tablefoottext{a}{Numbers(percentages) of the Total sample of the 42 objects with CANDELS counterparts.}
\tablefoottext{b}{Numbers(percentages) of the objects classified as H$\alpha$ emitters.}
}
\end{table*}

It is important to comment on the uncertainties of the morphological classification. It can be difficult to classify some of the objects as we are mainly dealing with small and faint sources. Some irregular features are hard to see or could remain undetected because of their intrinsic low surface-brightness. This effect could cause a bias in the Irregular category towards Disk.

\begin{figure}
\resizebox{\hsize}{!}{\includegraphics{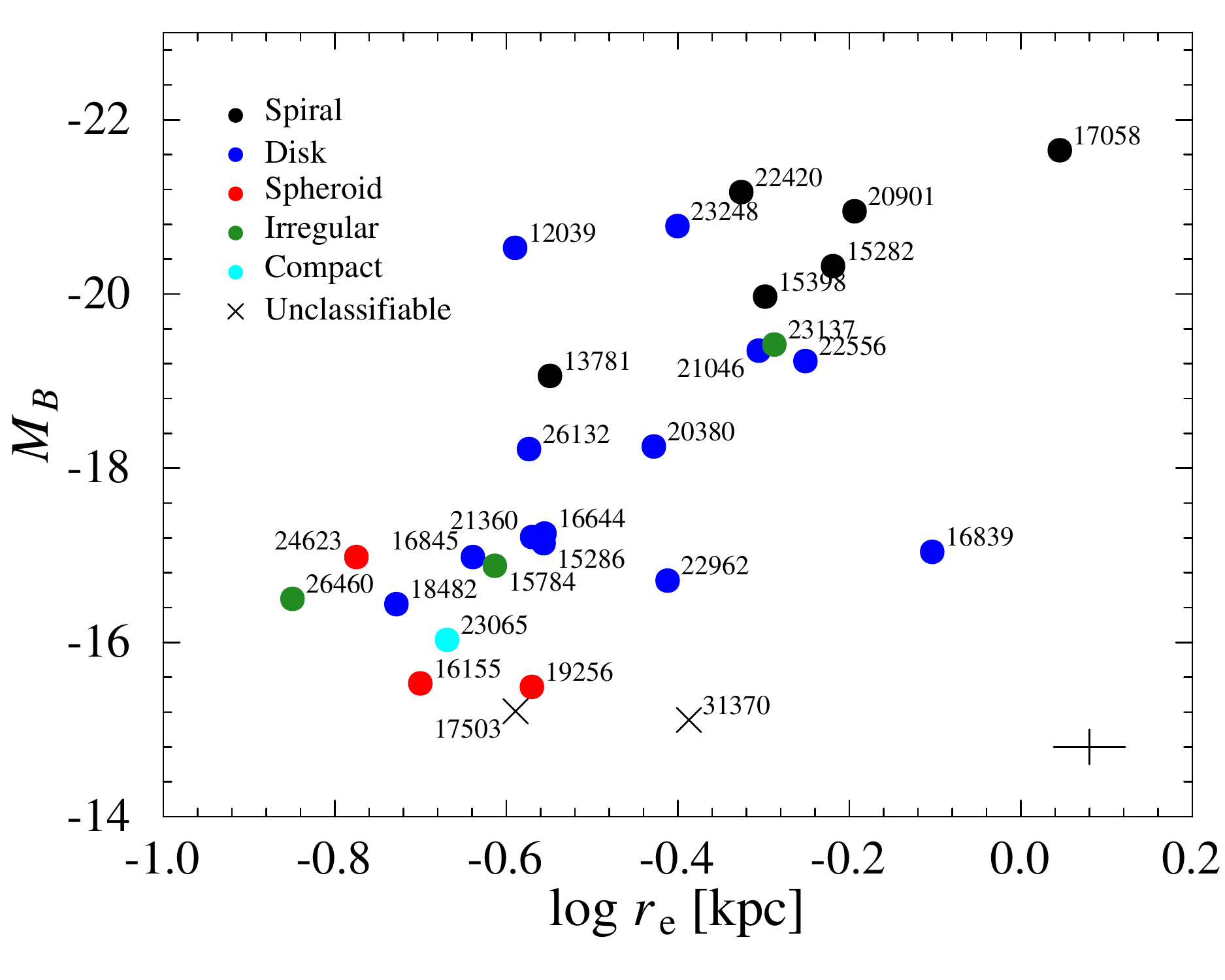}}
\caption{Rest-frame $B$-band absolute magnitude vs. effective radius measured in the F160W band for the 42 objects in our sample with CANDELS counterparts. We show the morphological classification of each object as it is indicated in the legend. Error bars are indicated in the bottom right corner.}
\label{fig:mb_reff_ha}
\end{figure}

In order to perform a more complete morphological characterisation of our sample, we studied the structural parameters. In Fig.~\ref{fig:mb_reff_ha} we show, for the 42 objects with CANDELS counterparts, the rest-frame $B$-band absolute magnitude versus the effective radius taken from the CANDELS stellar masses \citep{2015ApJ...801...97S} and photometric catalogues \citep{2013ApJS..207...24G}, respectively. The catalogues do not provide uncertainties for these parameters. We assumed a 0.2\,mag photometric uncertainty in the absolute magnitude as is indicated in Sect.~\ref{sec:haemitters}. In the case of the effective radius, we assumed 1\,pix uncertainty. We refer to \citet{1997ApJ...489..543P} for a similar diagram in which the authors studied the location of a sample of compact galaxies in the diagram in comparison with other local types. From a quick look at our plot, we note a tendency that Spheroid and Compact galaxies are fainter and more compact, and Spiral galaxies are brighter and larger (i.e. less compact). Moreover, the Unclassifiable CANDELS 17503 and 31370 are the two faintest objects, as is expected from their morphological category.

\begin{figure}
\resizebox{\hsize}{!}{\includegraphics{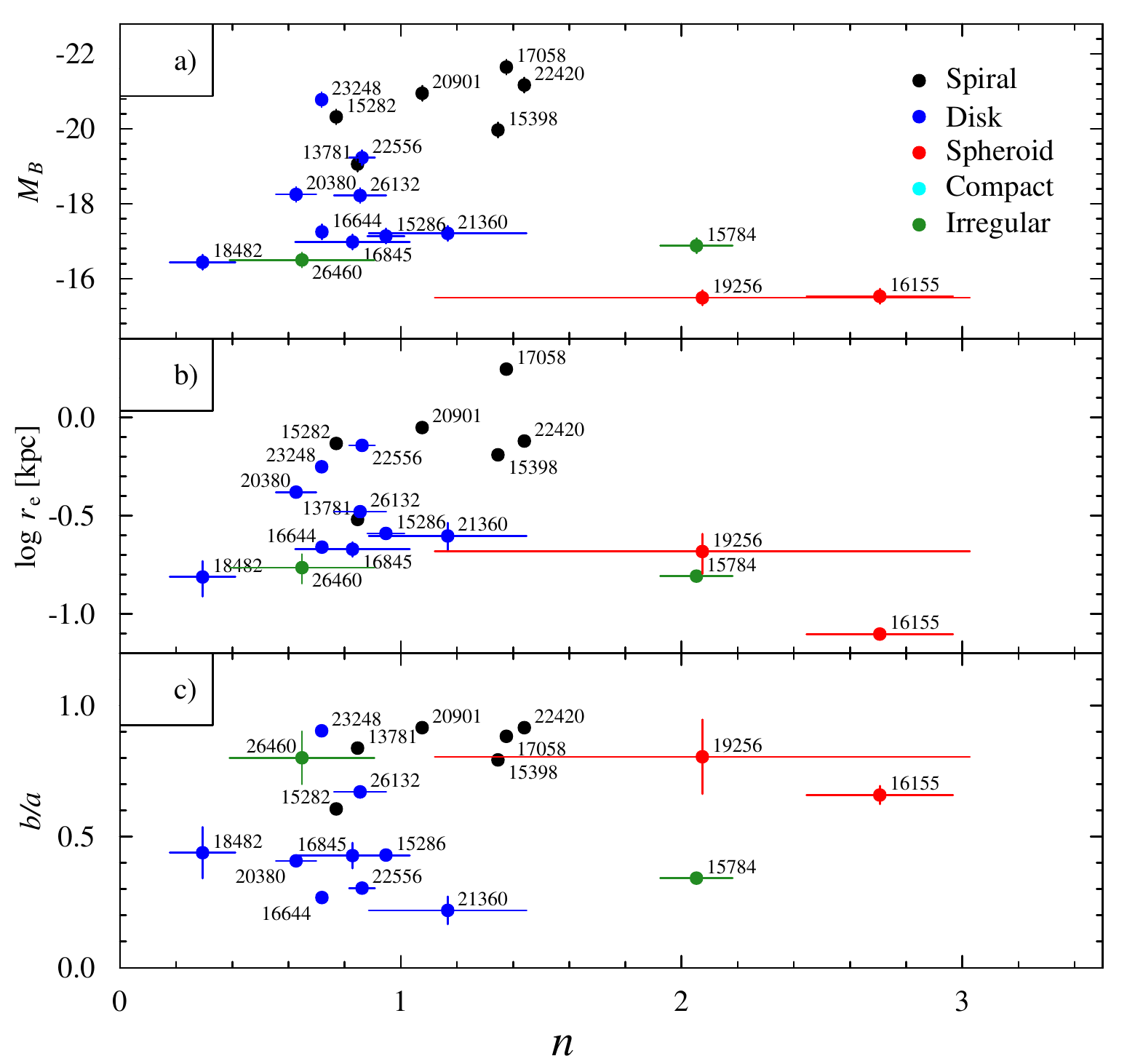}}
\caption{Structural parameters from \citet{2012ApJS..203...24V} measured in the F105W band. a) Rest-frame $B$-band absolute magnitude vs. S{\'e}rsic index. b) Effective radius vs. S{\'e}rsic index. c) Axis ratio vs. S{\'e}rsic index. We show the morphological classification with the colour code indicated in the legend.}
\label{fig:multi_morph_sersic_ha}
\end{figure}

In Fig.~\ref{fig:mb_reff_ha} we identified similarities between various visual categories. Different morphological classifications overlap in the same region of the plot. Compact and Spheroid objects tend towards low sizes and luminosities, whereas the Spiral category, where some of the Disk galaxies are located as well, is biased to large and bright objects. In order to better constrain this, we studied other available structural parameters for our sample in the CANDELS structural parameters catalogue \citep{2012ApJS..203...24V}. From this catalogue, we selected only the counterparts with a good GALFIT flag (flag = 0). In Fig.~\ref{fig:multi_morph_sersic_ha} we show several of these parameters along with the S{\'e}rsic index, a parameter that allows us to establish a continuous way to morphologically classify the galaxies. We were able to see the similarities and differences in terms of blue luminosities ($M_B$), compactness (effective radius, $r_\mathrm{e}$), and roundness (axis ratio, $b/a$) with morphology. In this case, we assumed a 0.2\,mag photometric uncertainty in the absolute magnitude with the values taken from \citet{2015ApJ...801...97S} shown in Fig.~\ref{fig:mb_reff_ha}. The uncertainties in the structural parameters were taken from \citet{2012ApJS..203...24V}. There is a clear contrast between the galaxies CANDELS 15784, 16155, and 19256, which have S{\'e}rsic indexes of $n\sim$ 2--3, and the rest, which have $n<1.5$. Spirals show the highest luminosities, the largest radii and axis ratios close to one. We note that the number of objects in Fig.~\ref{fig:mb_reff_ha} is greater than in Fig.~\ref{fig:multi_morph_sersic_ha} beacuse in the latter we only consider those sources with reliable parameters in the \citet{2012ApJS..203...24V} catalogue.

Taking into account all these structural parameters, we redistributed the galaxies in the different morphological categories in order to look for an even simpler approach. We looked at the structural similarities rather than the differences in the detailed features that we considered in the visual morphological classification. We can say that we had no reason to distinguish between Spheroid and Compact objects for our sample. They are compact and faint single starburst candidates that dominate the entire galaxy. The S{\'e}rsic index would be $n\sim$ 2--3 for galaxies of this kind. The Irregular CANDELS 15784 just presents a difference in terms of the axis ratio ($b/a=0.34\pm0.02$), compared to Compact(Spheroid) morphologies ($b/a>0.6$). It could be the star formation or its companion that is disrupting its shape. The Irregular CANDELS 26460 could be a pre- or post-stage Compact(Spheroid) class, as the difference with this category in terms of the S{\'e}rsic index ($n=0.65\pm0.26$) could be explained if the light is more disperse owing to a nuclear starburst that has not completely started or a disruption once the star formation has finished. The last Irregular CANDELS 23137 is a Merger with structural properties more related to Disk morphology, ($M_B=-19.42$, $r_\mathrm{e}=0.52$). In the case of Spiral and Disk classes, we only see a difference in terms of luminosities and sizes; the S{\'e}rsic indexes are $n<1.5$ in every case. The Spiral class is linked to observational effects. It is biased to face-on, brighter, and larger objects. It is easier to distinguish features in brighter and larger disks and we cannot distinguish spiral characteristics in edge-on objects. These galaxies tend to have larger S{\'e}rsic indexes. Within the Disk category we find a continuous distinction between the smallest disks closer to compact objects, in terms of luminosities and compactness, and larger disks related with spiral features. We should include all disks in the same Disk category regardless of their spiral structures since all these features are associated with biased interpretations rather than a difference in nature.

\subsection{Stellar masses}
\label{sec:mass}
The stellar masses are another product that we took from CANDELS in the GOODS-S field \citep{2015ApJ...801...97S}. The authors use the same data to compute masses through ten different codes and then they combine the results. We compared the CANDELS stellar masses with those released by 3D-HST \citep{2014ApJS..214...24S}. We measured an offset $\mathrm{median=-0.057}$ with a $rms=0.26$ once we eliminated the outliers CANDELS 12039 = 3D-HST 23397 and CANDELS 18444 = 3D-HST 3408. The offset is defined as $\log {(M^{\mathrm{CANDELS}}_{\mathrm{median}}/M^{\mathrm{3D-HST}})}$. Our offset perfectly matches the $\mathrm{median=-0.058\pm0.003}$ measured in \citet{2015ApJ...801...97S}. The authors of the CANDELS stellar mass catalogue argue that, because the comparison is satisfying at a first order, the differences between the two catalogues are due to different systematics in the SED fitting. The 3D-HST masses come from a single fit, but CANDELS values follow a median approach that deal better with the systematics because of the model assumptions. For the outliers we decided to substitute the CANDELS masses for the 3D-HST values as we indicated in Sect.~\ref{sec:specdata}. The 3D-HST catalogue does not provide uncertainties in the stellar masses, and so we assumed a typical 20\%. In Fig.~\ref{fig:hist} we present a histogram of the distribution of stellar masses in our 42 galaxies sample with CANDELS counterpart. In blue we indicate the 28 objects confirmed as H$\alpha$ emitters.

\subsection{Line flux and line luminosity estimation}
\label{sec:lineflux}
In order to obtain the emission-line fluxes from the NB, we followed the \citet{2007PASP..119...30P} approach as described and applied in \citet{2008ApJ...677..169V}. In a recent study by \citet{2015A&A...580A..47V}, the authors find that H$\alpha$ + [\ion{N}{ii}] fluxes obtained with the combination of two filters are underestimated by 28\% compared to the fluxes derived via SED fitting. It is important to take this into account when comparing fluxes calculated with different techniques.

Emission-line fluxes are computed using
\begin{equation}
 f_{l} = \Delta_{\mathrm{NB}} \left(f_{\mathrm{NB}} - f_{\mathrm{BB}}\right) \frac{1}{1-\epsilon} , \label{eq:flux}
\end{equation}
\noindent where $f_{\mathrm{NB}}$ and $f_{\mathrm{BB}}$ are the total fluxes in the NB and BB filters, $f_{l}$ is the line flux including the contribution of [\ion{N}{ii}]$\lambda\lambda$6548,6584, $\Delta_{\mathrm{NB}}$ is the NB filter width, and $\epsilon$ is the ratio of NB and BB widths.

To calculate the integrated emission-line flux of each galaxy, we used several apertures because the galaxies show a variety of sizes and shapes. The flux grows with the aperture diameter until the limit of the emission region is reached. Nitrogen contamination in the NB fluxes was removed following the indications in \citet{2008ApJ...677..169V}, where the authors study the trend between $\log {(I([\ion{N}{ii}]\lambda6584)/I(\mathrm{H\alpha}))}$ and $\log {(\mathrm{EW(H\alpha + [\ion{N}{ii}]\lambda6584})}$ in SDSS galaxies to get a relation to estimate [\ion{N}{ii}] contribution to the emission-line flux in the NB images for each source. In Fig.~\ref{fig:hist} we can see the flux distribution of the 46 candidates primarily selected. This value grows as we move to fainter magnitudes until the detection limit is reached.

Once the emission-line flux is calculated, the emission-line luminosity is immediate using
\begin{equation}
 L = 4\pi d_l^2(z) f_l , \label{eq:lum}
\end{equation}
\noindent where $d_l(z)$ is the distance luminosity. We used the cosmologial library \emph{Milia} \citep{sergio_pascual_2015_17810}\footnote{http://guaix.fis.ucm.es/projects/milia} to compute it. We employed the spec-$z$'s if available and the value $z=0.62$ otherwise.

\subsection{Reddening correction}
\label{sec:extinction}
The line fluxes and luminosities that we calculated in Sect.~\ref{sec:lineflux} are observed values, and so are uncorrected for dust extinction. We needed to develop a strategy to take into account the various sources of attenuation: the intrinsic extinction from the studied galaxy and the extinction from our own Galaxy. We followed the method employed by \citet{2008ApJ...677..169V}. The extinction-corrected fluxes and luminosities are then
\begin{equation}
 \log {f_{l,\mathrm{cor}}} = \log {f_l} + 0.4\left(A({\lambda_{\mathrm{rf}}}) + a({\lambda_z})\right) , \label{eq:flux_cor}
\end{equation}
\begin{equation}
 L = 4\pi d_l^2(z) f_{l,\mathrm{cor}} , \label{eq:lum_cor}
\end{equation}
\noindent where $A({\lambda_{\mathrm{rf}}})$ is the intrinsic extinction of the studied galaxy at the rest-frame wavelength and $a({\lambda_z})$ is the Galactic attenuation at the observed frame field.

For $a({\lambda_z})$ we took a constant value of 0.006\,mag as measured in UKIRT $J$ band (1.25\,$\mu$m) for the coordinates of the GOODS-S field \citep{2011ApJ...737..103S}. It is a completely negligible value compared with the intrinsic extinction.

On the other hand, it is known that dust absorbs an amount of the star emission. This dust emits it again in the IR. Therefore, a good approach for calculating the intrinsic extinction of the studied galaxy is to study the ratio between dust and FUV fluxes ($F_{\mathrm{dust}}/F_{\mathrm{FUV}}$), the IRX ratio. This value is related to the FUV extinction $A({\mathrm{FUV}})$ \citep{2005ApJ...619L..51B}:
\begin{equation}
 A({\mathrm{FUV}}) = -0.0333 y^3 + 0.3522 y^2 + 1.1960 y + 0.4967 , \label{eq:fuv_ext}
\end{equation}
where $y = \log {\left(\frac{F_{\mathrm{dust}}}{F_{\mathrm{FUV}}}\right)}$.

We obtained $F_{\mathrm{dust}}$ from the TIR luminosity synthetically calculated and $F_{\mathrm{FUV}}$ from synthetic FUV magnitude. Both of these values were taken from the Rainbow Database, as is indicated in Sect.~\ref{sec:photdata}. Then, we obtained the H$\alpha$ extinction ($A({\mathrm{H\alpha}})$) with the $A({\mathrm{FUV}})$ value applying a Calzetti extinction law \citep{2000ApJ...533..682C}:
\begin{equation}
 A({\mathrm{H\alpha}}) = A({\mathrm{FUV}})\frac{k_{\mathrm{H\alpha}}}{k_{\mathrm{FUV}}} , \label{eq:ha_ext}
\end{equation}
\noindent where $k_{\mathrm{H\alpha}} = 2.659\left(-1.857 + \frac{1.040}{\lambda_{\mathrm{H\alpha}}}\right) + R_V^{'}$, \\*$k_{\mathrm{FUV}} = 2.659\left(-2.156 + \frac{1.509}{\lambda_{\mathrm{FUV}}} - \frac{0.198}{\lambda_{\mathrm{FUV}}^2} + \frac{0.011}{\lambda_{\mathrm{FUV}}^3}\right) + R_V^{'}$ \\*and $R_V^{'} = 4.05$.

For CANDELS 15282 we got a value $A\mathrm{(H\alpha)}<0$, thus, we assumed $A\mathrm{(H\alpha)}=0$. In the case of CANDELS 12039 the synthetic TIR luminosity was not available and we calculated the extinction employing FUV and NUV synthetic magnitudes from the Rainbow Database. The $\mathrm{FUV-NUV}$ colour is related to the IRX ratio \citep{2009ApJ...701.1965M}:
\begin{equation}
 \frac{L_{\mathrm{TIR}}}{L_{\mathrm{FUV}}} = 10^{0.30 + 1.15\left(\mathrm{FUV-NUV}\right)} - 1.64 .  \label{eq:tirfuv}
\end{equation}

Once we have the IRX ratio from Eq.~\ref{eq:tirfuv}, we can substitute it in Eq.~\ref{eq:fuv_ext} and apply a Calzetti law (Eq.~\ref{eq:ha_ext}) to obtain the extinction for this object.

\begin{figure}
\resizebox{\hsize}{!}{\includegraphics{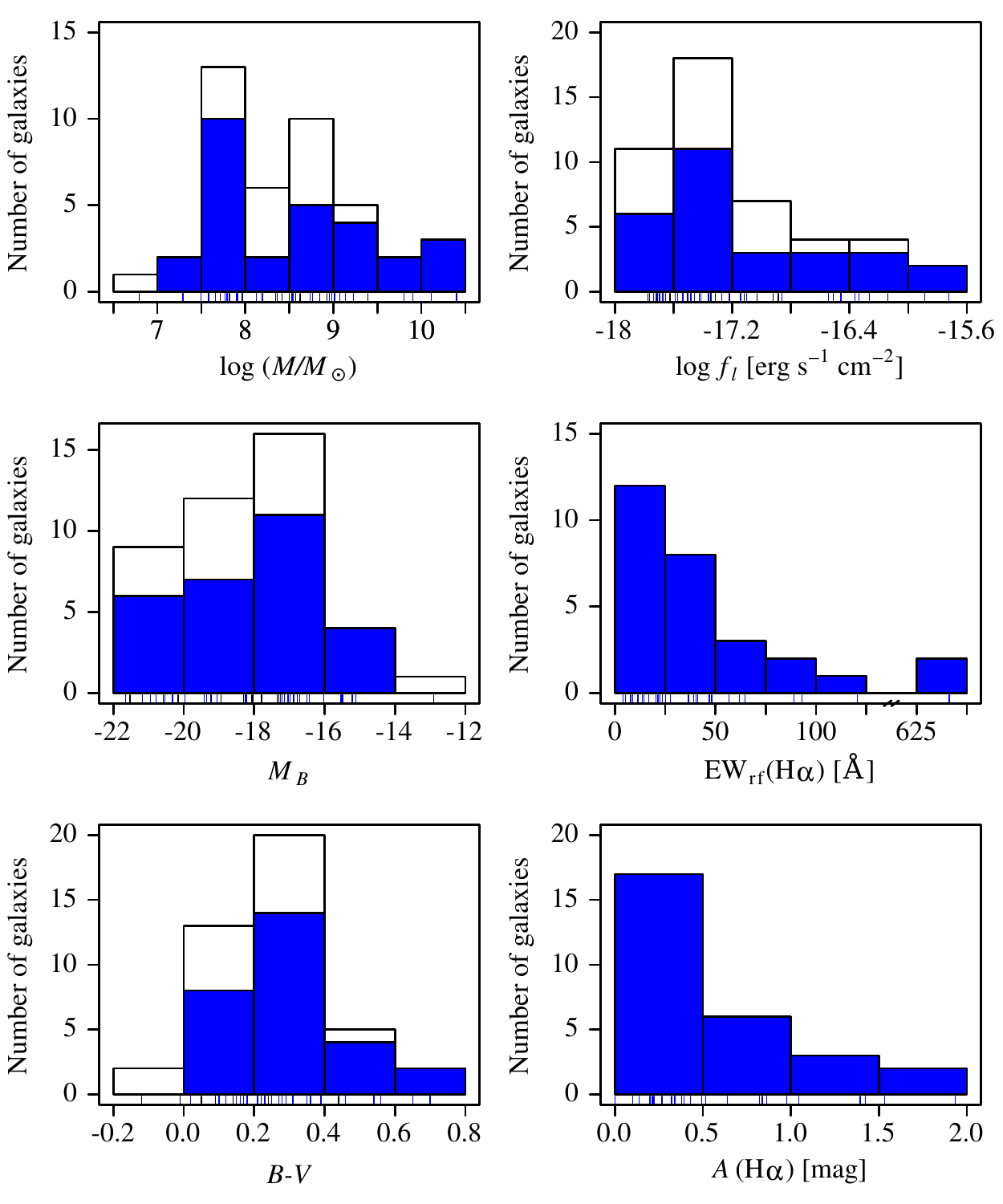}}
\caption{Histograms of several magnitudes. In white, the 42 galaxies with CANDELS counterparts and in blue the sample of 28 H$\alpha$ SFGs. \emph{Top left panel}: histogram of stellar masses. \emph{Middle left panel}: histogram of rest-frame $B$-band absolute magnitude. \emph{Bottom left panel}: histogram of $B-V$ colour. \emph{Top right panel}: histogram of line fluxes. In this case, we show in white the raw sample of 46 candidates and the 28 H$\alpha$ objects overlaped in blue. \emph{Middle right panel}: histogram of H$\alpha$ rest-frame EW for the 28 H$\alpha$ emitters. \emph{Bottom right panel}: histogram of H$\alpha$ extinction for the H$\alpha$ SFGs.}
\label{fig:hist}
\end{figure}

In Fig.~\ref{fig:hist} we show the distribution of the sample in terms of stellar masses, rest-frame $B$-band absolute magnitude and $B-V$ colour, which we took from the CANDELS stellar masses catalogue \citep{2015ApJ...801...97S}. We also show the H$\alpha$ line flux, H$\alpha$ rest-frame EW, and H$\alpha$ extinction that we calculated in this section. We separate the contribution of the 28 confirmed H$\alpha$ emitters in blue. We cannot see any significant difference between the H$\alpha$ emitters properties and the interlopers at other redshift that form the whole sample. We highlight the presence of very low line fluxes, luminosities, and stellar masses within the sample. In Table~\ref{tab:sample}, at the end of the document, we gather the complete information and calculations of the sample.

\begin{longtab}
\begin{landscape}
\begin{longtable}{c r r r r r r r r r r r}
\caption{\label{tab:sample}Sample data.}\\
\hline\hline
ID & $\alpha$(J2000) & $\delta$(J2000) & $z_\mathrm{phot}$ & $z_\mathrm{spec}$ & Line\tablefootmark{a} & $\log (M/M_{\odot})$ & $f_l$ & EW(H$\alpha$) & A(H$\alpha$) & SFR$_\mathrm{cor}$ & Morphology \\
 & (deg) & (deg) & & & & & (10$^{-18}$ erg s$^{-1}$ cm$^{-2}$) & (\AA) & (mag) & ($M_{\odot}$ yr$^{-1}$) & \\
\hline
\endfirsthead
\caption{continued.}\\
\hline\hline
ID & $\alpha$(J2000) & $\delta$(J2000) & $z_\mathrm{phot}$ & $z_\mathrm{spec}$ & Line & $\log (M/M_{\odot})$ & $f_l$ & EW(H$\alpha$) & A(H$\alpha$) & SFR$_\mathrm{cor}$ & Morphology \\
 & (deg) & (deg) & & & & & (10$^{-18}$ erg s$^{-1}$ cm$^{-2}$) & (\AA) & (mag) & ($M_{\odot}$ yr$^{-1}$) & \\
\hline
\endhead
\hline
\endfoot
12039 & 53.097336 & -27.800426 & 0.42 $\pm$ 0.15 & (...) & H$\alpha$ & 7.59 $\pm$ 0.09 & 2.7 $\pm$ 0.5 & 1038.9 $\pm$ 290.6 & 0.20 $\pm$ 0.31 & 0.04 $\pm$ 0.01 & Disk \\
13781 & 53.168017 & -27.789670 & 0.59 $\pm$ 0.30 & 0.619 & H$\alpha$ & 8.98 $\pm$ 0.04 & 34.5 $\pm$ 2.0 & 63.7 $\pm$ 4.6 & 0.14 $\pm$ 0.07 & 0.50 $\pm$ 0.04 & Spiral \\
15282 & 53.158169 & -27.781095 & 0.59 $\pm$ 0.30 & 0.620 & H$\alpha$ & 9.23 $\pm$ 0.05 & 54.5 $\pm$ 2.2 & 76.2 $\pm$ 4.1 & 0 & 0.70 $\pm$ 0.03 & Spiral \\
15286 & 53.172260 & -27.779516 & 0.65 $\pm$ 0.33 & (...) & H$\alpha$ & 8.19 $\pm$ 0.07 & 3.5 $\pm$ 2.2 & 38.5 $\pm$ 26.3 & 0.40 $\pm$ 0.06 & 0.06 $\pm$ 0.04 & Disk \\
15398 & 53.172609 & -27.780967 & 0.57 $\pm$ 0.29 & 0.619 & H$\alpha$ & 9.90 $\pm$ 0.08 & 43.3 $\pm$ 2.4 & 36.2 $\pm$2 .2 & 0.52 $\pm$ 0.06 & 0.89 $\pm$ 0.07 & Spiral \\
15784 & 53.178185 & -27.775894 & 0.60 $\pm$ 0.31 & (...) & H$\alpha$ & 7.82 $\pm$ 0.06 & 2.9 $\pm$ 2.1 & 33.1 $\pm$ 25.3 & 1.53 $\pm$ 0.09 & 0.15 $\pm$ 0.11 & Irregular \\
16155 & 53.168582 & -27.772930 & 0.56 $\pm$ 0.29 & (...) & H$\alpha$ & 7.79 $\pm$ 0.08 & 2.1 $\pm$ 1.9 & 65.4 $\pm$ 73.2 & 1.94 $\pm$ 0.07 & 0.16 $\pm$ 0.14 & Spheroid \\
16644 & 53.172101 & -27.770380 & 0.54 $\pm$ 0.28 & (...) & H$\alpha$ & 9.01 $\pm$ 0.06 & 2.2 $\pm$ 2.6 & 8.4 $\pm$ 10.0 & 0.98 $\pm$ 0.07 & 0.07 $\pm$ 0.08 & Disk \\
16839 & 53.153536 & -27.769445 & 0.51 $\pm$ 0.26 & (...) & H$\alpha$ & 8.96 $\pm$ 0.12 & 1.9 $\pm$ 2.1 & 14.3 $\pm$ 15.6 & 0.38 $\pm$ 0.13 & 0.03 $\pm$ 0.04 & Disk \\
16845 & 53.064759 & -27.767741 & 0.66 $\pm$ 0.34 & (...) & H$\alpha$ & 7.97 $\pm$ 0.05 & 4.6 $\pm$ 2.2 & 66.5 $\pm$ 40.7 & 0.64 $\pm$ 0.08 & 0.10 $\pm$ 0.05 & Disk \\
17058 & 53.169925 & -27.771024 & 0.61 $\pm$ 0.31 & 0.622 & H$\alpha$ & 10.39 $\pm$ 0.07 & 34.8 $\pm$ 2.6 & 18.6 $\pm$ 1.4 & 0.22 $\pm$ 0.03 & 0.55 $\pm$ 0.04 & Spiral \\
17503 & 53.114224 & -27.762464 & 0.62 $\pm$ 0.32 & (...) & H$\alpha$ & 7.58 $\pm$ 0.15 & 2.0 $\pm$ 1.4 & 195.1 $\pm$ 315.6 & 1.05 $\pm$ 0.14 & 0.07 $\pm$ 0.05 & Unclassifiable \\
18482 & 53.139078 & -27.753542 & 0.87 $\pm$ 0.44 & (...) & H$\alpha$ & 7.97 $\pm$ 0.07 & 3.1 $\pm$ 1.6 & 150.5 $\pm$ 155.6 & 0.84 $\pm$ 0.08 & 0.09 $\pm$ 0.04 & Disk \\
19256 & 53.110459 & -27.747339 & 0.38 $\pm$ 0.20 & (...) & H$\alpha$ & 7.77 $\pm$ 0.06 & 2.0 $\pm$ 2.1 & 40.9 $\pm$ 46.7 & 0.34 $\pm$ 0.14 & 0.04 $\pm$ 0.04 & Spheroid \\
20380 & 53.177126 & -27.737653 & 0.70 $\pm$ 0.36 & (...) & H$\alpha$ & 8.54 $\pm$ 0.10 & 7.9 $\pm$ 2.1 & 44.9 $\pm$ 13.3 & 0.22 $\pm$ 0.05 & 0.12 $\pm$ 0.03 & Disk \\
20901 & 53.190161 & -27.734945 & 0.60 $\pm$ 0.31 & 0.619 & H$\alpha$ & 10.11 $\pm$ 0.07 & 46.3 $\pm$ 2.8 & 23.3 $\pm$ 1.5 & 0.27 $\pm$ 0.04 & 0.75 $\pm$ 0.05 & Spiral \\
21046 & 53.116710 & -27.732614 & 0.54 $\pm$ 0.28 & 0.621 & H$\alpha$ & 9.14 $\pm$ 0.08 & 28.6 $\pm$ 2.3 & 59.4 $\pm$ 5.8 & 0.20 $\pm$ 0.04 & 0.44 $\pm$ 0.04 & Disk \\
21360 & 53.106294 & -27.727954 & 0.63 $\pm$ 0.32 & (...) & H$\alpha$ & 8.12 $\pm$ 0.10 & 2.6 $\pm$ 2.3 & 27.1 $\pm$ 25.3 & 0.43 $\pm$ 0.07 & 0.05 $\pm$ 0.04 & Disk \\
22420 & 53.097519 & -27.721271 & 0.58 $\pm$ 0.30 & 0.615 & H$\alpha$ & 10.40 $\pm$ 0.09 & 129.2 $\pm$ 2.4 & 34.5$ \pm$ 0.7 & 0.49 $\pm$ 0.04 & 2.55 $\pm$ 0.11 & Spiral \\
22556 & 53.063490 & -27.720247 & 0.59 $\pm$ 0.30 & 0.609 & H$\alpha$ & 9.07 $\pm$ 0.06 & 5.3 $\pm$ 2.3 & 12.6 $\pm$ 5.4 & 0.10 $\pm$ 0.11 & 0.07 $\pm$ 0.03 & Disk \\
22962 & 53.142643 & -27.708699 & 0.55 $\pm$ 0.28 & (...) & H$\alpha$ & 7.90 $\pm$ 0.10 & 7.2 $\pm$ 2.3 & 144.3 $\pm$ 90.1 & 0.34 $\pm$ 0.06 & 0.13 $\pm$ 0.04  & Disk \\
23065 & 53.152421 & -27.706515 & 0.68 $\pm$ 0.35 & (...) & H$\alpha$ & 7.29 $\pm$ 0.09 & 4.4 $\pm$ 0.9 & 1038.9 $\pm$ 289.5 & 0.86 $\pm$ 0.05 & 0.12 $\pm$ 0.02 & Compact \\
23137 & 53.077383 & -27.708195 & 0.57 $\pm$ 0.29 & 0.605 & H$\alpha$ & 8.73 $\pm$ 0.05 & 3.3 $\pm$ 2.9 & 6.5 $\pm$ 5.7 & 0.10 $\pm$ 0.04 & 0.04 $\pm$ 0.04 & Irregular \\
23248 & 53.141244 & -27.710543 & 0.61 $\pm$ 0.31 & 0.619 & H$\alpha$ & 9.80 $\pm$ 0.03 & 189.8 $\pm$ 2.9 & 77.9 $\pm$ 1.6 & 0.21 $\pm$ 0.04 & 2.93 $\pm$ 0.13 & Disk \\
24623 & 53.174256 & -27.678432 & 0.60 $\pm$ 0.31 & 0.616 & H$\alpha$ & 7.91 $\pm$ 0.08 & 13.8 $\pm$ 2.2 & 104.8 $\pm$ 25.4 & 0.32 $\pm$ 0.14 & 0.23 $\pm$ 0.05 & Spheroid \\
26132 & 53.131199 & -27.699450 & 0.60 $\pm$ 0.31 & 0.620 & H$\alpha$ & 8.76 $\pm$ 0.04 & 6.0 $\pm$ 2.3 & 22.0 $\pm$ 8.7 & 1.39 $\pm$ 0.11 & 0.28 $\pm$ 0.11 & Disk \\
26460 & 53.144189 & -27.703179 & 0.71 $\pm$ 0.36 & (...) & H$\alpha$ & 7.66 $\pm$ 0.14 & 3.9 $\pm$ 2.3 & 92.0 $\pm$ 81.5 & 0.82 $\pm$ 0.06 & 0.10 $\pm$ 0.06 & Irregular \\
31370 & 53.089858 & -27.781079 & 0.51 $\pm$ 0.26 & (...) & H$\alpha$ & 7.50 $\pm$ 0.17 & 1.8 $\pm$ 1.6 & 100.4 $\pm$ 131.7 & 1.42 $\pm$ 0.07 & 0.09 $\pm$ 0.08 & Unclassifiable \\
\cline{1-12}
11513 & 53.074157 & -27.804923 & 5.61 $\pm$ 2.81 & (...) & $z>2.5$ & 8.92 $\pm$ 0.14 & 2.4 $\pm$ 1.7 & (...) & (...) & (...) & Irregular \\
14122 & 53.115761 & -27.786059 & 1.65 $\pm$ 0.83 & (...) & [\ion{O}{ii}] & 8.50 $\pm$ 0.08 & 3.3 $\pm$ 1.6 & (...) & (...) & (...) & Irregular \\
15650 & 53.115894 & -27.776418 & 1.13 $\pm$ 0.57 & (...) & H$\beta$ & 8.41 $\pm$ 0.09 & 13.0 $\pm$ 2.3 & (...) & (...) & (...) & Compact \\
16483 & 53.160006 & -27.771002 & 1.91 $\pm$ 0.96 & 1.173 & H$\beta$ & 8.57 $\pm$ 0.12 & 2.4 $\pm$ 1.8 & (...) & (...) & (...) & Irregular \\
16798 & 53.125143 & -27.767847 & 1.42 $\pm$ 0.72 & (...) & H$\beta$ & 7.72 $\pm$ 0.11 & 9.3 $\pm$ 1.8 & (...) & (...) & (...) & Irregular \\
17075 & 53.164651 & -27.766606 & 1.65 $\pm$ 0.83 & (...) & [\ion{O}{ii}] & 8.85 $\pm$ 0.07 & 4.8 $\pm$ 1.8 & (...) & (...) & (...) & Disk \\
17183 & 53.111168 & -27.765216 & 1.14 $\pm$ 0.58 & (...) & H$\beta$ & 8.34 $\pm$ 0.07 & 12.0 $\pm$ 2.5 & (...) & (...) & (...) & Irregular \\
17393 & 53.206666 & -27.763450 & 1.08 $\pm$ 0.55 & (...) & H$\beta$ & 8.53 $\pm$ 0.06 & 7.7 $\pm$ 2.0 & (...) & (...) & (...) & Disk \\
17686 & 53.120235 & -27.760870 & 1.79 $\pm$ 0.90 & (...) & [\ion{O}{ii}] & 8.36 $\pm$ 0.08 & 3.8 $\pm$ 1.0 & (...) & (...) & (...) & Irregular \\
17704 & 53.115577 & -27.760782 & 0.14 $\pm$ 0.08 & (...) & [\ion{S}{iii}] & 6.80 $\pm$ 0.27 & 1.9 $\pm$ 2.0 & (...) & (...) & (...) & Irregular \\
18444 & 53.109247 & -27.754127 & 2.92 $\pm$ 0.13 & (...) & $z>2.5$ & 7.97 $\pm$ 0.09 & 1.7 $\pm$ 1.9 & (...) & (...) & (...) & Irregular \\
18684 & 53.130594 & -27.752070 & 1.26 $\pm$ 0.64 & (...) & H$\beta$ & 8.62 $\pm$ 0.04 & 1.7 $\pm$ 1.7 & (...) & (...) & (...) & Irregular \\
19214 & 53.058253 & -27.748481 & 0.89 $\pm$ 0.45 & (...) & H$\beta$ & 9.39 $\pm$ 0.07 & 3.2 $\pm$ 2.4 & (...) & (...) & (...) & Irregular \\
34118 & 53.132941 & -27.679610 & 1.35 $\pm$ 0.68 & (...) & H$\beta$ & 7.96 $\pm$ 0.15 & 3.3 $\pm$ 1.6 & (...) & (...) & (...) & Spheroid \\
\cline{1-12}
3148 & 53.179710 & -27.702845 & (...) & (...) & (...) & (...) & 30.9 $\pm$ 2.2 & (...) & (...) & (...) & (...) \\
3151 & 53.175771 & -27.704078 & (...) & (...) & (...) & (...) & 6.1 $\pm$ 1.9 & (...) & (...) & (...) & (...) \\
3361 & 53.150852 & -27.667993 & (...) & (...) & (...) & (...) & 72.7 $\pm$ 2.6 & (...) & (...) & (...) & (...) \\
3606 & 53.037469 & -27.710118 & (...) & (...) & (...) & (...) & 4.5 $\pm$ 0.6 & (...) & (...) & (...) & (...) \\                               
\end{longtable}
\tablefoot{
The table is separated into three parts with horizontal lines. First: H$\alpha$ emitters sample (28 galaxies; ID and coordinates from CANDELS). Second: other ELGs with CANDELS counterpart (14 galaxies; ID and coordinates from CANDELS). Third: Not-confirmed candidates from the raw sample (4 objects; ID and coordinates from our original survey; the complete original ID is HAWKI000 followed by the number in the table).\\
\tablefoottext{a}{Simplified notation: [\ion{S}{iii}]$\lambda$9069 emitters at $z\sim0.2$ are indicated as [\ion{S}{iii}]; H$\beta$ or [\ion{O}{iii}]$\lambda\lambda$4959,5007 at $z\sim1.1$ as H$\beta$; [\ion{O}{ii}]$\lambda$3727 at $z\sim1.8$ as [\ion{O}{ii}].}
}
\end{landscape}
\end{longtab}

\subsection{Spectroscopy}
\label{sec:spec}
As was described in Sect.~\ref{sec:specdata}, we collected the 12 spectra shown in Fig.~\ref{fig:spectra}. In this section, we present the outcome obtained through the spectra analysis. Making use of the software IRAF and the task \emph{splot}, we measured the emission lines that we identified in each spectrum. We performed ten measurements of each line, taking the average value of them with the standard deviation as the uncertainty. The spectroscopic properties of ELGs are generally characterised using line-ratio diagnostic diagrams \citep{1981PASP...93....5B,1987ApJS...63..295V}. We focus the analysis on two diagrams: excitation versus luminosity and rest-frame EW([\ion{O}{ii}]) versus luminosity. These diagrams are common in studies of this kind \citep{1997ApJ...475..502G,1997ApJ...489..559G}.

\begin{figure}
\resizebox{\hsize}{!}{\includegraphics{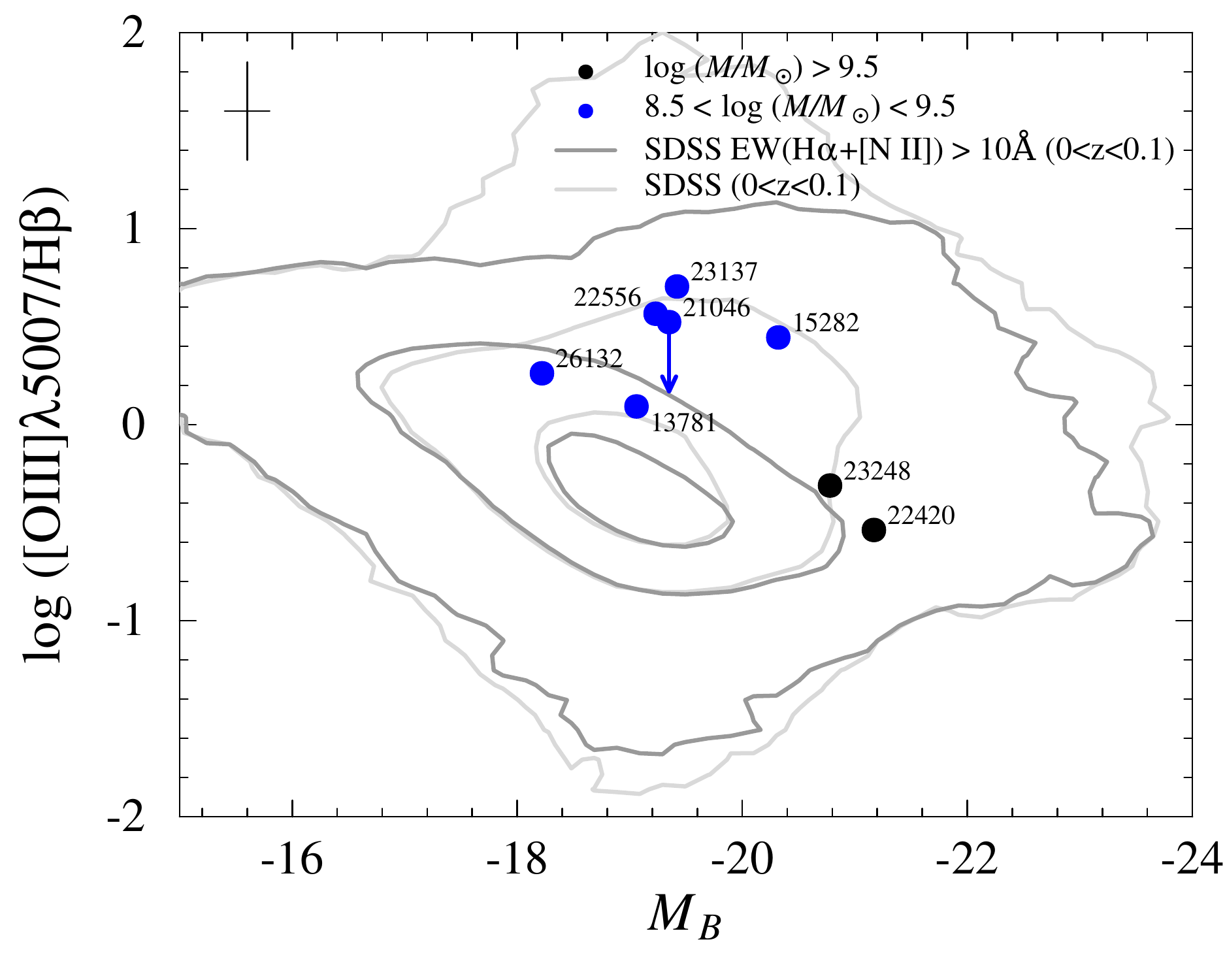}}
\caption{Excitation vs. rest-frame $B$-band absolute magnitude. The sample is represented as filled circles with the mass differentiation in different colours as indicated in the legend. The SDSS reference is shown as grey lines with the contours corresponding to 99\%, 75\%, and 25\%. Average error bars are located in the upper left corner.}
\label{fig:diag_el}
\end{figure}

In Fig.~\ref{fig:diag_el} we present the [\ion{O}{iii}]$\lambda$5007/H$\beta$ versus $M_B$ diagram. The ratio [\ion{O}{iii}] to H$\beta$ is an indicator of the excitation of the ionized gas. As noted in \citet{1997ApJ...489..559G}, the galaxies in this diagram are distributed following a sequence analogous to the \ion{H}{ii} sequence in the [\ion{O}{iii}]/H$\beta$ versus [\ion{N}{ii}]/H$\alpha$ diagram \citep{1987ApJS...63..295V}. This sequence indicates a variation in the metallicity of the gas \citep{1986ApJ...307..431D} through the [\ion{O}{iii}]/[\ion{N}{ii}] ratio \citep{2004MNRAS.348L..59P,2013A&A...559A.114M}. Along the sequence, the metallicity grows with luminosity (i.e. lower [\ion{O}{iii}]/[\ion{N}{ii}] ratio). We separate the galaxies into three mass bins: $\log {(M/M_{\odot})}<8.5$, $8.5<\log {(M/M_{\odot})}<9.5$, and $\log {(M/M_{\odot})}>9.5$. In this case none of the objects is in the first bin. We can see a wide range in excitation and metallicity and a clear difference in terms of mass. The intermediate-mass galaxies show higher excitation (i.e. lower metallicity) than the high-mass galaxies. We include references to SDSS-DR8 data \citep{2011ApJS..193...29A} with galaxies in the redshift bin $0.0<z<0.1$ (light grey contours). We mark separately the distribution of those galaxies that satisfy the criterion $\mathrm{EW(H\alpha+[\ion{N}{ii}])>10\,\AA}$ (dark grey contours) to guarantee star-forming objects similar to our sample. As the SDSS sample does not include $B$-band absolute magnitudes, we obtain them from $g'-B$ colours in \citet{1995PASP..107..945F}. The uncertainties that we considered were a 25\% in the [\ion{O}{iii}]/H$\beta$ ratio \citep{1997ApJ...489..559G} and 0.2\,mag in the $B$-band absolute magnitude as in previous sections.

\begin{figure}
\resizebox{\hsize}{!}{\includegraphics{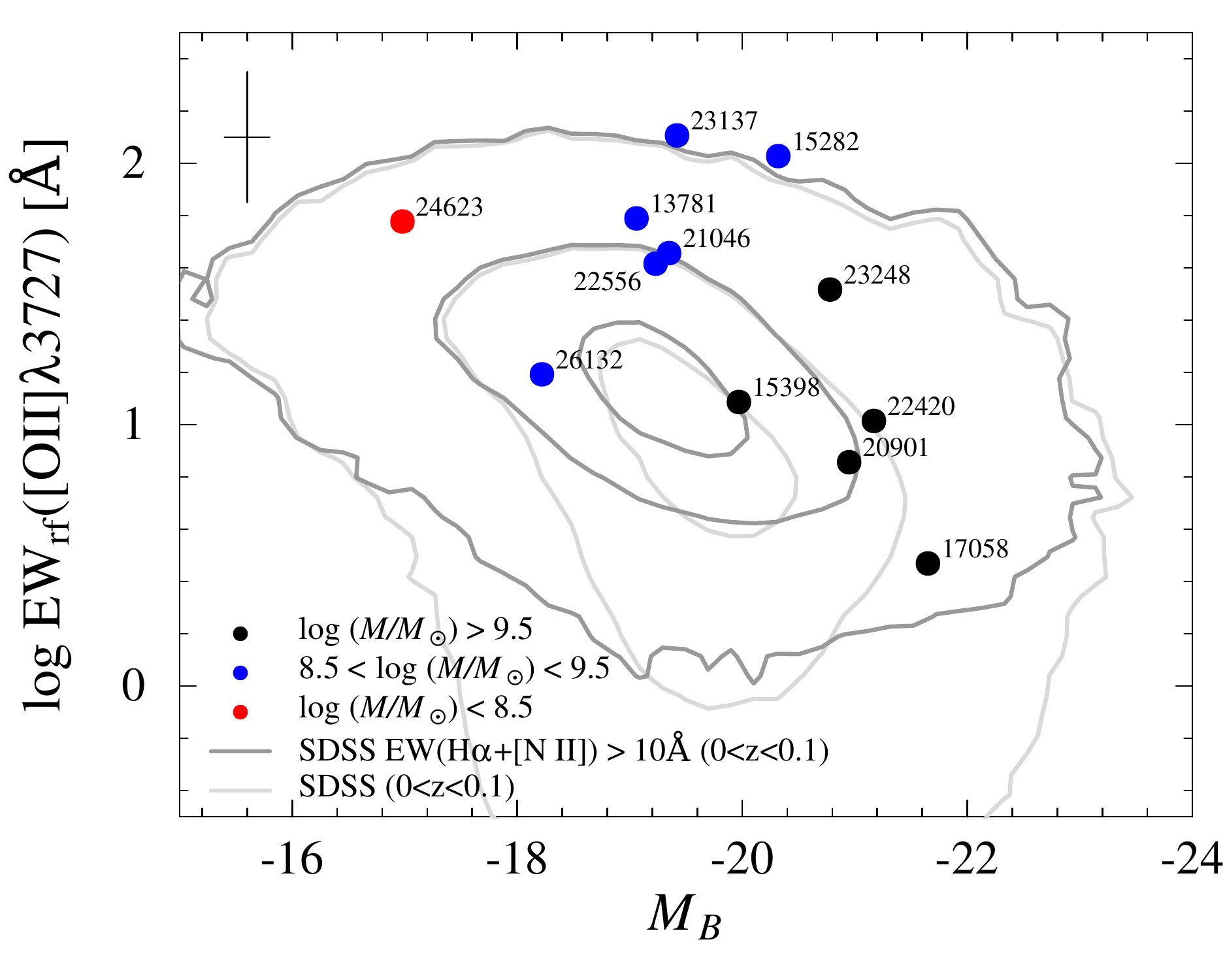}}
\caption{Rest-frame [\ion{O}{ii}] EW vs. rest-frame $B$-band absolute magnitude. The sample is represented as filled circles. The mass bins are shown in different colours as indicated in the legend. The SDSS reference is shown as grey lines with contour levels 99\%, 75\%, and 25\%. Average error bars are in the upper left corner.}
\label{fig:diag_ewl}
\end{figure}

We show the rest-frame EW([\ion{O}{ii}]$\lambda$3727) versus luminosity diagram in Fig.~\ref{fig:diag_ewl}. In this case, we employ the same mass bins and SDSS references as in Fig.~\ref{fig:diag_el}. We assumed a typical 15\% uncertainty in $\mathrm{EW_{rf}([\ion{O}{ii}])}$ \citep{1997ApJ...489..559G}. This diagram provides information about the star formation of the galaxies. The EW([\ion{O}{ii}]) is used to obtain the [\ion{O}{ii}] luminosities \citep{1997ApJ...489..559G}, a good tracer of the SFR \citep{1998ARA&A..36..189K}. In addition, the EW is directly related with the burst strength. The higher the EW([\ion{O}{ii}]), the higher the burst strength. We can see that the lowest mass system has one of the highest EW([\ion{O}{ii}]). Low-mass systems of this kind are candidates for being galaxies dominated by the starburst. Intermediate-mass galaxies have similar EW([\ion{O}{ii}]) to the low-mass galaxies, but they have great dispersion. These intermediate-mass objects receive an important contribution from the burst. The high-mass galaxies present lower EW([\ion{O}{ii}]), they have an important contribution from the continuum, and have lower burst strengths.

In terms of the mass distinction, we find only one galaxy with $\log {(M/M_{\odot})}<8.5$, namely CANDELS 24623. This object has a mass $\log {(M/M_{\odot})}=7.91\pm0.08$, which is one of the lowest values in our sample. It is a clear star-forming object with a very prominent [\ion{O}{ii}]$\lambda$3727 line and a very flat continuum (see Fig.~\ref{fig:spectra}). Its Spheroid(Compact) morphology and its high $\mathrm{EW_{rf}([\ion{O}{ii}])=96.8\,\AA}$ indicate that it could be completely dominated by the starburst. This low-mass, low-luminosity ($M_B=-16.98$) galaxy is our best low metallicity candidate. It is the only object in our spectroscopic sample with these characteristics and has rest-frame colours $U-V=0.58$ and $V-J=0.11$.

In the intermediate-mass interval $8.5<\log {(M/M_{\odot})}<9.5$, we find seven galaxies. These objects all show spectroscopic features corresponding to an \ion{H}{ii}-like class \citep{1997ApJ...489..559G} (see Fig.~\ref{fig:spectra}). They have intense emission lines with a flat continuum. In these cases [\ion{O}{iii}]$\lambda$5007 > H$\beta$. They show low to intermediate metallicities. The best example of this class is CANDELS 23137, with prominent emission lines and featureless continuum. Its excitation is high ($\mathrm{\log {([\ion{O}{iii}]/H\beta)}=0.71}$), as is the $\mathrm{EW_{rf}([\ion{O}{ii}])=205.3\,\AA}$, which is above the value of the low-mass galaxy 24623. CANDELS 15282 is the most massive of the galaxies in this mass bin with $\log {(M/M_{\odot})}=9.23\pm0.05$. Its properties are closer to the high-mass galaxies. CANDELS 21046 has an excitation value that does not correspond to its spectrum. The value that we get is higher than expected. With a detailed examination of its spectrum in Fig.~\ref{fig:spectra} we discovered that a bad sky subtraction has eliminated a pixel in the H$\beta$ line. The $\mathrm{FWHM(H\beta)=11.4\,\AA}$ is below $\mathrm{FWHM([\ion{O}{iii}]\lambda5007)=28.3\,\AA}$, which corresponds to the instrument profile width because the emission lines are not resolved. The estimated real value is indicated with an arrow in Fig.~\ref{fig:diag_el}. CANDELS 21046 and 13781 have the lowest excitation in this category, a signature of an older burst in star formation. The two remaining objects CANDELS 22556 and 26132 have a very noisy spectra and we cannot give further information. \ion{H}{ii}-like galaxies show average rest-frame colours $U-V=0.83\pm0.29$ and $V-J=0.44\pm0.30$.

The rest of the galaxies in the spectroscopic sample are in the high-mass bin $\log {(M/M_{\odot})}>9.5$. These five objects share characteristics related with a Disk-like class \citep{1997ApJ...489..559G} as we can see in the Fig.~\ref{fig:spectra} spectra. They have stronger H$\beta$ than [\ion{O}{iii}] and some absorption lines. These galaxies have higher metallicities than those in the previous categories. CANDELS 22420 and 23248 are the best cases in this Disk-like category. Their spectra show lower excitation with H$\beta$ > [\ion{O}{iii}]. In addition, a clear Balmer break appears with absorption lines, indicative of an older stellar population. All of these features are well connected with Disk morphologies and redder postage stamps. The rest of the spectra in this class are very noisy and we had found it very difficult to perform a proper measurement of the lines. These galaxies are the most luminous ones with luminosities $M_B=[-19.97,-21.65]$ and masses ranging $\log {(M/M_{\odot})}=[9.8,10.4]$. Disk-like galaxies have average rest-frame colours $U-V=1.13\pm0.18$ and $V-J=1.00\pm0.18$.

\section{H$\alpha$ luminosity function at $z\sim0.62$}
\label{sec:half}
In this section we show our calculation of the H$\alpha$-based LF at $z\sim0.62$. This is a very important redshift in H$\alpha$-based studies because no surveys cover the gap $0.5<z<0.8$. The LF determines how the SFR is distributed among the galaxy sample. First, we determined the observed LF. After that, we presented the extinction-corrected LF. We corrected both of them for incompleteness. The objects employed for this purpose are those classified as H$\alpha$ emitters by their spec-$z$ or photo-$z$ at $z\sim0.62$ as we indicated in Sect.~\ref{sec:haemitters}. We have a H$\alpha$ SFGs sample of 28 objects.

To build our H$\alpha$ LF we employed the $V/V_{\mathrm{max}}$ method \citep{1968ApJ...151..393S}. The number of galaxies per unit volume and interval $\Delta \log L$ is
\begin{equation}
 \phi (\log L_i) = \frac{1}{\Delta \log L \cdot \Omega}\sum\limits_j {\frac{1}{V_j^{\mathrm{max}}(z)}} , \label{eq:vvmax}
\end{equation}
\noindent where $L_i$ is the central luminosity in bin $i$ of interval $\Delta \log L$, $\Omega$ is the solid angle of the survey, and $V_j^{\mathrm{max}}(z)$ is the maximum volume in which we can detect the $j$ object per unit solid angle.

We needed to obtain $V_j^{\mathrm{max}}(z)$. The galaxies can be detected in the redshift interval of the NB filter effective width. This interval is $0.6098<z<0.6263$. The maximum volume of detection is the difference between the comoving volume at these redshifts. To calculate the comoving volume we used the \emph{Milia} library again, which provided us this value per solid angle. As the survey field is 7.5\arcmin$\times$7.5\arcmin, the solid angle is $\Omega=4.76 \times 10^{-6}$\,srad. The maximum volume of detection for each galaxy is then $V_j^{\mathrm{max}}=2.57 \times 10^{8}$\,Mpc$^3$ srad$^{-1}$. Hence, the volume covered by this survey is 1222\,Mpc$^3$. It is a smaller value than those used in other similar studies; for example \citet{2008ApJ...677..169V} at $z\sim0.84$ has a 15.4\arcmin$\times$15.4\arcmin field.

We used these values to build the LF. For the observed LF we chose a luminosity bin $\Delta \log L=0.4$\,erg s$^{-1}$ and a first limit $\log L=39.6$\,erg s$^{-1}$. These two values provide an optimal agreement between LF resolution and reliable statistics in the number of objects per bin. To determine the uncertainties in each bin we consider the process as Poissonian, so if the number of objects is $N$, the uncertainty is $\sqrt{N}$.

Next, we performed an extinction correction of the LF. A canonical extinction $A(\mathrm{H\alpha})=1$\,mag \citep{1992ApJ...388..310K,1998ARA&A..36..189K} is often applied in similar studies \citep[e.g.,][]{2008MNRAS.388.1473G,2010A&A...509L...5H,2012MNRAS.420.1926S,2013MNRAS.428.1128S,2013MNRAS.433..796D}. Other authors argue that an extinction correction done individually for each object is a better approach \citep[e.g.,][]{2007ApJ...657..738L,2008MNRAS.383..339W,2008ApJ...677..169V,2010ApJ...708..534W,2011ApJ...726..109L,2013MNRAS.433.2764G,2014ApJ...784..152A} as the extinction depends on redshift and on the luminosity of the galaxy, and affects the overall shape of the LF. Most luminous galaxies suffer higher extinctions flattening the shape of the faint end of the extinction-corrected LF \citep{2014ApJ...784..152A}. We decided to apply the most precise approach with the calculation of the extinction for each object individually as explained in Sect.~\ref{sec:extinction}. In this case, we operated with a bin $\Delta \log L=0.4$\,erg s$^{-1}$ and a first limit $\log L=40.0$\,erg s$^{-1}$.

For both observed and extinction-corrected LF we performed a correction for incompleteness. We followed the results obtained by our group in \citet{2008ApJ...677..169V}, as indicated in Sect.~\ref{sec:incompleteness}. To take this effect into account, the fraction of simulated galaxies that meet our selection criteria is calculated. This is the so-called completeness fraction. The correction is based on assuming that this fraction is the probability that a galaxy wil be detected and selected by our method. For each galaxy, we expect that the inverse of this fraction is the real number of galaxies. This is equivalent to multiplying the detection volume of each object by its completeness fraction. The factor is greater for the fainter sources.

\begin{figure}
\resizebox{\hsize}{!}{\includegraphics{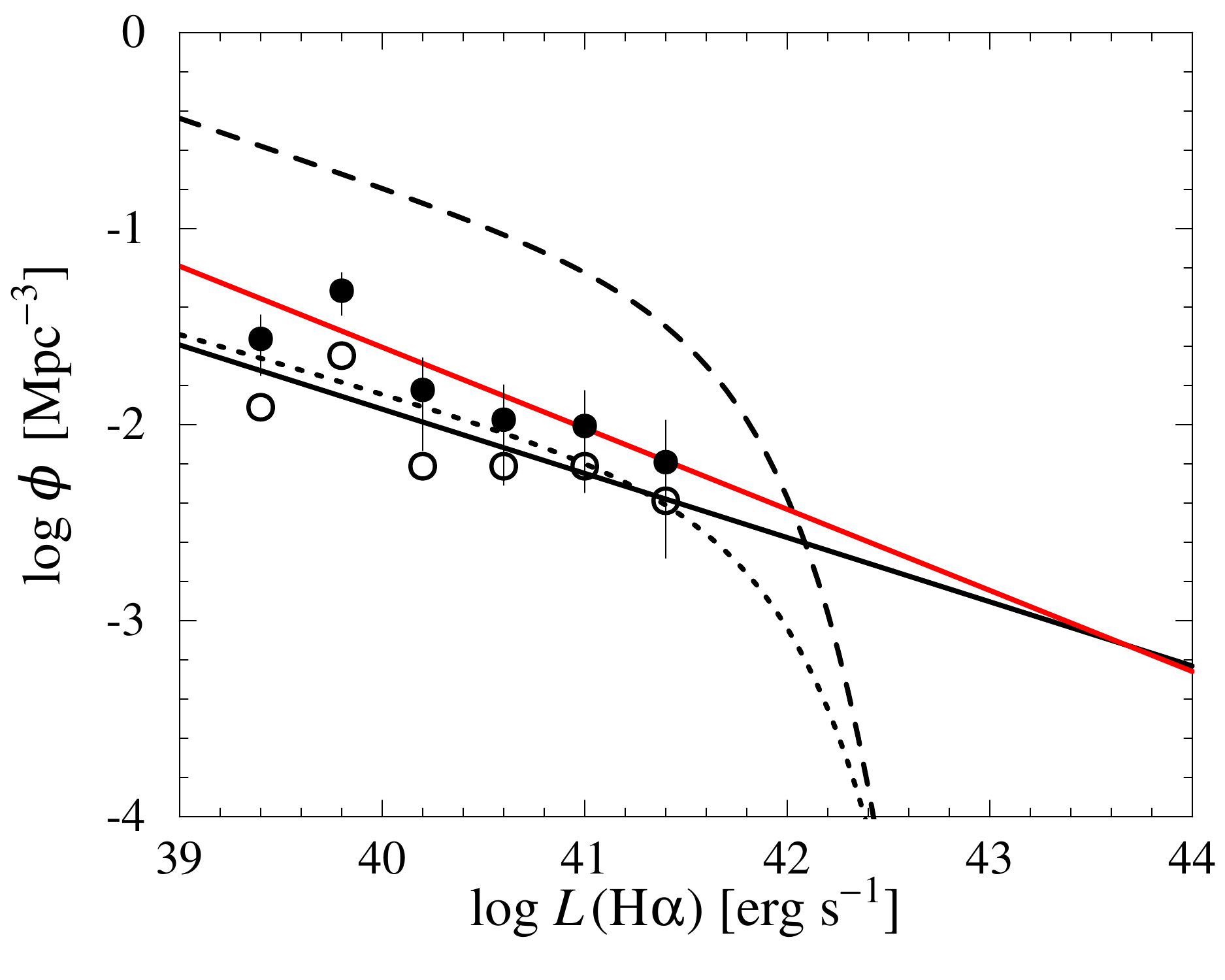}}
\caption{Observed H$\alpha$ luminosity function at $z\sim0.62$. We represent the LF corrected for incompleteness in black filled circles with the best linear fit as a red solid line. The open symbols indicate the LF values before applying this correction with the best linear fit as a black solid line. We also include as a reference the LF at $z\sim0$ from \citet{1995ApJ...455L...1G} (black dotted line) and the observed LF at $z\sim0.84$ from \citet{2008ApJ...677..169V} (black dashed line).}
\label{fig:lf_obs}
\end{figure}

\begin{figure}
\resizebox{\hsize}{!}{\includegraphics{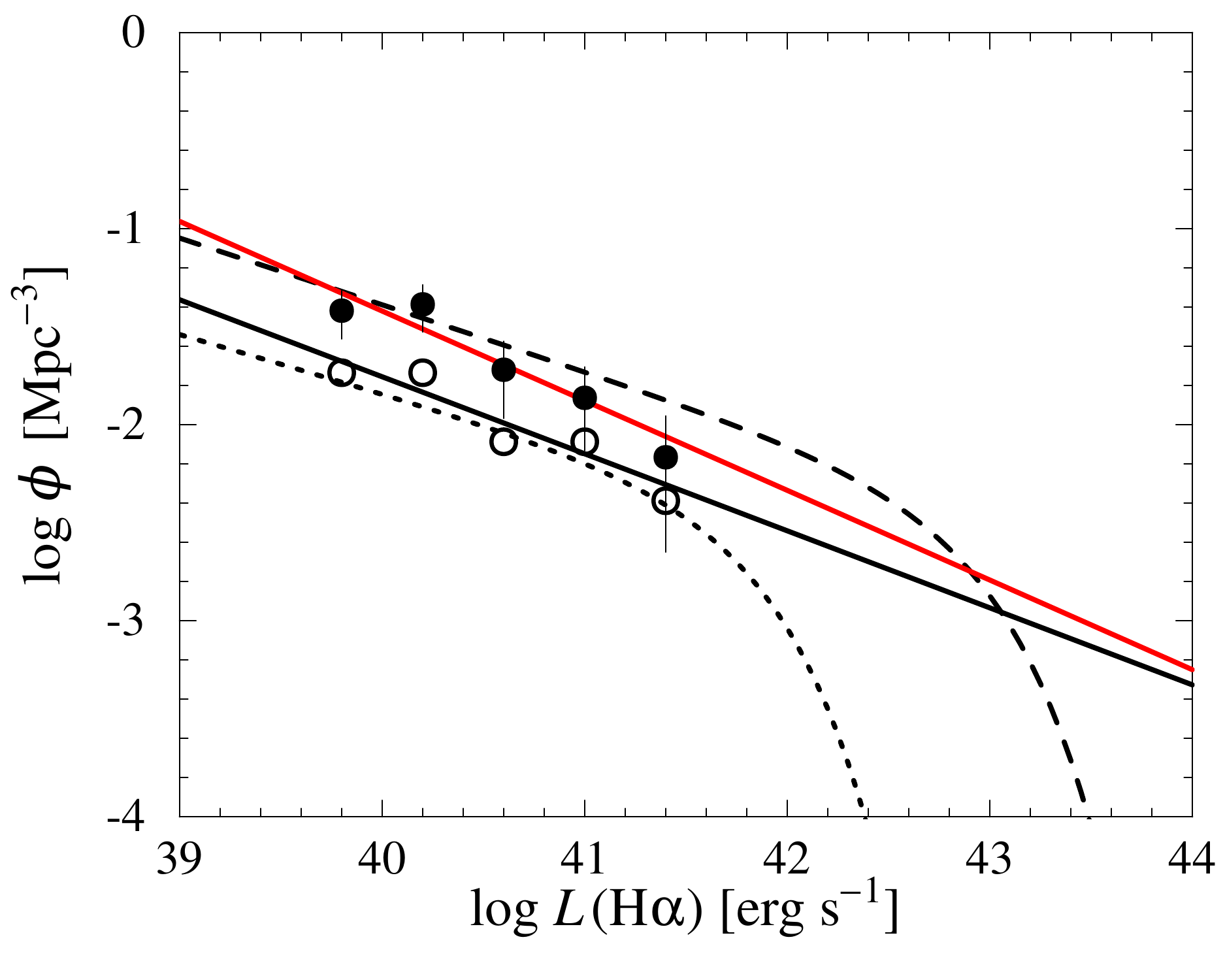}}
\caption{Extinction-corrected H$\alpha$ luminosity function at $z\sim0.62$. The LF corrected for incompleteness is shown as black filled circles along with the best linear fit as a red solid line. The LF values before applying the incompleteness correction are included as open symbols with the best linear fit as a black solid line. For comparison the LF at $z\sim0$ from \citet{1995ApJ...455L...1G} (black dotted line) and the extinction-corrected LF at $z\sim0.84$ from \citet{2008ApJ...677..169V} (black dashed line) are also shown.}
\label{fig:lf_ext}
\end{figure}

In Figs.~\ref{fig:lf_obs} and \ref{fig:lf_ext} we show both observed and extinction-corrected LFs along with the result once the incompleteness correction is applied. In Table~\ref{tab:lf} we gather the number densities of the emitters for each luminosity bin. We see that the reddening correction moves the LF towards brighter values. Moreover, an individual treatment for each galaxy causes a redistribution of the objects in the luminosity bins. Because of this, we have a five-point LF in the corrected case instead of the six-point LF for the observed one. The incompleteness correction makes the LF move up, because it is more important for the faintest bins, and makes the LF steeper.

\begin{table}
\caption{Number densities of H$\alpha$ emitters for each luminosity bin.}
\label{tab:lf}
\centering
\begin{tabular}{c c c}
\hline\hline
 & Extinction-corrected & Observed \\
$\log L(\mathrm{H\alpha})$ & $\log \phi$ & $\log \phi$ \\
(erg s$^{-1}$) & (Mpc$^{-3}$) & (Mpc$^{-3}$) \\
\hline
39.4 & (...) & -1.56(-1.91)$_{-0.23}^{+0.15}$ \\
39.8 & -1.42(-1.74)$_{-0.18}^{+0.12}$ & -1.32(-1.65)$_{-0.16}^{+0.11}$ \\
40.2 & -1.39(-1.74)$_{-0.18}^{+0.12}$ & -1.82(-2.21)$_{-0.37}^{+0.20}$ \\
40.6 & -1.72(-2.09)$_{-0.30}^{+0.18}$ & -1.97(-2.21)$_{-0.37}^{+0.20}$ \\
41.0 & -1.86(-2.09)$_{-0.30}^{+0.18}$ & -2.01(-2.21)$_{-0.37}^{+0.20}$ \\
41.4 & -2.17(-2.39)$_{-0.53}^{+0.23}$ & -2.19(-2.39)$_{-0.53}^{+0.23}$ \\
\hline
\end{tabular}
\tablefoot{
We show both extinction-corrected and observed values after applying the incompleteness correction. In parentheses we also include the values before applying the incompleteness correction.
}
\end{table}

We were able to model the LF to a typical Schechter function \citep{1976ApJ...203..297S,1995ApJ...455L...1G}
\begin{equation}
 \phi (L)dL = \phi^*\left(\frac{L}{L^*}\right)^\alpha \exp \left(-\frac{L}{L^*}\right)\frac{dL}{L^*} , \label{eq:schechter}
\end{equation}
\noindent where $L^*$ is the characteristic galaxy luminosity where the power law form of the function cuts off. The parameter $\phi^*$, in units of number density, provides the normalisation, and $\alpha$ is the faint-end slope of the LF. However, we note that there are no objects in the most luminous bins, the bright-end. We cannot trace the whole shape of the LF including the knee and the bright-end. We are limited to the faint end, nicely described by a power law. This issue is also present in other studies \citep{2008PASJ...60.1219M,2010A&A...509L...5H,2014ApJ...784..152A}. The brighter the galaxy, the fewer there are. As a result, a higher number of brighter galaxies are gathered when covering larger areas. Our field is smaller than the fields used in similar studies. For instance, \citet{2008ApJ...677..169V} have a field approximately four times larger and they find around four times more objects. Consequently, they reach an extinction-corrected maximum luminosity bin of $\log L=43.65$\,erg s$^{-1}$ and trace the LF knee properly. This gain of two orders of magnitude is due to the larger volume that has been sampled and also because larger extinctions corrections for their brightest galaxies have been applied.

\subsection{Faint end of the luminosity function}
\label{sec:felf}
Our survey is extremely deep. A total integration time in the NB of 31.9\,h led us to detect objects with fluxes $>1.7 \times 10^{-18}$\,erg s$^{-1}$ cm$^{-2}$. Considering only those objects classified as H$\alpha$ emitters at $z\sim0.62$, the intrinsic H$\alpha$ luminosity reached is $>2.9 \times 10^{39}$\,erg s$^{-1}$. We were able to reach a very faint population of galaxies. Consequently, we measured extremely deep luminosity bins in the LF faint end. This part of the LF is a power law whose slope is directly $1+\alpha$. We were able to measure a robust value of $\alpha$ with a simple linear fit. Robust values of $\alpha$ are only achievable through extremely deep surveys, like the one we conducted. In Figs.~\ref{fig:lf_obs} and \ref{fig:lf_ext} we show a wide coverage of the LF faint end with six points in the observed case and five points in the extinction-corrected case.

We obtained four measurements of the $\alpha$ parameter. The values are listed in Table~\ref{tab:alpha}. First, we obtained the observed and extinction-corrected values as the result of a weighted linear fit. The $\alpha$ reddening-corrected value is larger than the observed one. Following an individual extinction correction for each galaxy, other studies suggest the opposite \citep{2008ApJ...677..169V,2014ApJ...784..152A}. The highest attenuated sources move to their actual high luminosities flattening the LF and leading to a smaller $\alpha$ value. The reason for this discrepancy could depend on a different trend in terms of attenuation at the low luminostites that we trace in our work, that cover different ranges to those traced in the other studies (see Fig. \ref{fig:ext_lum}). Another reason could be the use of LFs divided into luminosity bins. The objects move to higher luminosities when the extinction correction is applied, but not enough to populate a new and more luminous bin and thus to lead to a shallower result. Second, we obtained the $\alpha$ values corrected for incompleteness. The results are steeper owing to the nature of this correction as the faintest bins are more affected. In all cases, considering the uncertainties we measure, all the different values obtained here would be indistinguishable. We discuss these matters further in Sect.~\ref{sec:disc}.

We calculated the uncertainties of the $\alpha$ parameters from Monte Carlo simulations. We computed $\sim$10000 values. We varied each point of the LF with a Poisson probability distribution characterised by the $\lambda$ parameter as the object detection follows a Poissonian process. The value of $\lambda$ for each case was the number of objects in each bin. We performed a linear fit for every case and obtained the probability distribution of the parameter. The final uncertainties in the $\alpha$ parameters are the percentiles in the distributions within 68\%, 1$\sigma$ confidence level. This process should account for the binnig effect on the $\alpha$ measurement.

\begin{table}
\caption{$\alpha$ values from LF faint-end linear fit.}
\label{tab:alpha}
\centering
\begin{tabular}{l c}
\hline\hline
 & $\alpha$ \\
\hline
Extinction-corrected + incompleteness & -1.46$_{-0.08}^{+0.16}$ \\
Extinction-corrected & -1.39$_{-0.12}^{+0.18}$ \\
Observed + incompleteness & -1.41$_{-0.08}^{+0.12}$ \\
Observed & -1.33$_{-0.10}^{+0.15}$ \\
\hline
\end{tabular}
\tablefoot{
We collect the values of both the extinction-corrected and observed LFs, after and before applying the incompleteness correction.
}
\end{table}

\section{Discussion}
\label{sec:disc}
We discuss several considerations of our results in relation with similar studies. First of all, we want to treat the characterisation of the sample. Our ultra-deep observations led us to detect a wide variety of objects as we show through our morphological classification. These different categories of galaxies have different properties. In Table~\ref{tab:param_morpho} we include the average values and deviations for several properties within the morphological categories. In this case, we only use the H$\alpha$ sample and the simplest version of the classification as we indicated in Sect.~\ref{sec:morph}. We include Spheroid galaxies and the Irregular CANDELS 15784 and 26460 in the Compact category. We combine the Spiral and Disk classes along with CANDELS 23137 into the same Disk. With this reduced classification 72\% are Disk, 21\% are Compact, 7\% are Unclassifiable.

\begin{table*}
\caption{Average values and deviations of several physical properties for the `H$\alpha$' sample.}
\label{tab:param_morpho}
\centering
\begin{tabular}{l r r r r r r r r}
\hline\hline
Category & $\log (M/M_{\odot})$ & $\log r_e$ & $n$\tablefootmark{a} & $M_B$ & $\mathrm{EW_{rf}(H\alpha)}$ & $A(\mathrm{H\alpha})$ & SFR$_\mathrm{cor}$ \\
 & & (kpc) & & & (\AA) & (mag) & ($M_{\odot}$ yr$^{-1}$) \\
\hline
Disk (72\%) & 8.94 $\pm$ 0.85 & -0.40 $\pm$ 0.20 & 0.92 $\pm$ 0.31 & -18.88 $\pm$ 1.69 & 61 $\pm$ 140 & 0.40 $\pm$ 0.34 & 0.52 $\pm$ 0.81 \\
Compact (21\%) & 7.71 $\pm$ 0.22 & -0.70 $\pm$ 0.10 & 1.87 $\pm$ 0.87 & -16.24 $\pm$ 0.65 & 141 $\pm$ 250 & 0.97 $\pm$ 0.65 & 0.13 $\pm$ 0.07 \\
Unclassifiable (7\%) & 7.54 $\pm$ 0.06 & -0.49 $\pm$ 0.14 & (...) & -15.16 $\pm$ 0.07 & 91 $\pm$ 40 & 1.24 $\pm$ 0.27 & 0.08 $\pm$ 0.01 \\
\hline
\end{tabular}
\tablefoot{
\tablefoottext{a}{The S{\'e}rsic index values only consider the objects with structural parameters available in \citet{2012ApJS..203...24V}.}
}
\end{table*}

In terms of mass and size the Disk class shows the highest values with large deviations because this category includes a wide variety of disks. Disk is also the category with the highest luminosity and SFR. Thus, these galaxies dominate the brighter part of our LF. In contrast, Compact galaxies have very low masses and sizes, with low luminosites and SFRs. However, the higher EW(H$\alpha$) and extinction that they present compared to the Disk galaxies is indicative and shows a more important contribution of the burst and higher dust content. In addition, as we noted in Sect.~\ref{sec:morph}, the S{\'e}rsic index $n\sim2$ is very distinctive of this Compact category. In addition to the two main classes, we find two galaxies marked as Unclassifiable (CANDELS 17503 and 31370). These are very interesting objects as they show the lowest masses, luminosities, and SFRs with the highest extinctions. In terms of specific star formation rate (sSFR), the galaxies with the highest values are Compact galaxies with $\mathrm{sSFR(Compact)}=2.8\pm1.9$\,Gyr$^{-1}$, a much higher value than $\mathrm{sSFR(Disk)}=0.7\pm1.1$\,Gyr$^{-1}$ ($\mathrm{sSFR(Unclassifiable)}=2.3\pm0.7$\,Gyr$^{-1}$). The contribution of star formation in Compact galaxies relative to their mass is much more important than in the case of Disk galaxies. They are low-mass, compact, starburst-dominated systems. The epoch when low-mass SFGs form the bulk of their stellar mass is controversial. \citet{2015ApJ...799...36R} find results that suggest a recent stellar mass assembly in agreement with cosmological downsizing \citep{1996AJ....112..839C} from a sample of 31 spectroscopically confirmed galaxies at $0.3<z<0.9$ with masses $7.3 \leq \log (M/M_{\odot}) \leq 8.0$.

We can compare our sample with other works at different redshifts. Locally, \citet{1996A&AS..120..385V} morphologically classified the UCM Survey sample. The results were 10\% early types, 83\% disks, 5\% compacts, and 3\% irregulars. These percentages are roughly similar to our results at $z\sim0.62$. The population of SFGs in the universe at $z\sim0.62$ is similar to local galaxies, morphologically speaking. The bulk of star formation appears to be contained within disks at these redshifts $0<z<0.62$. Previous studies have already shown that SFGs were dominated by disks at $z\sim0.84$ \citep{2008ApJ...677..169V,2009MNRAS.398...75S}. For instance, the sample of SFGs from \citet{2009MNRAS.398...75S} is composed of 80\% of disks. At $z\sim1$ \citet{2011PASJ...63S.363K} found that star formation is mainly located on massive disk galaxies that are not triggered by early-phase galaxy-galaxy interactions disturbing the morphology in the NIR. The change from the locally similar scenario is found in studies at $z>1$. \citet{2012ApJ...745...85L} suggested that the picture of SFGs at $z\sim2$ is consistent with gas-rich, compact, and triaxial systems dominated by velocity dispersion, rather than an axisymmetric disk view. The sample of ELGs at $z\sim2.24$ from \citet{2014ApJ...784..152A} have a distribution of 45\% that show signs of merger/interaction and just 23\% of disks. This survey also has 18\% of compact systems similar to ours, indicating that morphologies of this kind are present in a long stage of the cosmic history, with number percentages showing slight changes. Our ultra-deep survey allows us to detect very faint star formation regimes (i.e., $\mathrm{SFR}<0.1\,M_{\odot}$ yr$^{-1}$) that are dominated by compact morphologies. \citet{2011MNRAS.413.1236B} studied a sample of faint H$\alpha$ emitters at $0.002<z<0.13$ with similar low stellar masses obtaining similar SFR and compact radii to ours. The difference in the percentages in the morphologies of these galaxies in other surveys is probably due to the depth of the study, but in general, the morphologies are similar in studies at $z<1$. This is also an argument in favour of narrow-band surveys like ours. Covering smaller volumes and detecting a smaller number of galaxies, it detects a representative sample of the morphologies in the universe at the studied redshift.

\begin{figure}
\resizebox{\hsize}{!}{\includegraphics{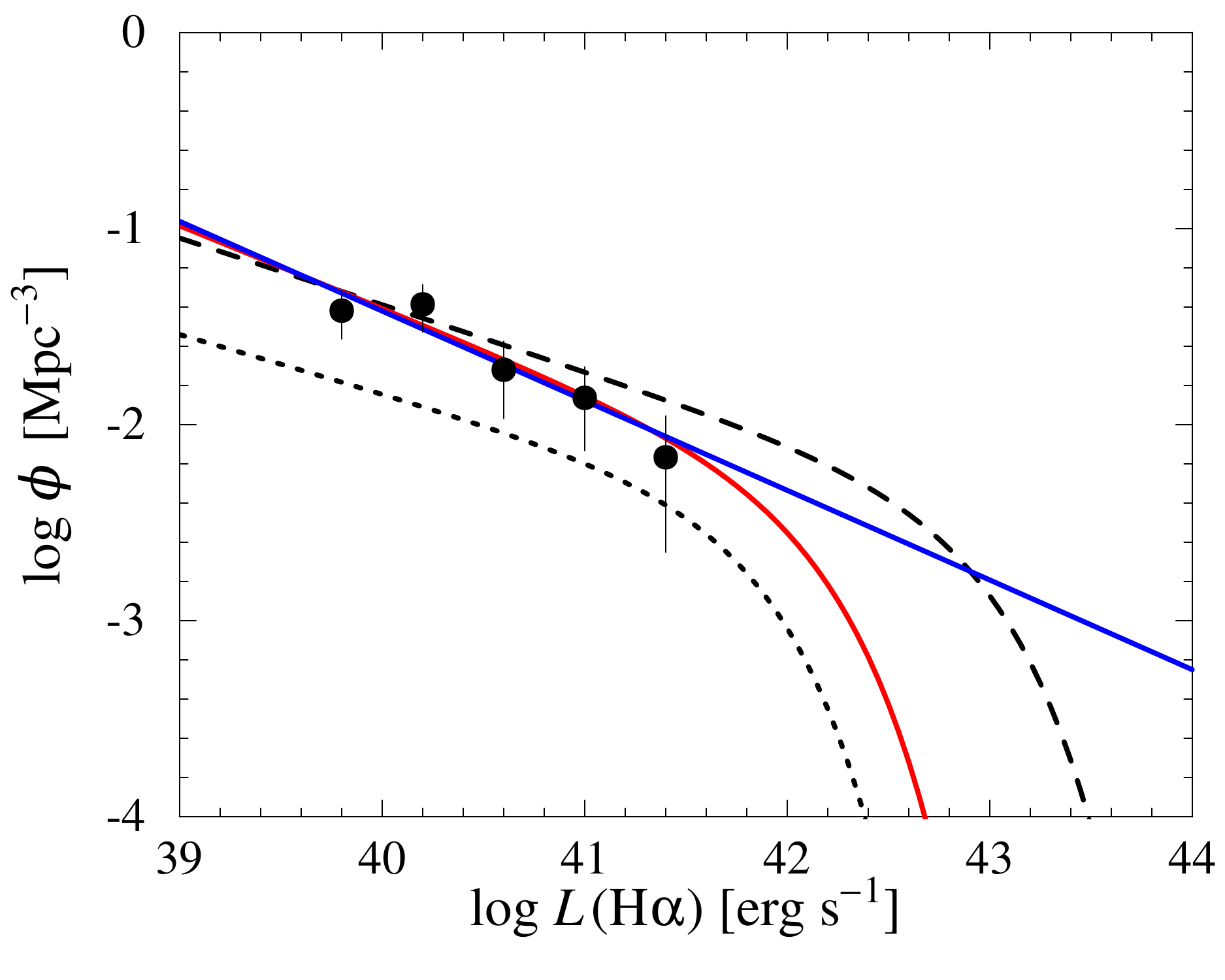}}
\caption{Extinction and incompleteness corrected H$\alpha$ luminosity function at $z\sim0.62$. The best Schechter function fit is shown as a red solid line and the best linear fit as a blue solid line. LF at $z\sim0$ from \citet{1995ApJ...455L...1G} (black dotted line) and the extinction-corrected LF at $z\sim0.84$ from \citet{2008ApJ...677..169V} (black dashed line) are also included.}
\label{fig:lf_sch}
\end{figure}

In Sect.~\ref{sec:felf} we calculated the LF faint-end slope by performing a linear fit. This is done under the assumption that none of the LF points are tracing the LF knee or the bright-end. They all lie in the LF regime that is accurately described by a power law. According to the results from \citet{1995ApJ...455L...1G} and \citet{2008ApJ...677..169V} this is a safe assumption (see Figs.~\ref{fig:lf_obs} and \ref{fig:lf_ext}). However, it is worth discussing what the result would be if we did not make this previous supposition and if we fit a typical Schechter function \citep{1976ApJ...203..297S}. A simple way to do this with our data is by fixing the characteristic luminosity $L^*$ and by fitting $\phi^*$ and $\alpha$. \citet{2013MNRAS.428.1128S} propose $\log L^*\mathrm{(H\alpha)}(z)=0.45z+41.87$ as an evolution law for $L^*\mathrm{(H\alpha)}$ with redshift. At $z\sim0.62$, $\log L^*\mathrm{(H\alpha)}=42.15$\,erg s$^{-1}$. A Schechter fit leads us to the following parameters where the uncertainties come from Monte Carlo realisations, as in the $\alpha$ linear fit: $\phi^*=10^{-2.67_{-0.18}^{+0.29}}\,\mathrm{Mpc^{-3}}$ and $\alpha=-1.42_{-0.09}^{+0.15}$. The $\alpha$ value obtained in this way is similar to those presented in Table~\ref{tab:alpha} calculated with a linear fit. This confirms that the assumption made in Sect.~\ref{sec:felf} is correct. In Fig.~\ref{fig:lf_sch} we show the output of this approach.

\begin{figure}
\resizebox{\hsize}{!}{\includegraphics{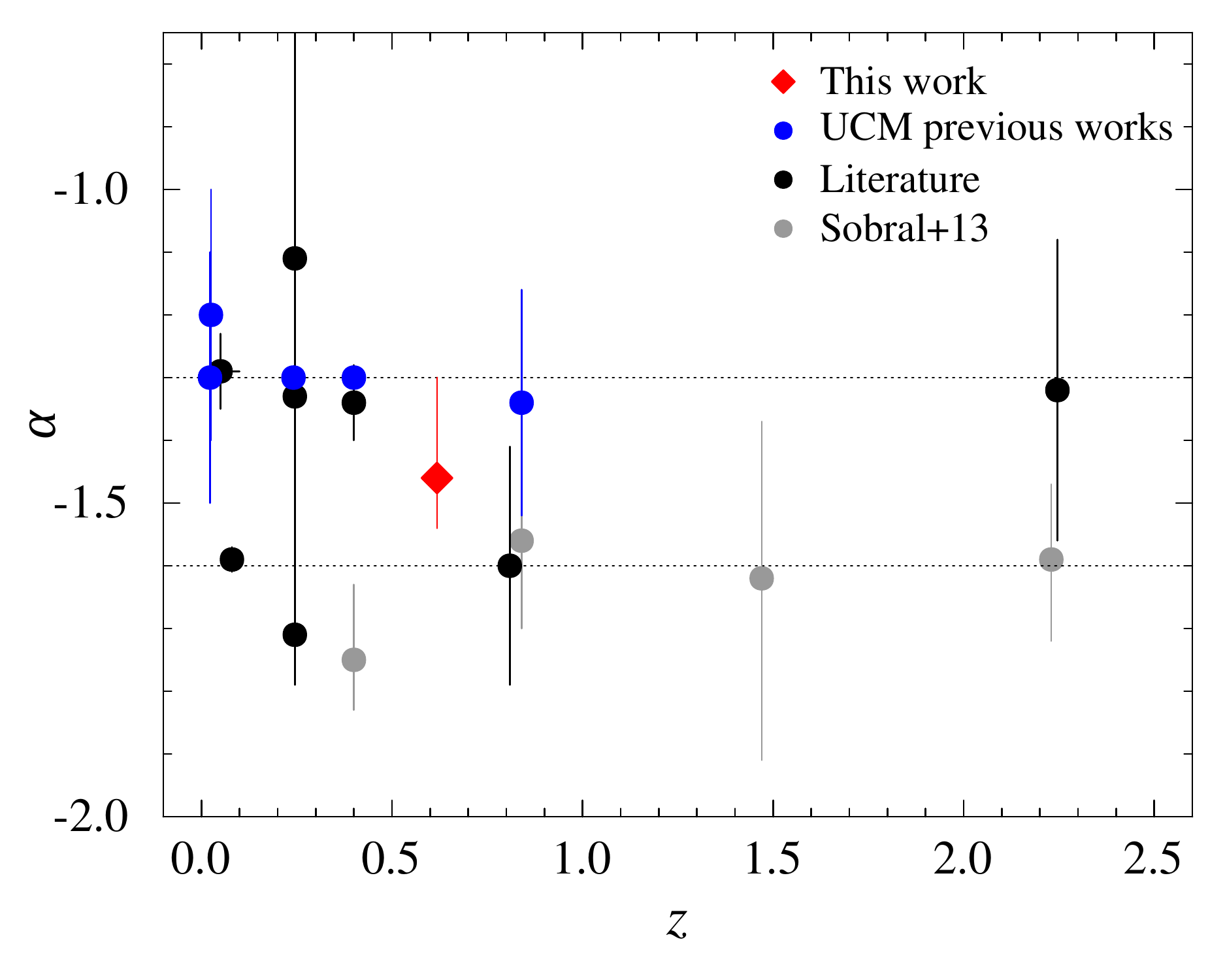}}
\caption{Evolution of the $\alpha$ parameter as calculated by several studies in the literature along with our result. This work is shown as red filled diamonds. We present the extinction + incompleteness corrected value obtained through a linear fit. We emphasise previous UCM works as blue filled circles. (\citet{2001A&A...379..798P,2005PASP..117..120P} fix the $\alpha=-1.3$ value). The black circles designate several studies from the literature: at $z<0.1$ \citet{2015MNRAS.447..875G}; at $z\sim0.24$ \citet{2008MNRAS.383..339W}, \citet{2007ApJ...657..738L} $z<0.4$, and \citet{2011ApJ...726..109L} $z\sim0.8$; and at $z\sim2.24$ \citet{2014ApJ...784..152A}. The evolution from \citet{2013MNRAS.428.1128S} at $z\sim0.4,0.84,1.47,2.23$ is shown as grey filled circles.}
\label{fig:alpha_evol}
\end{figure}

We can compare the $\alpha$ values we calculate in this work with the values in the literature. We detect objects with fluxes $>1.7 \times 10^{-18}$\,erg s$^{-1}$ cm$^{-2}$, thus reaching luminosity bins $>\log L=39.4$\,erg s$^{-1}$ ($>\log L=39.8$\,erg s$^{-1}$ extinction-corrected). We can robustly obtain a value of the $\alpha$ parameter. In Fig.~\ref{fig:alpha_evol} we plot the evolution of $\alpha$ with redshift from several literature studies along with our result from a linear fit and corrected for extinction and incompleteness. We clearly see two trends: a non-evolving $\alpha=-1.3$ \citep{1995ApJ...455L...1G,2003ApJ...591..827P,2008MNRAS.383..339W,2008ApJ...677..169V,2014ApJ...784..152A} and a non-evolving $\alpha=-1.6$ \citep{2007ApJ...657..738L,2011ApJ...726..109L,2012MNRAS.420.1926S} with redshift. \citet{2014ApJ...784..152A} suggest that this difference is mostly explained as the result of applying an individual extinction correction for each galaxy ($\alpha=-1.3$) or using a canonical value $A\mathrm{(H\alpha)=1}$\,mag ($\alpha=-1.6$). These authors argue that an individual extinction correction is vital in the shape of the H$\alpha$ LF, as a constant value tends to overestimate the effect in the faint end of the LF and to underestimate it in the bright end, which causes a steeper $\alpha$.

\begin{figure}
\resizebox{\hsize}{!}{\includegraphics{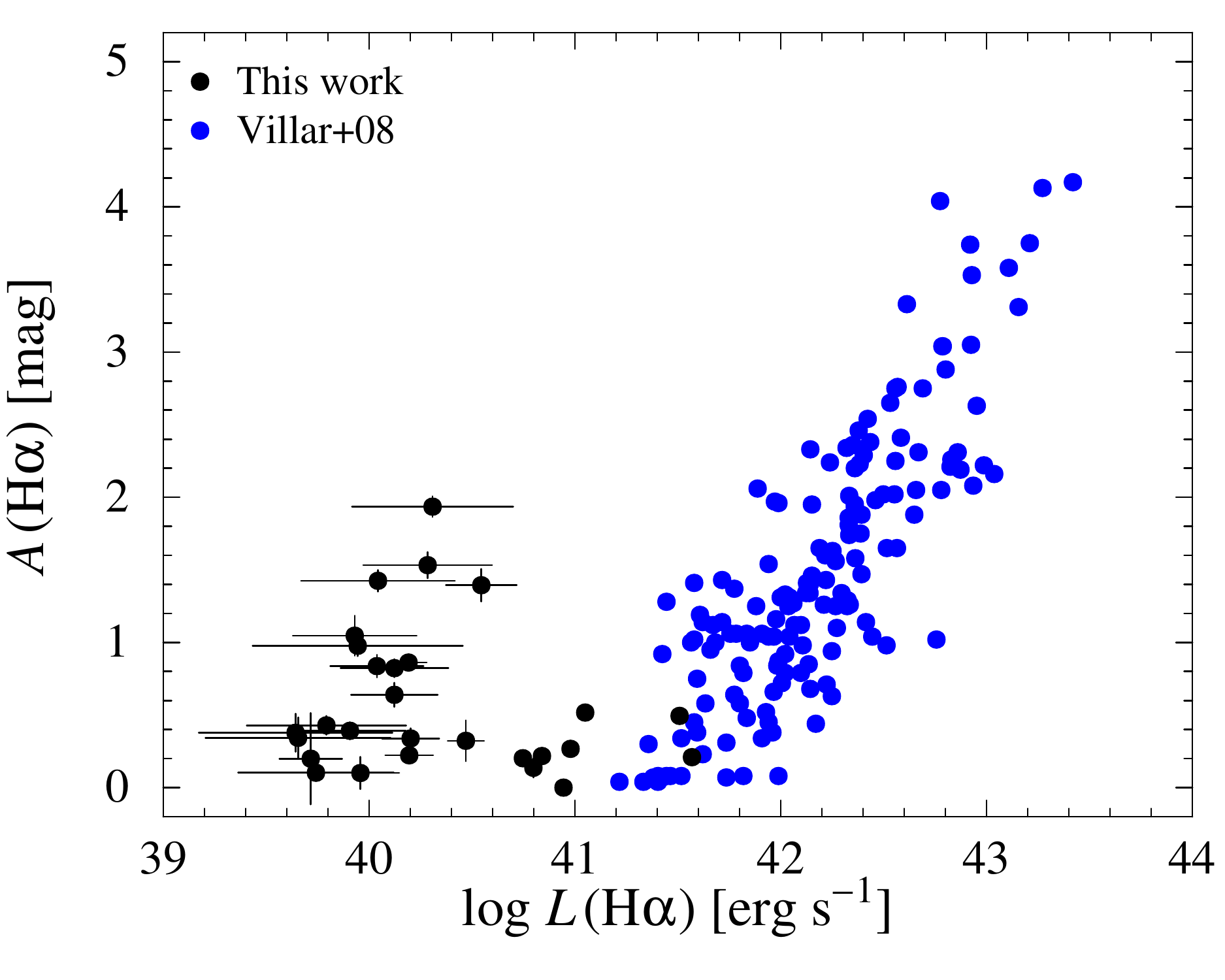}}
\caption{H$\alpha$ extinction vs. H$\alpha$ luminosity. The objects in our sample are shown as black filled circles and the ones from \citet{2008ApJ...677..169V} at $z\sim0.84$ as blue filled circles.}
\label{fig:ext_lum}
\end{figure}

Our results suggest that correcting for extinction for each object individually goes in the opposite direction and produces a steeper $\alpha$ (see Table~\ref{tab:alpha}). The reason for this discrepancy could be a real physical difference in the galaxy population in terms of attenuation at the lowest luminosities. Nevertheless, we have to take into account the resolution in our LF that is due to the choice of the luminosity bins. The objects move to higher luminosities when the extinction correction is applied, but not enough to populate a new and more luminous bin leading to a steeper result. By performing Monte Carlo realisations to get the uncertainties we are accounting for this effect. It is important to note that the $\alpha$ results presented in Table~\ref{tab:alpha} and the values given by \citet{2014ApJ...784..152A} ($\alpha=-1.32\pm0.24$) are indistinguishable within the uncertainties. If there is a physical origin in the discrepancy, in order to explain the steeper $\alpha$ after applying an individual extinction correction the galaxies with lower luminosities would be the ones with higher extinctions (see Table~\ref{tab:param_morpho}). The objects with lower luminosities as shown by $M_B$ are the ones with higher extinctions. In Fig.~\ref{fig:ext_lum} we explore this further by plotting the extinction as a function of the H$\alpha$ luminosity. Covering different luminosity ranges to those used in our work, we do not see the same trend in the sample from \citet{2008ApJ...677..169V} and Fig.~10 in \citet{2014ApJ...784..152A}. In these works the higher the luminosity, the higher the extinction. However, in our work we see that the galaxies with the highest luminosities do not present the highest extinctions. In fact, the objects with the highest extinctions have much lower luminosities. This could explain why we obtain a steeper value in the LF faint-end slope. As a sanity check we verify that we get the same behaviour for our objects when comparing the H$\alpha$ extinction with $M_B$ because the $M_B$ values are independent from our measurements.

\begin{figure}
\resizebox{\hsize}{!}{\includegraphics{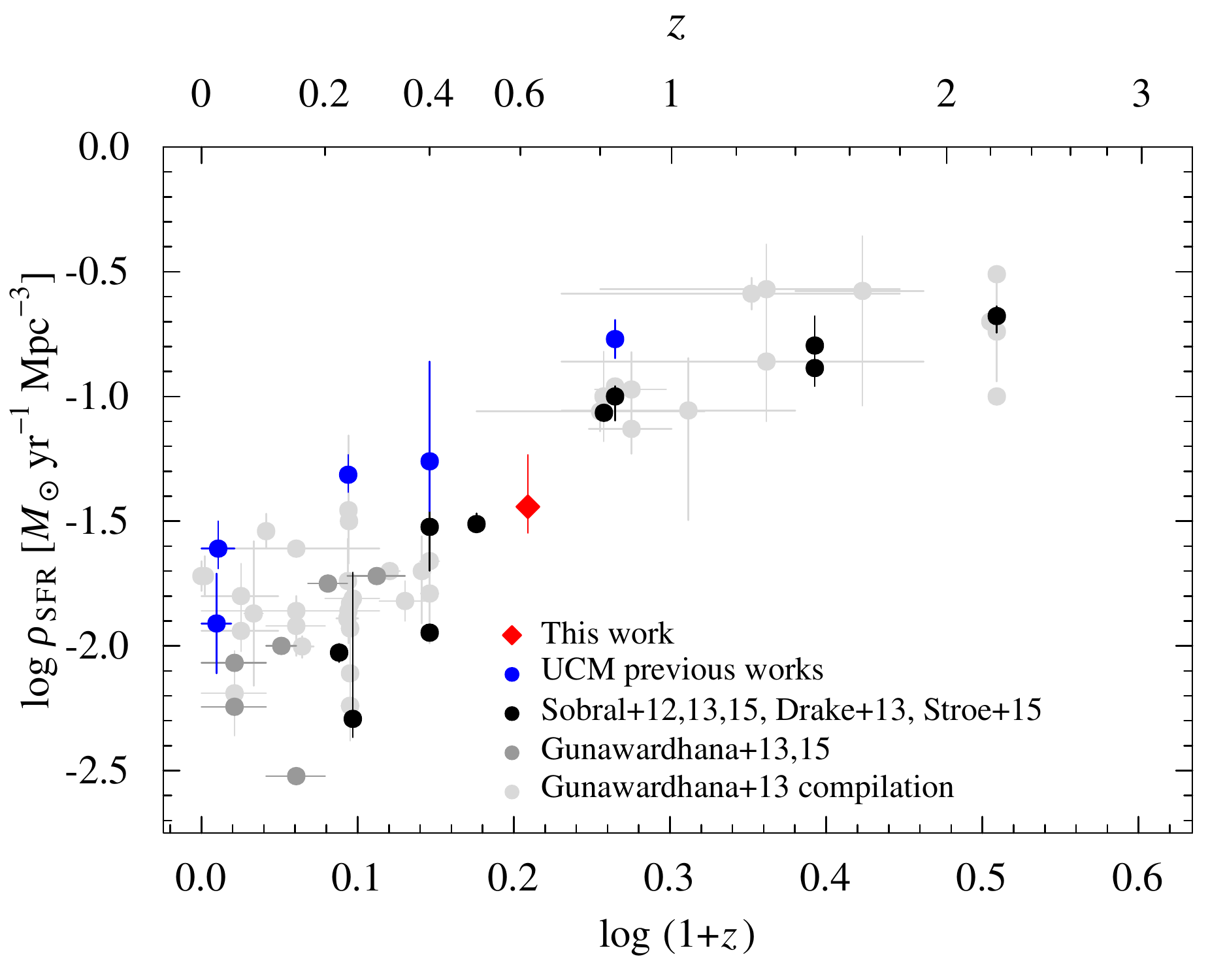}}
\caption{Evolution of the SFRD with redshift from H$\alpha$ studies. This work is shown as red filled diamond. We indicate previous UCM works as blue filled circles. The black circles correspond to the studies of \citet{2012MNRAS.420.1926S} at $z\sim1.47$; \citet{2013MNRAS.428.1128S} at $z\sim0.4,0.84,1.47,2.23$; \citet{2013MNRAS.433..796D} at $z\sim0.25,0.4,0.5$; \citet{2015MNRAS.451.2303S} at $z\sim0.8$; and \citet{2015MNRAS.453..242S} at $z\sim0.2$. The compilation in Table B1 of \citet{2013MNRAS.433.2764G} is included as grey filled circles. The results in Table 2 of \citet{2013MNRAS.433.2764G} from both GAMA and SDSS-DR7 data are shown as dark grey filled circles, substituted by the more recent values in \citet{2015MNRAS.447..875G} for the redshift bins $0.1<z<0.15$, $0.17<z<0.24$, and $0.24<z<0.35$.}
\label{fig:madau}
\end{figure}

Additionally, we can calculate the H$\alpha$ luminosity density,
\begin{equation}
 \rho_L\left({\mathrm{H}\alpha}\right) = {\phi^*}{L^*}\Gamma\left({2+\alpha}\right) , \label{eq:lumdens}
\end{equation}
from the Schechter parameters previously obtained. This result can be converted to SFRD through the \citet{1998ARA&A..36..189K} calibration. We get a value of $\rho_\mathrm{SFR} = 0.036_{-0.008}^{+0.012}\,M_{\odot}~\mathrm{yr^{-1}~Mpc^{-3}}$. We put our result in the context of the SFRD evolution in Fig.~\ref{fig:madau}. The result follows the trend established by previous works in the literature. We include the H$\alpha$-based SFRD complitation in Table B1 of \citet{2013MNRAS.433.2764G}, as well as the values calculated by these authors in the redshift range $z<0.35$ \citep{2013MNRAS.433.2764G,2015MNRAS.447..875G}. We substitute the values from GAMA data at $z>0.1$ for the newly calculated values in \citet{2015MNRAS.447..875G}. In addition, we add the most recent works not included in the mentioned compilation \citep{2012MNRAS.420.1926S,2013MNRAS.428.1128S,2013MNRAS.433..796D,2015MNRAS.451.2303S,2015MNRAS.453..242S}. All studies here are compared assuming a common concordance cosmology, Salpeter IMF, and applying a extinction correction \citep{2013MNRAS.433.2764G}. Our work is the first conducted in H$\alpha$ at $z\sim0.6$, connecting and filling the gap between $0.5<z<0.8$ studies.

\section{Summary}
\label{sec:sum}
In this project we selected a sample of 28 star-forming galaxies at $z\sim0.62$ from their H$\alpha$ emission. We employed narrow-band ultra-deep VLT/HAWK-I observations, for a total of 31.9\,h in this filter centred at 1.06\,$\mu$m. The main goal of this work is to study the properties of the resulting sample of faint H$\alpha$ emitters at $z\sim0.62$. We mainly focus on their contribution to the faint end of the luminosity function and we derived the star formation rate density. We also characterised their morphologies and basic photometric and spectroscopic properties. In studies of this kind $z\sim0.62$ is an important redshift region; there is a gap in $0.5<z<0.8$ H$\alpha$ works due to the efficiency problems in the detector transition from optical to near-infrared.

\begin{itemize}
  \item We characterised the sample. Morphologically, we collected a disk-dominated sample (79\%) with an important contribution from compact galaxies (21\%). This is consistent with a local scenario of disk systems dominating the bulk of star-forming galaxies at $0<z<1$.
  \item The star formation rate is higher in disks, but the specific star formation rate is greater in compact morphologies. These galaxies are low-mass, compact, starburst-dominated systems. The structural parameter analysis shows light profiles in an intermediate stage from early- to late-types because they have S{\'e}rsic indexes $n\sim2$.
  \item We calculated a robust value for the faint-end slope of the luminosity function $\alpha=-1.46_{-0.08}^{+0.16}$, applying an individual extinction correction for each galaxy in the sample and taking into account the incompleteness factor.
    \item The extinction is lower in the emitters with the highest H$\alpha$ luminosity than in fainter sources within the sample. This could explain the steeper $\alpha$ value obtained after individually correcting each object for extinction.
  \item We obtained an extinction-corrected star formation rate density $\rho_\mathrm{SFR} = 0.036_{-0.008}^{+0.012}\,M_{\odot}~\mathrm{yr^{-1}~Mpc^{-3}}$ at $z\sim0.62$. This result is consistent with the evolution of this parameter with cosmic history as shown in the literature.
\end{itemize}

\begin{acknowledgements}
We would like to thank the anonymous referee whose valuables comments have improved this work. We acknowledge support from the Spanish Programa Nacional de Astronom\'ia y Astrof\'isica: projects AYA2013-46724-P, AYA2012-30717, and AYA2009-10368. This work has made use of the Rainbow Cosmological Surveys Database, which is operated by the Universidad Complutense de Madrid (UCM), partnered with the University of California Observatories at Santa Cruz (UCO/Lick,UCSC). We are very grateful to Pablo G. P\'erez-Gonz\'alez for his technical advice on Rainbow usage and its data. We thank Guillermo Barro for helping with the stellar masses catalogue, discussions, and last details and Michael Cooper for kindly providing his available spectra for our sample. We thank Nicol\'as Cardiel, Omaira Gonz\'alez-Mart\'in, and Sune Toft for useful comments and advice. C.G.G acknowledges funding by the Spanish Beca de Colaboraci\'on 2013-2014 programme.
\end{acknowledgements}


\bibliographystyle{aa}{}
\bibliography{fha_z062.bib}{}

\begin{thebibliography}{110}
\expandafter\ifx\csname natexlab\endcsname\relax\def\natexlab#1{#1}\fi

\bibitem[{{Aihara} {et~al.}(2011){Aihara}, {Allende Prieto}, {An}, {Anderson},
  {Aubourg}, {Balbinot}, {Beers}, {Berlind}, {Bickerton}, {Bizyaev}, {Blanton},
  {Bochanski}, {Bolton}, {Bovy}, {Brandt}, {Brinkmann}, {Brown}, {Brownstein},
  {Busca}, {Campbell}, {Carr}, {Chen}, {Chiappini}, {Comparat}, {Connolly},
  {Cortes}, {Croft}, {Cuesta}, {da Costa}, {Davenport}, {Dawson}, {Dhital},
  {Ealet}, {Ebelke}, {Edmondson}, {Eisenstein}, {Escoffier}, {Esposito},
  {Evans}, {Fan}, {Femen{\'{\i}}a Castell{\'a}}, {Font-Ribera}, {Frinchaboy},
  {Ge}, {Gillespie}, {Gilmore}, {Gonz{\'a}lez Hern{\'a}ndez}, {Gott}, {Gould},
  {Grebel}, {Gunn}, {Hamilton}, {Harding}, {Harris}, {Hawley}, {Hearty}, {Ho},
  {Hogg}, {Holtzman}, {Honscheid}, {Inada}, {Ivans}, {Jiang}, {Johnson},
  {Jordan}, {Jordan}, {Kazin}, {Kirkby}, {Klaene}, {Knapp}, {Kneib},
  {Kochanek}, {Koesterke}, {Kollmeier}, {Kron}, {Lampeitl}, {Lang}, {Le Goff},
  {Lee}, {Lin}, {Long}, {Loomis}, {Lucatello}, {Lundgren}, {Lupton}, {Ma},
  {MacDonald}, {Mahadevan}, {Maia}, {Makler}, {Malanushenko}, {Malanushenko},
  {Mandelbaum}, {Maraston}, {Margala}, {Masters}, {McBride}, {McGehee},
  {McGreer}, {M{\'e}nard}, {Miralda-Escud{\'e}}, {Morrison}, {Mullally},
  {Muna}, {Munn}, {Murayama}, {Myers}, {Naugle}, {Neto}, {Nguyen}, {Nichol},
  {O'Connell}, {Ogando}, {Olmstead}, {Oravetz}, {Padmanabhan},
  {Palanque-Delabrouille}, {Pan}, {Pandey}, {P{\^a}ris}, {Percival},
  {Petitjean}, {Pfaffenberger}, {Pforr}, {Phleps}, {Pichon}, {Pieri}, {Prada},
  {Price-Whelan}, {Raddick}, {Ramos}, {Reyl{\'e}}, {Rich}, {Richards}, {Rix},
  {Robin}, {Rocha-Pinto}, {Rockosi}, {Roe}, {Rollinde}, {Ross}, {Ross},
  {Rossetto}, {S{\'a}nchez}, {Sayres}, {Schlegel}, {Schlesinger}, {Schmidt},
  {Schneider}, {Sheldon}, {Shu}, {Simmerer}, {Simmons}, {Sivarani}, {Snedden},
  {Sobeck}, {Steinmetz}, {Strauss}, {Szalay}, {Tanaka}, {Thakar}, {Thomas},
  {Tinker}, {Tofflemire}, {Tojeiro}, {Tremonti}, {Vandenberg}, {Vargas
  Maga{\~n}a}, {Verde}, {Vogt}, {Wake}, {Wang}, {Weaver}, {Weinberg}, {White},
  {White}, {Yanny}, {Yasuda}, {Yeche}, \& {Zehavi}}]{2011ApJS..193...29A}
{Aihara}, H., {Allende Prieto}, C., {An}, D., {et~al.} 2011, \apjs, 193, 29

\bibitem[{{An} {et~al.}(2014){An}, {Zheng}, {Wang}, {Huang}, {Kong}, {Wang},
  {Fang}, {Zhu}, {Gu}, {Wu}, {Hao}, \& {Xia}}]{2014ApJ...784..152A}
{An}, F.~X., {Zheng}, X.~Z., {Wang}, W.-H., {et~al.} 2014, \apj, 784, 152

\bibitem[{{Baldwin} {et~al.}(1981){Baldwin}, {Phillips}, \&
  {Terlevich}}]{1981PASP...93....5B}
{Baldwin}, J.~A., {Phillips}, M.~M., \& {Terlevich}, R. 1981, \pasp, 93, 5

\bibitem[{{Balestra} {et~al.}(2010){Balestra}, {Mainieri}, {Popesso},
  {Dickinson}, {Nonino}, {Rosati}, {Teimoorinia}, {Vanzella}, {Cristiani},
  {Cesarsky}, {Fosbury}, {Kuntschner}, \& {Rettura}}]{2010A&A...512A..12B}
{Balestra}, I., {Mainieri}, V., {Popesso}, P., {et~al.} 2010, \aap, 512, A12

\bibitem[{{Barro} {et~al.}(2011{\natexlab{a}}){Barro},
  {P{\'e}rez-Gonz{\'a}lez}, {Gallego}, {Ashby}, {Kajisawa}, {Miyazaki},
  {Villar}, {Yamada}, \& {Zamorano}}]{2011ApJS..193...13B}
{Barro}, G., {P{\'e}rez-Gonz{\'a}lez}, P.~G., {Gallego}, J., {et~al.}
  2011{\natexlab{a}}, \apjs, 193, 13

\bibitem[{{Barro} {et~al.}(2011{\natexlab{b}}){Barro},
  {P{\'e}rez-Gonz{\'a}lez}, {Gallego}, {Ashby}, {Kajisawa}, {Miyazaki},
  {Villar}, {Yamada}, \& {Zamorano}}]{2011ApJS..193...30B}
{Barro}, G., {P{\'e}rez-Gonz{\'a}lez}, P.~G., {Gallego}, J., {et~al.}
  2011{\natexlab{b}}, \apjs, 193, 30

\bibitem[{{Beckwith}(2005)}]{2005yCat.2258....0B}
{Beckwith}, S.~V.~W. 2005, VizieR Online Data Catalog, 2258, 0

\bibitem[{{Behroozi} {et~al.}(2013){Behroozi}, {Wechsler}, \&
  {Conroy}}]{2013ApJ...770...57B}
{Behroozi}, P.~S., {Wechsler}, R.~H., \& {Conroy}, C. 2013, \apj, 770, 57

\bibitem[{{Brinchmann} {et~al.}(2004){Brinchmann}, {Charlot}, {White},
  {Tremonti}, {Kauffmann}, {Heckman}, \& {Brinkmann}}]{2004MNRAS.351.1151B}
{Brinchmann}, J., {Charlot}, S., {White}, S.~D.~M., {et~al.} 2004, \mnras, 351,
  1151

\bibitem[{{Brough} {et~al.}(2011){Brough}, {Hopkins}, {Sharp}, {Gunawardhana},
  {Wijesinghe}, {Robotham}, {Driver}, {Baldry}, {Bamford}, {Liske}, {Loveday},
  {Norberg}, {Peacock}, {Bland-Hawthorn}, {Brown}, {Cameron}, {Croom}, {Frenk},
  {Foster}, {Hill}, {Jones}, {Kelvin}, {Kuijken}, {Nichol}, {Parkinson},
  {Pimbblet}, {Popescu}, {Prescott}, {Sutherland}, {Taylor}, {Thomas}, {Tuffs},
  \& {van Kampen}}]{2011MNRAS.413.1236B}
{Brough}, S., {Hopkins}, A.~M., {Sharp}, R.~G., {et~al.} 2011, \mnras, 413,
  1236

\bibitem[{{Buat} {et~al.}(2005){Buat}, {Iglesias-P{\'a}ramo}, {Seibert},
  {Burgarella}, {Charlot}, {Martin}, {Xu}, {Heckman}, {Boissier}, {Boselli},
  {Barlow}, {Bianchi}, {Byun}, {Donas}, {Forster}, {Friedman}, {Jelinski},
  {Lee}, {Madore}, {Malina}, {Milliard}, {Morissey}, {Neff}, {Rich},
  {Schiminovitch}, {Siegmund}, {Small}, {Szalay}, {Welsh}, \&
  {Wyder}}]{2005ApJ...619L..51B}
{Buat}, V., {Iglesias-P{\'a}ramo}, J., {Seibert}, M., {et~al.} 2005, \apjl,
  619, L51

\bibitem[{{Calzetti} {et~al.}(2000){Calzetti}, {Armus}, {Bohlin}, {Kinney},
  {Koornneef}, \& {Storchi-Bergmann}}]{2000ApJ...533..682C}
{Calzetti}, D., {Armus}, L., {Bohlin}, R.~C., {et~al.} 2000, \apj, 533, 682

\bibitem[{{Casali} {et~al.}(2006){Casali}, {Pirard}, {Kissler-Patig},
  {Moorwood}, {Bedin}, {Biereichel}, {Delabre}, {Dorn}, {Finger}, {Gojak},
  {Huster}, {Jung}, {Koch}, {Lizon}, {Mehrgan}, {Pozna}, {Silber}, {Sokar}, \&
  {Stegmeier}}]{2006SPIE.6269E..0WC}
{Casali}, M., {Pirard}, J.-F., {Kissler-Patig}, M., {et~al.} 2006, in Society
  of Photo-Optical Instrumentation Engineers (SPIE) Conference Series, Vol.
  6269, Society of Photo-Optical Instrumentation Engineers (SPIE) Conference
  Series, 0

\bibitem[{{Cl{\'e}ment} {et~al.}(2012){Cl{\'e}ment}, {Cuby}, {Courbin},
  {Fontana}, {Freudling}, {Fynbo}, {Gallego}, {Hibon}, {Kneib}, {Le F{\`e}vre},
  {Lidman}, {McMahon}, {Milvang-Jensen}, {Moller}, {Moorwood}, {Nilsson},
  {Pentericci}, {Venemans}, {Villar}, \& {Willis}}]{2012A&A...538A..66C}
{Cl{\'e}ment}, B., {Cuby}, J.-G., {Courbin}, F., {et~al.} 2012, \aap, 538, A66

\bibitem[{{Cooper} {et~al.}(2012){Cooper}, {Yan}, {Dickinson}, {Juneau},
  {Lotz}, {Newman}, {Papovich}, {Salim}, {Walth}, {Weiner}, \&
  {Willmer}}]{2012MNRAS.425.2116C}
{Cooper}, M.~C., {Yan}, R., {Dickinson}, M., {et~al.} 2012, \mnras, 425, 2116

\bibitem[{{Cowie} {et~al.}(1996){Cowie}, {Songaila}, {Hu}, \&
  {Cohen}}]{1996AJ....112..839C}
{Cowie}, L.~L., {Songaila}, A., {Hu}, E.~M., \& {Cohen}, J.~G. 1996, \aj, 112,
  839

\bibitem[{{Dahlen} {et~al.}(2013){Dahlen}, {Mobasher}, {Faber}, {Ferguson},
  {Barro}, {Finkelstein}, {Finlator}, {Fontana}, {Gruetzbauch}, {Johnson},
  {Pforr}, {Salvato}, {Wiklind}, {Wuyts}, {Acquaviva}, {Dickinson}, {Guo},
  {Huang}, {Huang}, {Newman}, {Bell}, {Conselice}, {Galametz}, {Gawiser},
  {Giavalisco}, {Grogin}, {Hathi}, {Kocevski}, {Koekemoer}, {Koo}, {Lee},
  {McGrath}, {Papovich}, {Peth}, {Ryan}, {Somerville}, {Weiner}, \&
  {Wilson}}]{2013ApJ...775...93D}
{Dahlen}, T., {Mobasher}, B., {Faber}, S.~M., {et~al.} 2013, \apj, 775, 93

\bibitem[{{Dale} {et~al.}(2010){Dale}, {Barlow}, {Cohen}, {Cook}, {Johnson},
  {Kattner}, {Moore}, {Schuster}, \& {Staudaher}}]{2010ApJ...712L.189D}
{Dale}, D.~A., {Barlow}, R.~J., {Cohen}, S.~A., {et~al.} 2010, \apjl, 712, L189

\bibitem[{{Doherty} {et~al.}(2006){Doherty}, {Bunker}, {Sharp}, {Dalton},
  {Parry}, \& {Lewis}}]{2006MNRAS.370..331D}
{Doherty}, M., {Bunker}, A., {Sharp}, R., {et~al.} 2006, \mnras, 370, 331

\bibitem[{{Dopita} \& {Evans}(1986)}]{1986ApJ...307..431D}
{Dopita}, M.~A. \& {Evans}, I.~N. 1986, \apj, 307, 431

\bibitem[{{Drake} {et~al.}(2013){Drake}, {Simpson}, {Collins}, {James},
  {Baldry}, {Ouchi}, {Jarvis}, {Bonfield}, {Ono}, {Best}, {Dalton}, {Dunlop},
  {McLure}, \& {Smith}}]{2013MNRAS.433..796D}
{Drake}, A.~B., {Simpson}, C., {Collins}, C.~A., {et~al.} 2013, \mnras, 433,
  796

\bibitem[{{Elbaz} {et~al.}(2011){Elbaz}, {Dickinson}, {Hwang},
  {D{\'{\i}}az-Santos}, {Magdis}, {Magnelli}, {Le Borgne}, {Galliano},
  {Pannella}, {Chanial}, {Armus}, {Charmandaris}, {Daddi}, {Aussel}, {Popesso},
  {Kartaltepe}, {Altieri}, {Valtchanov}, {Coia}, {Dannerbauer}, {Dasyra},
  {Leiton}, {Mazzarella}, {Alexander}, {Buat}, {Burgarella}, {Chary}, {Gilli},
  {Ivison}, {Juneau}, {Le Floc'h}, {Lutz}, {Morrison}, {Mullaney}, {Murphy},
  {Pope}, {Scott}, {Brodwin}, {Calzetti}, {Cesarsky}, {Charlot}, {Dole},
  {Eisenhardt}, {Ferguson}, {F{\"o}rster Schreiber}, {Frayer}, {Giavalisco},
  {Huynh}, {Koekemoer}, {Papovich}, {Reddy}, {Surace}, {Teplitz}, {Yun}, \&
  {Wilson}}]{2011A&A...533A.119E}
{Elbaz}, D., {Dickinson}, M., {Hwang}, H.~S., {et~al.} 2011, \aap, 533, A119

\bibitem[{{Fujita} {et~al.}(2003){Fujita}, {Ajiki}, {Shioya}, {Nagao},
  {Murayama}, {Taniguchi}, {Umeda}, {Yamada}, {Yagi}, {Okamura}, \&
  {Komiyama}}]{2003ApJ...586L.115F}
{Fujita}, S.~S., {Ajiki}, M., {Shioya}, Y., {et~al.} 2003, \apjl, 586, L115

\bibitem[{{Fukugita} {et~al.}(1995){Fukugita}, {Shimasaku}, \&
  {Ichikawa}}]{1995PASP..107..945F}
{Fukugita}, M., {Shimasaku}, K., \& {Ichikawa}, T. 1995, \pasp, 107, 945

\bibitem[{{Gallego} {et~al.}(1995){Gallego}, {Zamorano}, {Aragon-Salamanca}, \&
  {Rego}}]{1995ApJ...455L...1G}
{Gallego}, J., {Zamorano}, J., {Aragon-Salamanca}, A., \& {Rego}, M. 1995,
  \apjl, 455, L1

\bibitem[{{Gallego} {et~al.}(1997){Gallego}, {Zamorano}, {Rego}, \&
  {Vitores}}]{1997ApJ...475..502G}
{Gallego}, J., {Zamorano}, J., {Rego}, M., \& {Vitores}, A.~G. 1997, \apj, 475,
  502

\bibitem[{{Geach} {et~al.}(2008){Geach}, {Smail}, {Best}, {Kurk}, {Casali},
  {Ivison}, \& {Coppin}}]{2008MNRAS.388.1473G}
{Geach}, J.~E., {Smail}, I., {Best}, P.~N., {et~al.} 2008, \mnras, 388, 1473

\bibitem[{{Glazebrook} {et~al.}(1999){Glazebrook}, {Blake}, {Economou},
  {Lilly}, \& {Colless}}]{1999MNRAS.306..843G}
{Glazebrook}, K., {Blake}, C., {Economou}, F., {Lilly}, S., \& {Colless}, M.
  1999, \mnras, 306, 843

\bibitem[{{Glazebrook} {et~al.}(2004){Glazebrook}, {Tober}, {Thomson},
  {Bland-Hawthorn}, \& {Abraham}}]{2004AJ....128.2652G}
{Glazebrook}, K., {Tober}, J., {Thomson}, S., {Bland-Hawthorn}, J., \&
  {Abraham}, R. 2004, \aj, 128, 2652

\bibitem[{{Grogin} {et~al.}(2011){Grogin}, {Kocevski}, {Faber}, {Ferguson},
  {Koekemoer}, {Riess}, {Acquaviva}, {Alexander}, {Almaini}, {Ashby}, {Barden},
  {Bell}, {Bournaud}, {Brown}, {Caputi}, {Casertano}, {Cassata}, {Castellano},
  {Challis}, {Chary}, {Cheung}, {Cirasuolo}, {Conselice}, {Roshan Cooray},
  {Croton}, {Daddi}, {Dahlen}, {Dav{\'e}}, {de Mello}, {Dekel}, {Dickinson},
  {Dolch}, {Donley}, {Dunlop}, {Dutton}, {Elbaz}, {Fazio}, {Filippenko},
  {Finkelstein}, {Fontana}, {Gardner}, {Garnavich}, {Gawiser}, {Giavalisco},
  {Grazian}, {Guo}, {Hathi}, {H{\"a}ussler}, {Hopkins}, {Huang}, {Huang},
  {Jha}, {Kartaltepe}, {Kirshner}, {Koo}, {Lai}, {Lee}, {Li}, {Lotz}, {Lucas},
  {Madau}, {McCarthy}, {McGrath}, {McIntosh}, {McLure}, {Mobasher},
  {Moustakas}, {Mozena}, {Nandra}, {Newman}, {Niemi}, {Noeske}, {Papovich},
  {Pentericci}, {Pope}, {Primack}, {Rajan}, {Ravindranath}, {Reddy}, {Renzini},
  {Rix}, {Robaina}, {Rodney}, {Rosario}, {Rosati}, {Salimbeni}, {Scarlata},
  {Siana}, {Simard}, {Smidt}, {Somerville}, {Spinrad}, {Straughn}, {Strolger},
  {Telford}, {Teplitz}, {Trump}, {van der Wel}, {Villforth}, {Wechsler},
  {Weiner}, {Wiklind}, {Wild}, {Wilson}, {Wuyts}, {Yan}, \&
  {Yun}}]{2011ApJS..197...35G}
{Grogin}, N.~A., {Kocevski}, D.~D., {Faber}, S.~M., {et~al.} 2011, \apjs, 197,
  35

\bibitem[{{Gunawardhana} {et~al.}(2013){Gunawardhana}, {Hopkins},
  {Bland-Hawthorn}, {Brough}, {Sharp}, {Loveday}, {Taylor}, {Jones},
  {Lara-L{\'o}pez}, {Bauer}, {Colless}, {Owers}, {Baldry},
  {L{\'o}pez-S{\'a}nchez}, {Foster}, {Bamford}, {Brown}, {Driver},
  {Drinkwater}, {Liske}, {Meyer}, {Norberg}, {Robotham}, {Ching}, {Cluver},
  {Croom}, {Kelvin}, {Prescott}, {Steele}, {Thomas}, \&
  {Wang}}]{2013MNRAS.433.2764G}
{Gunawardhana}, M.~L.~P., {Hopkins}, A.~M., {Bland-Hawthorn}, J., {et~al.}
  2013, \mnras, 433, 2764

\bibitem[{{Gunawardhana} {et~al.}(2015){Gunawardhana}, {Hopkins}, {Taylor},
  {Bland-Hawthorn}, {Norberg}, {Baldry}, {Loveday}, {Owers}, {Wilkins},
  {Colless}, {Brown}, {Driver}, {Alpaslan}, {Brough}, {Cluver}, {Croom},
  {Kelvin}, {Lara-L{\'o}pez}, {Liske}, {L{\'o}pez-S{\'a}nchez}, \&
  {Robotham}}]{2015MNRAS.447..875G}
{Gunawardhana}, M.~L.~P., {Hopkins}, A.~M., {Taylor}, E.~N., {et~al.} 2015,
  \mnras, 447, 875

\bibitem[{{Guo} {et~al.}(2013){Guo}, {Ferguson}, {Giavalisco}, {Barro},
  {Willner}, {Ashby}, {Dahlen}, {Donley}, {Faber}, {Fontana}, {Galametz},
  {Grazian}, {Huang}, {Kocevski}, {Koekemoer}, {Koo}, {McGrath}, {Peth},
  {Salvato}, {Wuyts}, {Castellano}, {Cooray}, {Dickinson}, {Dunlop}, {Fazio},
  {Gardner}, {Gawiser}, {Grogin}, {Hathi}, {Hsu}, {Lee}, {Lucas}, {Mobasher},
  {Nandra}, {Newman}, \& {van der Wel}}]{2013ApJS..207...24G}
{Guo}, Y., {Ferguson}, H.~C., {Giavalisco}, M., {et~al.} 2013, \apjs, 207, 24

\bibitem[{{Guzm{\'a}n} {et~al.}(1997){Guzm{\'a}n}, {Gallego}, {Koo},
  {Phillips}, {Lowenthal}, {Faber}, {Illingworth}, \&
  {Vogt}}]{1997ApJ...489..559G}
{Guzm{\'a}n}, R., {Gallego}, J., {Koo}, D.~C., {et~al.} 1997, \apj, 489, 559

\bibitem[{{Hanish} {et~al.}(2006){Hanish}, {Meurer}, {Ferguson}, {Zwaan},
  {Heckman}, {Staveley-Smith}, {Bland-Hawthorn}, {Kilborn}, {Koribalski},
  {Putman}, {Ryan-Weber}, {Oey}, {Kennicutt}, {Knezek}, {Meyer}, {Smith},
  {Webster}, {Dopita}, {Doyle}, {Drinkwater}, {Freeman}, \&
  {Werk}}]{2006ApJ...649..150H}
{Hanish}, D.~J., {Meurer}, G.~R., {Ferguson}, H.~C., {et~al.} 2006, \apj, 649,
  150

\bibitem[{{Hao} {et~al.}(2011){Hao}, {Kennicutt}, {Johnson}, {Calzetti},
  {Dale}, \& {Moustakas}}]{2011ApJ...741..124H}
{Hao}, C.-N., {Kennicutt}, R.~C., {Johnson}, B.~D., {et~al.} 2011, \apj, 741,
  124

\bibitem[{{Hayes} {et~al.}(2010){Hayes}, {Schaerer}, \&
  {{\"O}stlin}}]{2010A&A...509L...5H}
{Hayes}, M., {Schaerer}, D., \& {{\"O}stlin}, G. 2010, \aap, 509, L5

\bibitem[{{Hippelein} {et~al.}(2003){Hippelein}, {Maier}, {Meisenheimer},
  {Wolf}, {Fried}, {von Kuhlmann}, {K{\"u}mmel}, {Phleps}, \&
  {R{\"o}ser}}]{2003A&A...402...65H}
{Hippelein}, H., {Maier}, C., {Meisenheimer}, K., {et~al.} 2003, \aap, 402, 65

\bibitem[{{Hopkins}(2004)}]{2004ApJ...615..209H}
{Hopkins}, A.~M. 2004, \apj, 615, 209

\bibitem[{{Hopkins} \& {Beacom}(2006)}]{2006ApJ...651..142H}
{Hopkins}, A.~M. \& {Beacom}, J.~F. 2006, \apj, 651, 142

\bibitem[{{Hopkins} {et~al.}(2000){Hopkins}, {Connolly}, \&
  {Szalay}}]{2000AJ....120.2843H}
{Hopkins}, A.~M., {Connolly}, A.~J., \& {Szalay}, A.~S. 2000, \aj, 120, 2843

\bibitem[{{James} {et~al.}(2008){James}, {Knapen}, {Shane}, {Baldry}, \& {de
  Jong}}]{2008A&A...482..507J}
{James}, P.~A., {Knapen}, J.~H., {Shane}, N.~S., {Baldry}, I.~K., \& {de Jong},
  R.~S. 2008, \aap, 482, 507

\bibitem[{{Karachentsev} \& {Kaisin}(2010)}]{2010AJ....140.1241K}
{Karachentsev}, I.~D. \& {Kaisin}, S.~S. 2010, \aj, 140, 1241

\bibitem[{{Kartaltepe} {et~al.}(2015){Kartaltepe}, {Mozena}, {Kocevski},
  {McIntosh}, {Lotz}, {Bell}, {Faber}, {Ferguson}, {Koo}, {Bassett}, {Bernyk},
  {Blancato}, {Bournaud}, {Cassata}, {Castellano}, {Cheung}, {Conselice},
  {Croton}, {Dahlen}, {de Mello}, {DeGroot}, {Donley}, {Guedes}, {Grogin},
  {Hathi}, {Hilton}, {Hollon}, {Koekemoer}, {Liu}, {Lucas}, {Martig},
  {McGrath}, {McPartland}, {Mobasher}, {Morlock}, {O'Leary}, {Peth}, {Pforr},
  {Pillepich}, {Rosario}, {Soto}, {Straughn}, {Telford}, {Sunnquist}, {Trump},
  {Weiner}, \& {Wuyts}}]{2015ApJS..221...11K}
{Kartaltepe}, J.~S., {Mozena}, M., {Kocevski}, D., {et~al.} 2015, \apjs, 221,
  11

\bibitem[{{Kennicutt} \& {Evans}(2012)}]{2012ARA&A..50..531K}
{Kennicutt}, R.~C. \& {Evans}, N.~J. 2012, \araa, 50, 531

\bibitem[{{Kennicutt}(1992)}]{1992ApJ...388..310K}
{Kennicutt}, Jr., R.~C. 1992, \apj, 388, 310

\bibitem[{{Kennicutt}(1998)}]{1998ARA&A..36..189K}
{Kennicutt}, Jr., R.~C. 1998, \araa, 36, 189

\bibitem[{{Kissler-Patig} {et~al.}(2008){Kissler-Patig}, {Pirard}, {Casali},
  {Moorwood}, {Ageorges}, {Alves de Oliveira}, {Baksai}, {Bedin}, {Bendek},
  {Biereichel}, {Delabre}, {Dorn}, {Esteves}, {Finger}, {Gojak}, {Huster},
  {Jung}, {Kiekebush}, {Klein}, {Koch}, {Lizon}, {Mehrgan}, {Petr-Gotzens},
  {Pritchard}, {Selman}, \& {Stegmeier}}]{2008A&A...491..941K}
{Kissler-Patig}, M., {Pirard}, J.-F., {Casali}, M., {et~al.} 2008, \aap, 491,
  941

\bibitem[{{Kochiashvili} {et~al.}(2015){Kochiashvili}, {M{\o}ller},
  {Milvang-Jensen}, {Christensen}, {Fynbo}, {Freudling}, {Cl{\'e}ment}, {Cuby},
  {Zabl}, \& {Zibetti}}]{2015A&A...580A..42K}
{Kochiashvili}, I., {M{\o}ller}, P., {Milvang-Jensen}, B., {et~al.} 2015, \aap,
  580, A42

\bibitem[{{Koekemoer} {et~al.}(2011){Koekemoer}, {Faber}, {Ferguson}, {Grogin},
  {Kocevski}, {Koo}, {Lai}, {Lotz}, {Lucas}, {McGrath}, {Ogaz}, {Rajan},
  {Riess}, {Rodney}, {Strolger}, {Casertano}, {Castellano}, {Dahlen},
  {Dickinson}, {Dolch}, {Fontana}, {Giavalisco}, {Grazian}, {Guo}, {Hathi},
  {Huang}, {van der Wel}, {Yan}, {Acquaviva}, {Alexander}, {Almaini}, {Ashby},
  {Barden}, {Bell}, {Bournaud}, {Brown}, {Caputi}, {Cassata}, {Challis},
  {Chary}, {Cheung}, {Cirasuolo}, {Conselice}, {Roshan Cooray}, {Croton},
  {Daddi}, {Dav{\'e}}, {de Mello}, {de Ravel}, {Dekel}, {Donley}, {Dunlop},
  {Dutton}, {Elbaz}, {Fazio}, {Filippenko}, {Finkelstein}, {Frazer}, {Gardner},
  {Garnavich}, {Gawiser}, {Gruetzbauch}, {Hartley}, {H{\"a}ussler},
  {Herrington}, {Hopkins}, {Huang}, {Jha}, {Johnson}, {Kartaltepe},
  {Khostovan}, {Kirshner}, {Lani}, {Lee}, {Li}, {Madau}, {McCarthy},
  {McIntosh}, {McLure}, {McPartland}, {Mobasher}, {Moreira}, {Mortlock},
  {Moustakas}, {Mozena}, {Nandra}, {Newman}, {Nielsen}, {Niemi}, {Noeske},
  {Papovich}, {Pentericci}, {Pope}, {Primack}, {Ravindranath}, {Reddy},
  {Renzini}, {Rix}, {Robaina}, {Rosario}, {Rosati}, {Salimbeni}, {Scarlata},
  {Siana}, {Simard}, {Smidt}, {Snyder}, {Somerville}, {Spinrad}, {Straughn},
  {Telford}, {Teplitz}, {Trump}, {Vargas}, {Villforth}, {Wagner}, {Wandro},
  {Wechsler}, {Weiner}, {Wiklind}, {Wild}, {Wilson}, {Wuyts}, \&
  {Yun}}]{2011ApJS..197...36K}
{Koekemoer}, A.~M., {Faber}, S.~M., {Ferguson}, H.~C., {et~al.} 2011, \apjs,
  197, 36

\bibitem[{{Konishi} {et~al.}(2011){Konishi}, {Akiyama}, {Kajisawa}, {Ichikawa},
  {Suzuki}, {Tokoku}, {Katsuno Uchimoto}, {Yoshikawa}, {Tanaka}, {Onodera},
  {Ouchi}, {Omata}, {Nishimura}, \& {Yamada}}]{2011PASJ...63S.363K}
{Konishi}, M., {Akiyama}, M., {Kajisawa}, M., {et~al.} 2011, \pasj, 63, 363

\bibitem[{{Law} {et~al.}(2012){Law}, {Steidel}, {Shapley}, {Nagy}, {Reddy}, \&
  {Erb}}]{2012ApJ...745...85L}
{Law}, D.~R., {Steidel}, C.~C., {Shapley}, A.~E., {et~al.} 2012, \apj, 745, 85

\bibitem[{{Le F{\`e}vre} {et~al.}(2004){Le F{\`e}vre}, {Vettolani}, {Paltani},
  {Tresse}, {Zamorani}, {Le Brun}, {Moreau}, {Bottini}, {Maccagni}, {Picat},
  {Scaramella}, {Scodeggio}, {Zanichelli}, {Adami}, {Arnouts}, {Bardelli},
  {Bolzonella}, {Cappi}, {Charlot}, {Contini}, {Foucaud}, {Franzetti},
  {Garilli}, {Gavignaud}, {Guzzo}, {Ilbert}, {Iovino}, {McCracken}, {Mancini},
  {Marano}, {Marinoni}, {Mathez}, {Mazure}, {Meneux}, {Merighi}, {Pell{\`o}},
  {Pollo}, {Pozzetti}, {Radovich}, {Zucca}, {Arnaboldi}, {Bondi}, {Bongiorno},
  {Busarello}, {Ciliegi}, {Gregorini}, {Mellier}, {Merluzzi}, {Ripepi}, \&
  {Rizzo}}]{2004A&A...428.1043L}
{Le F{\`e}vre}, O., {Vettolani}, G., {Paltani}, S., {et~al.} 2004, \aap, 428,
  1043

\bibitem[{{Lilly} {et~al.}(1996){Lilly}, {Le Fevre}, {Hammer}, \&
  {Crampton}}]{1996ApJ...460L...1L}
{Lilly}, S.~J., {Le Fevre}, O., {Hammer}, F., \& {Crampton}, D. 1996, \apjl,
  460, L1

\bibitem[{{Ly} {et~al.}(2011){Ly}, {Lee}, {Dale}, {Momcheva}, {Salim},
  {Staudaher}, {Moore}, \& {Finn}}]{2011ApJ...726..109L}
{Ly}, C., {Lee}, J.~C., {Dale}, D.~A., {et~al.} 2011, \apj, 726, 109

\bibitem[{{Ly} {et~al.}(2007){Ly}, {Malkan}, {Kashikawa}, {Shimasaku}, {Doi},
  {Nagao}, {Iye}, {Kodama}, {Morokuma}, \& {Motohara}}]{2007ApJ...657..738L}
{Ly}, C., {Malkan}, M.~A., {Kashikawa}, N., {et~al.} 2007, \apj, 657, 738

\bibitem[{{Madau} \& {Dickinson}(2014)}]{2014ARA&A..52..415M}
{Madau}, P. \& {Dickinson}, M. 2014, \araa, 52, 415

\bibitem[{{Madau} {et~al.}(1996){Madau}, {Ferguson}, {Dickinson}, {Giavalisco},
  {Steidel}, \& {Fruchter}}]{1996MNRAS.283.1388M}
{Madau}, P., {Ferguson}, H.~C., {Dickinson}, M.~E., {et~al.} 1996, \mnras, 283,
  1388

\bibitem[{{Magnelli} {et~al.}(2013){Magnelli}, {Popesso}, {Berta}, {Pozzi},
  {Elbaz}, {Lutz}, {Dickinson}, {Altieri}, {Andreani}, {Aussel},
  {B{\'e}thermin}, {Bongiovanni}, {Cepa}, {Charmandaris}, {Chary}, {Cimatti},
  {Daddi}, {F{\"o}rster Schreiber}, {Genzel}, {Gruppioni}, {Harwit}, {Hwang},
  {Ivison}, {Magdis}, {Maiolino}, {Murphy}, {Nordon}, {Pannella}, {P{\'e}rez
  Garc{\'{\i}}a}, {Poglitsch}, {Rosario}, {Sanchez-Portal}, {Santini}, {Scott},
  {Sturm}, {Tacconi}, \& {Valtchanov}}]{2013A&A...553A.132M}
{Magnelli}, B., {Popesso}, P., {Berta}, S., {et~al.} 2013, \aap, 553, A132

\bibitem[{{Marino} {et~al.}(2013){Marino}, {Rosales-Ortega}, {S{\'a}nchez},
  {Gil de Paz}, {V{\'{\i}}lchez}, {Miralles-Caballero}, {Kehrig},
  {P{\'e}rez-Montero}, {Stanishev}, {Iglesias-P{\'a}ramo}, {D{\'{\i}}az},
  {Castillo-Morales}, {Kennicutt}, {L{\'o}pez-S{\'a}nchez}, {Galbany},
  {Garc{\'{\i}}a-Benito}, {Mast}, {Mendez-Abreu}, {Monreal-Ibero}, {Husemann},
  {Walcher}, {Garc{\'{\i}}a-Lorenzo}, {Masegosa}, {Del Olmo Orozco},
  {Mour{\~a}o}, {Ziegler}, {Moll{\'a}}, {Papaderos},
  {S{\'a}nchez-Bl{\'a}zquez}, {Gonz{\'a}lez Delgado}, {Falc{\'o}n-Barroso},
  {Roth}, {van de Ven}, \& {Califa Team}}]{2013A&A...559A.114M}
{Marino}, R.~A., {Rosales-Ortega}, F.~F., {S{\'a}nchez}, S.~F., {et~al.} 2013,
  \aap, 559, A114

\bibitem[{{Moorwood} {et~al.}(2000){Moorwood}, {van der Werf}, {Cuby}, \&
  {Oliva}}]{2000A&A...362....9M}
{Moorwood}, A.~F.~M., {van der Werf}, P.~P., {Cuby}, J.~G., \& {Oliva}, E.
  2000, \aap, 362, 9

\bibitem[{{Morioka} {et~al.}(2008){Morioka}, {Nakajima}, {Taniguchi}, {Shioya},
  {Murayama}, \& {Sasaki}}]{2008PASJ...60.1219M}
{Morioka}, T., {Nakajima}, A., {Taniguchi}, Y., {et~al.} 2008, \pasj, 60, 1219

\bibitem[{{Mu{\~n}oz-Mateos} {et~al.}(2009){Mu{\~n}oz-Mateos}, {Gil de Paz},
  {Boissier}, {Zamorano}, {Dale}, {P{\'e}rez-Gonz{\'a}lez}, {Gallego},
  {Madore}, {Bendo}, {Thornley}, {Draine}, {Boselli}, {Buat}, {Calzetti},
  {Moustakas}, \& {Kennicutt}}]{2009ApJ...701.1965M}
{Mu{\~n}oz-Mateos}, J.~C., {Gil de Paz}, A., {Boissier}, S., {et~al.} 2009,
  \apj, 701, 1965

\bibitem[{{Nakamura} {et~al.}(2004){Nakamura}, {Fukugita}, {Brinkmann}, \&
  {Schneider}}]{2004AJ....127.2511N}
{Nakamura}, O., {Fukugita}, M., {Brinkmann}, J., \& {Schneider}, D.~P. 2004,
  \aj, 127, 2511

\bibitem[{{Oke}(1974)}]{1974ApJS...27...21O}
{Oke}, J.~B. 1974, \apjs, 27, 21

\bibitem[{{Oteo} {et~al.}(2015){Oteo}, {Sobral}, {Ivison}, {Smail}, {Best},
  {Cepa}, \& {P{\'e}rez-Garc{\'{\i}}a}}]{2015MNRAS.452.2018O}
{Oteo}, I., {Sobral}, D., {Ivison}, R.~J., {et~al.} 2015, \mnras, 452, 2018

\bibitem[{{Pascual}(2005)}]{2005PASP..117..120P}
{Pascual}, S. 2005, \pasp, 117, 120

\bibitem[{Pascual(2015)}]{sergio_pascual_2015_17810}
Pascual, S. 2015, milia: version 1.0.0

\bibitem[{{Pascual} {et~al.}(2001){Pascual}, {Gallego}, {Arag{\'o}n-Salamanca},
  \& {Zamorano}}]{2001A&A...379..798P}
{Pascual}, S., {Gallego}, J., {Arag{\'o}n-Salamanca}, A., \& {Zamorano}, J.
  2001, \aap, 379, 798

\bibitem[{{Pascual} {et~al.}(2007){Pascual}, {Gallego}, \&
  {Zamorano}}]{2007PASP..119...30P}
{Pascual}, S., {Gallego}, J., \& {Zamorano}, J. 2007, \pasp, 119, 30

\bibitem[{{Patel} {et~al.}(2012){Patel}, {Holden}, {Kelson}, {Franx}, {van der
  Wel}, \& {Illingworth}}]{2012ApJ...748L..27P}
{Patel}, S.~G., {Holden}, B.~P., {Kelson}, D.~D., {et~al.} 2012, \apjl, 748,
  L27

\bibitem[{{P{\'e}rez-Gonz{\'a}lez} {et~al.}(2010){P{\'e}rez-Gonz{\'a}lez},
  {Egami}, {Rex}, {Rawle}, {Kneib}, {Richard}, {Johansson}, {Altieri}, {Blain},
  {Bock}, {Boone}, {Bridge}, {Chung}, {Cl{\'e}ment}, {Clowe}, {Combes}, {Cuby},
  {Dessauges-Zavadsky}, {Dowell}, {Espino-Briones}, {Fadda}, {Fiedler},
  {Gonzalez}, {Horellou}, {Ilbert}, {Ivison}, {Jauzac}, {Lutz}, {Pell{\'o}},
  {Pereira}, {Rieke}, {Rodighiero}, {Schaerer}, {Smith}, {Valtchanov}, {Walth},
  {van der Werf}, {Werner}, \& {Zemcov}}]{2010A&A...518L..15P}
{P{\'e}rez-Gonz{\'a}lez}, P.~G., {Egami}, E., {Rex}, M., {et~al.} 2010, \aap,
  518, L15

\bibitem[{{P{\'e}rez-Gonz{\'a}lez} {et~al.}(2005){P{\'e}rez-Gonz{\'a}lez},
  {Rieke}, {Egami}, {Alonso-Herrero}, {Dole}, {Papovich}, {Blaylock}, {Jones},
  {Rieke}, {Rigby}, {Barmby}, {Fazio}, {Huang}, \&
  {Martin}}]{2005ApJ...630...82P}
{P{\'e}rez-Gonz{\'a}lez}, P.~G., {Rieke}, G.~H., {Egami}, E., {et~al.} 2005,
  \apj, 630, 82

\bibitem[{{P{\'e}rez-Gonz{\'a}lez}
  {et~al.}(2008{\natexlab{a}}){P{\'e}rez-Gonz{\'a}lez}, {Rieke}, {Villar},
  {Barro}, {Blaylock}, {Egami}, {Gallego}, {Gil de Paz}, {Pascual}, {Zamorano},
  \& {Donley}}]{2008ApJ...675..234P}
{P{\'e}rez-Gonz{\'a}lez}, P.~G., {Rieke}, G.~H., {Villar}, V., {et~al.}
  2008{\natexlab{a}}, \apj, 675, 234

\bibitem[{{P{\'e}rez-Gonz{\'a}lez}
  {et~al.}(2008{\natexlab{b}}){P{\'e}rez-Gonz{\'a}lez}, {Trujillo}, {Barro},
  {Gallego}, {Zamorano}, \& {Conselice}}]{2008ApJ...687...50P}
{P{\'e}rez-Gonz{\'a}lez}, P.~G., {Trujillo}, I., {Barro}, G., {et~al.}
  2008{\natexlab{b}}, \apj, 687, 50

\bibitem[{{P{\'e}rez-Gonz{\'a}lez} {et~al.}(2003){P{\'e}rez-Gonz{\'a}lez},
  {Zamorano}, {Gallego}, {Arag{\'o}n-Salamanca}, \& {Gil de
  Paz}}]{2003ApJ...591..827P}
{P{\'e}rez-Gonz{\'a}lez}, P.~G., {Zamorano}, J., {Gallego}, J.,
  {Arag{\'o}n-Salamanca}, A., \& {Gil de Paz}, A. 2003, \apj, 591, 827

\bibitem[{{Pettini} \& {Pagel}(2004)}]{2004MNRAS.348L..59P}
{Pettini}, M. \& {Pagel}, B.~E.~J. 2004, \mnras, 348, L59

\bibitem[{{Phillips} {et~al.}(1997){Phillips}, {Guzm{\'a}n}, {Gallego}, {Koo},
  {Lowenthal}, {Vogt}, {Faber}, \& {Illingworth}}]{1997ApJ...489..543P}
{Phillips}, A.~C., {Guzm{\'a}n}, R., {Gallego}, J., {et~al.} 1997, \apj, 489,
  543

\bibitem[{{Pirard} {et~al.}(2004){Pirard}, {Kissler-Patig}, {Moorwood},
  {Biereichel}, {Delabre}, {Dorn}, {Finger}, {Gojak}, {Huster}, {Jung}, {Koch},
  {Le Louarn}, {Lizon}, {Mehrgan}, {Pozna}, {Silber}, {Sokar}, \&
  {Stegmeier}}]{2004SPIE.5492.1763P}
{Pirard}, J.-F., {Kissler-Patig}, M., {Moorwood}, A., {et~al.} 2004, in Society
  of Photo-Optical Instrumentation Engineers (SPIE) Conference Series, Vol.
  5492, Ground-based Instrumentation for Astronomy, ed. A.~F.~M. {Moorwood} \&
  M.~{Iye}, 1763--1772

\bibitem[{{Rodr{\'{\i}}guez-Mu{\~n}oz}
  {et~al.}(2015){Rodr{\'{\i}}guez-Mu{\~n}oz}, {Gallego}, {Pacifici}, {Tresse},
  {Charlot}, {Gil de Paz}, {Barro}, \& {Villar}}]{2015ApJ...799...36R}
{Rodr{\'{\i}}guez-Mu{\~n}oz}, L., {Gallego}, J., {Pacifici}, C., {et~al.} 2015,
  \apj, 799, 36

\bibitem[{{Santini} {et~al.}(2015){Santini}, {Ferguson}, {Fontana}, {Mobasher},
  {Barro}, {Castellano}, {Finkelstein}, {Grazian}, {Hsu}, {Lee}, {Lee},
  {Pforr}, {Salvato}, {Wiklind}, {Wuyts}, {Almaini}, {Cooper}, {Galametz},
  {Weiner}, {Amorin}, {Boutsia}, {Conselice}, {Dahlen}, {Dickinson},
  {Giavalisco}, {Grogin}, {Guo}, {Hathi}, {Kocevski}, {Koekemoer},
  {Kurczynski}, {Merlin}, {Mortlock}, {Newman}, {Paris}, {Pentericci},
  {Simons}, \& {Willner}}]{2015ApJ...801...97S}
{Santini}, P., {Ferguson}, H.~C., {Fontana}, A., {et~al.} 2015, \apj, 801, 97

\bibitem[{{Schechter}(1976)}]{1976ApJ...203..297S}
{Schechter}, P. 1976, \apj, 203, 297

\bibitem[{{Schlafly} \& {Finkbeiner}(2011)}]{2011ApJ...737..103S}
{Schlafly}, E.~F. \& {Finkbeiner}, D.~P. 2011, \apj, 737, 103

\bibitem[{{Schmidt}(1968)}]{1968ApJ...151..393S}
{Schmidt}, M. 1968, \apj, 151, 393

\bibitem[{{Shim} {et~al.}(2009){Shim}, {Colbert}, {Teplitz}, {Henry}, {Malkan},
  {McCarthy}, \& {Yan}}]{2009ApJ...696..785S}
{Shim}, H., {Colbert}, J., {Teplitz}, H., {et~al.} 2009, \apj, 696, 785

\bibitem[{{Shioya} {et~al.}(2008){Shioya}, {Taniguchi}, {Sasaki}, {Nagao},
  {Murayama}, {Takahashi}, {Ajiki}, {Ideue}, {Mihara}, {Nakajima}, {Scoville},
  {Mobasher}, {Aussel}, {Giavalisco}, {Guzzo}, {Hasinger}, {Impey}, {Le
  F{\`e}vre}, {Lilly}, {Renzini}, {Rich}, {Sanders}, {Schinnerer}, {Shopbell},
  {Leauthaud}, {Kneib}, {Rhodes}, \& {Massey}}]{2008ApJS..175..128S}
{Shioya}, Y., {Taniguchi}, Y., {Sasaki}, S.~S., {et~al.} 2008, \apjs, 175, 128

\bibitem[{{Siebenmorgen} {et~al.}(2011){Siebenmorgen}, {Carraro}, {Valenti},
  {Petr-Gotzens}, {Brammer}, {Garcia}, \& {Casali}}]{2011Msngr.144....9S}
{Siebenmorgen}, R., {Carraro}, G., {Valenti}, E., {et~al.} 2011, The Messenger,
  144, 9

\bibitem[{{Skelton} {et~al.}(2014){Skelton}, {Whitaker}, {Momcheva}, {Brammer},
  {van Dokkum}, {Labb{\'e}}, {Franx}, {van der Wel}, {Bezanson}, {Da Cunha},
  {Fumagalli}, {F{\"o}rster Schreiber}, {Kriek}, {Leja}, {Lundgren}, {Magee},
  {Marchesini}, {Maseda}, {Nelson}, {Oesch}, {Pacifici}, {Patel}, {Price},
  {Rix}, {Tal}, {Wake}, \& {Wuyts}}]{2014ApJS..214...24S}
{Skelton}, R.~E., {Whitaker}, K.~E., {Momcheva}, I.~G., {et~al.} 2014, \apjs,
  214, 24

\bibitem[{{Sobral} {et~al.}(2009){Sobral}, {Best}, {Geach}, {Smail}, {Kurk},
  {Cirasuolo}, {Casali}, {Ivison}, {Coppin}, \& {Dalton}}]{2009MNRAS.398...75S}
{Sobral}, D., {Best}, P.~N., {Geach}, J.~E., {et~al.} 2009, \mnras, 398, 75

\bibitem[{{Sobral} {et~al.}(2012){Sobral}, {Best}, {Matsuda}, {Smail}, {Geach},
  \& {Cirasuolo}}]{2012MNRAS.420.1926S}
{Sobral}, D., {Best}, P.~N., {Matsuda}, Y., {et~al.} 2012, \mnras, 420, 1926

\bibitem[{{Sobral} {et~al.}(2015){Sobral}, {Matthee}, {Best}, {Smail},
  {Khostovan}, {Milvang-Jensen}, {Kim}, {Stott}, {Calhau}, {Nayyeri}, \&
  {Mobasher}}]{2015MNRAS.451.2303S}
{Sobral}, D., {Matthee}, J., {Best}, P.~N., {et~al.} 2015, \mnras, 451, 2303

\bibitem[{{Sobral} {et~al.}(2013){Sobral}, {Smail}, {Best}, {Geach}, {Matsuda},
  {Stott}, {Cirasuolo}, \& {Kurk}}]{2013MNRAS.428.1128S}
{Sobral}, D., {Smail}, I., {Best}, P.~N., {et~al.} 2013, \mnras, 428, 1128

\bibitem[{{Stroe} \& {Sobral}(2015)}]{2015MNRAS.453..242S}
{Stroe}, A. \& {Sobral}, D. 2015, \mnras, 453, 242

\bibitem[{{Sullivan} {et~al.}(2000){Sullivan}, {Treyer}, {Ellis}, {Bridges},
  {Milliard}, \& {Donas}}]{2000MNRAS.312..442S}
{Sullivan}, M., {Treyer}, M.~A., {Ellis}, R.~S., {et~al.} 2000, \mnras, 312,
  442

\bibitem[{{Tadaki} {et~al.}(2011){Tadaki}, {Kodama}, {Koyama}, {Hayashi},
  {Tanaka}, \& {Tokoku}}]{2011PASJ...63S.437T}
{Tadaki}, K.-I., {Kodama}, T., {Koyama}, Y., {et~al.} 2011, \pasj, 63, 437

\bibitem[{{Tresse} \& {Maddox}(1998)}]{1998ApJ...495..691T}
{Tresse}, L. \& {Maddox}, S.~J. 1998, \apj, 495, 691

\bibitem[{{Tresse} {et~al.}(2002){Tresse}, {Maddox}, {Le F{\`e}vre}, \&
  {Cuby}}]{2002MNRAS.337..369T}
{Tresse}, L., {Maddox}, S.~J., {Le F{\`e}vre}, O., \& {Cuby}, J.-G. 2002,
  \mnras, 337, 369

\bibitem[{{van der Wel} {et~al.}(2012){van der Wel}, {Bell}, {H{\"a}ussler},
  {McGrath}, {Chang}, {Guo}, {McIntosh}, {Rix}, {Barden}, {Cheung}, {Faber},
  {Ferguson}, {Galametz}, {Grogin}, {Hartley}, {Kartaltepe}, {Kocevski},
  {Koekemoer}, {Lotz}, {Mozena}, {Peth}, \& {Peng}}]{2012ApJS..203...24V}
{van der Wel}, A., {Bell}, E.~F., {H{\"a}ussler}, B., {et~al.} 2012, \apjs,
  203, 24

\bibitem[{{Vanzella} {et~al.}(2008){Vanzella}, {Cristiani}, {Dickinson},
  {Giavalisco}, {Kuntschner}, {Haase}, {Nonino}, {Rosati}, {Cesarsky},
  {Ferguson}, {Fosbury}, {Grazian}, {Moustakas}, {Rettura}, {Popesso},
  {Renzini}, {Stern}, \& {GOODS Team}}]{2008A&A...478...83V}
{Vanzella}, E., {Cristiani}, S., {Dickinson}, M., {et~al.} 2008, \aap, 478, 83

\bibitem[{{Veilleux} \& {Osterbrock}(1987)}]{1987ApJS...63..295V}
{Veilleux}, S. \& {Osterbrock}, D.~E. 1987, \apjs, 63, 295

\bibitem[{{Vilella-Rojo} {et~al.}(2015){Vilella-Rojo}, {Viironen},
  {L{\'o}pez-Sanjuan}, {Cenarro}, {Varela}, {D{\'{\i}}az-Garc{\'{\i}}a},
  {Crist{\'o}bal-Hornillos}, {Ederoclite}, {Mar{\'{\i}}n-Franch}, \&
  {Moles}}]{2015A&A...580A..47V}
{Vilella-Rojo}, G., {Viironen}, K., {L{\'o}pez-Sanjuan}, C., {et~al.} 2015,
  \aap, 580, A47

\bibitem[{{Villar} {et~al.}(2008){Villar}, {Gallego}, {P{\'e}rez-Gonz{\'a}lez},
  {Pascual}, {Noeske}, {Koo}, {Barro}, \& {Zamorano}}]{2008ApJ...677..169V}
{Villar}, V., {Gallego}, J., {P{\'e}rez-Gonz{\'a}lez}, P.~G., {et~al.} 2008,
  \apj, 677, 169

\bibitem[{{Vitores} {et~al.}(1996){Vitores}, {Zamorano}, {Rego}, {Gallego}, \&
  {Alonso}}]{1996A&AS..120..385V}
{Vitores}, A.~G., {Zamorano}, J., {Rego}, M., {Gallego}, J., \& {Alonso}, O.
  1996, \aaps, 120, 385

\bibitem[{{Westra} {et~al.}(2010){Westra}, {Geller}, {Kurtz}, {Fabricant}, \&
  {Dell'Antonio}}]{2010ApJ...708..534W}
{Westra}, E., {Geller}, M.~J., {Kurtz}, M.~J., {Fabricant}, D.~G., \&
  {Dell'Antonio}, I. 2010, \apj, 708, 534

\bibitem[{{Westra} \& {Jones}(2008)}]{2008MNRAS.383..339W}
{Westra}, E. \& {Jones}, D.~H. 2008, \mnras, 383, 339

\bibitem[{{Williams} {et~al.}(2009){Williams}, {Quadri}, {Franx}, {van Dokkum},
  \& {Labb{\'e}}}]{2009ApJ...691.1879W}
{Williams}, R.~J., {Quadri}, R.~F., {Franx}, M., {van Dokkum}, P., \&
  {Labb{\'e}}, I. 2009, \apj, 691, 1879

\bibitem[{{Xue} {et~al.}(2011){Xue}, {Luo}, {Brandt}, {Bauer}, {Lehmer},
  {Broos}, {Schneider}, {Alexander}, {Brusa}, {Comastri}, {Fabian}, {Gilli},
  {Hasinger}, {Hornschemeier}, {Koekemoer}, {Liu}, {Mainieri}, {Paolillo},
  {Rafferty}, {Rosati}, {Shemmer}, {Silverman}, {Smail}, {Tozzi}, \&
  {Vignali}}]{2011ApJS..195...10X}
{Xue}, Y.~Q., {Luo}, B., {Brandt}, W.~N., {et~al.} 2011, \apjs, 195, 10

\bibitem[{{Yan} {et~al.}(1999){Yan}, {McCarthy}, {Freudling}, {Teplitz},
  {Malumuth}, {Weymann}, \& {Malkan}}]{1999ApJ...519L..47Y}
{Yan}, L., {McCarthy}, P.~J., {Freudling}, W., {et~al.} 1999, \apjl, 519, L47

\bibitem[{{Zamorano} {et~al.}(1996){Zamorano}, {Gallego}, {Rego}, {Vitores}, \&
  {Alonso}}]{1996ApJS..105..343Z}
{Zamorano}, J., {Gallego}, J., {Rego}, M., {Vitores}, A.~G., \& {Alonso}, O.
  1996, \apjs, 105, 343

\bibitem[{{Zamorano} {et~al.}(1994){Zamorano}, {Rego}, {Gallego}, {Vitores},
  {Gonzalez-Riestra}, \& {Rodriguez-Caderot}}]{1994ApJS...95..387Z}
{Zamorano}, J., {Rego}, M., {Gallego}, J.~G., {et~al.} 1994, \apjs, 95, 387

\end{thebibliography}

\end{document}